\documentclass[12pt]{article}
\usepackage{latexsym}
\usepackage{amsmath}
\usepackage{amssymb}
\usepackage{epsfig,graphics}
\usepackage{graphicx}

\newcommand{\be}{\begin{equation}}
\newcommand{\ee}{\end{equation}}

\begin{document}

\begin{titlepage}

\vspace*{0.6in}

\begin{center}
  {\large\bf The glueball spectrum of SU(3) gauge theory in 3+1 dimensions}\\
\vspace*{0.75in}
{Andreas Athenodorou$^{a}$ and Michael Teper$^{b}$\\
  \vspace*{.25in}
  $^{a}$ Dipartimento di Fisica, Universit\`a di Pisa and INFN, Sezione di Pisa, Largo Pontecorvo 3, 56127 Pisa, Italy. \\
  \centerline{and}
Computation-based Science and Technology Research Center, The Cyprus Institute, 20 Kavafi Str., Nicosia 2121, Cyprus \\
\vspace*{.1in}
$^{b}$Rudolf Peierls Centre for Theoretical Physics, University of Oxford,\\
Parks Road, Oxford OX1 3PU, UK\\
\centerline{and}
All Souls College, University of Oxford,\\
High Street, Oxford OX1 4AL, UK}
\end{center}

\vspace*{0.6in}

\begin{center}
{\bf Abstract}
\end{center}

We calculate the low-lying glueball spectrum of the $SU(3)$ lattice gauge theory
in $3+1$ dimensions for the range $\beta\leq 6.50$ using the standard plaquette action.
We do so for states in all the representations $R$ of the cubic rotation group,
and for both values of parity $P$ and charge conjugation $C$. We extrapolate
these results to the continuum limit of the theory using the confining string tension
$\sigma$ as our energy scale. We also present our results in units of the $r_0$ scale
and, from that, in terms of physical `GeV' units. For a number of these states
we are able to identify their continuum spins $J$ with very little ambiguity.
We also calculate the topological charge $Q$ of the lattice gauge fields so as to
show that we have sufficient ergodicity throughout our range of $\beta$,
and we calculate the multiplicative renormalisation of $Q$ as a function of $\beta$.
We also obtain the continuum limit of the $SU(3)$ topological susceptibility.

\vspace*{0.95in}

\leftline{{\it E-mail:} andreas.athinodorou@df.unipi.it, mike.teper@physics.ox.ac.uk}

\end{titlepage}

\setcounter{page}{1}
\newpage
\pagestyle{plain}

\tableofcontents

\section{Introduction}
\label{section_intro}

The spectrum of the $SU(3)$ gauge theory in 3+1 dimensions is of interest for both phenomenological
and theoretical reasons. On the phenomenological side it may inform us about the approximate location of
glueballs in QCD if the mixing with $q\bar{q}$ states is as modest as suggested by the OZI rule.
Here it is the lighter glueballs, with masses below say 3 GeV, that are of greatest interest given
the experimental constraints in observing and identifying glueballs. The masses of such glueballs,
some 4 or 5 in total, have been reasonably well determined for a long time: see for example
\cite{CMMT-1989}.
Subsequent calculations, for example
\cite{CM-UKQCD-1993,MP-1999,MP-2005,HM-2005},
have remained consistent with these earlier estimates, albeit with much reduced errors, indicating
that this part of the spectrum is well understood. Slight shifts in the mass estimates are
mostly due to differences in importing the physical `GeV' scale into the unphysical pure gauge
theory. Since the two theories are not the same this reflects an intrinsic ambiguity.

On the theoretical side the spectrum represents the obvious challenge for attempts to obtain
an analytic control of the gauge theory. (This is also of indirect phenomenological importance
since much of the difficulty in obtaining analytic control of the long distance physics of QCD
arises from the non-perturbative physics of the underlying gauge theory.) To use the glueball
spectrum as a test of theoretical approaches it is important to have a substantial number of
states for a variety of spins $J$, parities $P$ and charge conjugations $C$, with
excited as well as ground states. Many of these states will be quite heavy and calculating
their masses on the lattice is more challenging. Moreover here we need accurate mass estimates,
since the errors cannot be hidden within the uncertainty associated with importing a physical
`GeV' scale. Such calculations are still open to considerable improvement and the purpose of
the present paper is to take some further steps in that direction.

The first step is to try to improve the reliability of our calculation of lattice masses. Since
the usual way to calculate masses on the lattice is from the time decay of correlators of
lattice operators, we use a large basis of operators so as to decrease the probability that
we miss intermediate states in the spectrum. To improve our extrapolation of these masses
to the continuum limit we perform calculations over a wide range of lattice spacings, $a$,
down to $a\surd\sigma \sim 0.10$ where $\sigma$ is the confining string tension (which we also
calculate). We also perform a finite volume comparison at one of our smaller lattice spacings
so as to identify any finite volume corrections to our calculations. The calculations at
the smallest values of $a$ are of course the most important, and here one needs to
be concerned about the ergodicity of the Monte Carlo and in particular about the possible
freezing of the fields into sectors of unchanging topological charge $Q$. To investigate
this we calculate the decorrelation of $Q$ for all our values of $a$. In the process we
need to demonstrate that our method of calculating $Q$ is reliable and this also allows
us to provide a precise estimate of the continuum topological susceptibility $\chi_t$.

The glueball states that we calculate will belong to the representations $R$ of the subgroup
of rotations that leave the cubic lattice invariant. That is to say, they are labelled
by $R^{PC}$ rather than the continuum $J^{PC}$. Since one knows how the $2J+1$ dimensional
vector space associated with a glueball of spin $J$ is partitioned amongst the various
cubic representations $R$, and since these $2J+1$ states should be nearly degenerate
at small $a$, one can attempt to identify the value of $J$ through the observation
of the appropriate near-degeneracies. This is limited by the accuracy of our calculation
of enough excited states as well as the ground states but we are still able to identify $J$
for a substantial number of states. It is however worth emphasising that if one wants to
compare the predicted spectrum of a theoretical approach or model to our lattice data,
it may make sense to convert the predictions from those for $J^{PC}$ to those for $R^{PC}$.
This is simple to do using the conversion that we will present later in this paper.

Our paper is as follows. In the next section we give a brief overview of our lattice
setup and how we calculate energies. We then move on to present our calculations of the
confining string tension $\sigma$ which we shall use as our scale for the calculated
glueball masses in our extrapolation of the latter to the continuum limit.
We then turn to the calculation of the glueball masses $m_G$ on our cubic lattices.
Here we discuss the representations of the cubic group and present the relation between 
these and the continuum rotation group. We also provide a detailed discussion and
check of finite volume effects. 
In the following section we present our extrapolations to the continuum limit of the
dimensionless ratios $m_G/\surd\sigma$ and the conversion of the masses into `GeV'
units. Where possible we also identify the continuum  spin $J$ of the states.
We follow this section with a comparison of our results to those
of some of the most comprehensive earlier spectrum calculations.
We then return to the issue of the freezing of the topological
charge at the smallest lattice spacings. We calculate $Q$ on large ensembles of
lattice fields, to check for ergodicity, and as a byproduct obtain a value for
the topological susceptibility in the continuum limit. We then turn
to pinpointing the ways in which our calculations
need to be improved. We summarise our findings in the concluding section.

\section{Calculating on a lattice}
\label{section_lattice_energies} 

Our lattice setup is entirely standard.
We work on hypercubic lattices of size $L_s^3L_t$ with lattice spacing $a$ and with
periodic boundary conditions on the fields. Our fields are $SU(3)$ matrices, $U_l$,
assigned to the links $l$ of the lattice, The Euclidean path integral is 
\begin{equation}
Z=\int {\cal{D}}U \exp\{- \beta S[U]\},
\label{eqn_Z}
\end{equation}
where ${\cal{D}}U$ is the Haar measure and we use the standard plaquette action,
\begin{equation}
\beta S = \beta \sum_p \left\{1-\frac{1}{3} {\text{ReTr}} U_p\right\}  
\quad ; \quad \beta=\frac{6}{g^2}.
\label{eqn_S}
\end{equation}
Here $U_p$ is the ordered product of link matrices around the plaquette $p$. We write
$\beta=6/g^2$, where $g^2$ is the bare coupling and this provides a definition of
the running coupling  on the length scale $a$. Since the theory is asymptotically
free $g^2\to 0$ and hence $\beta\to \infty$ as $a\to 0$. Monte Carlo calculations are
performed using a mixture of standard Cabibbo-Marinari heat bath and over-relaxation
sweeps through the lattice in the ratio 1:4. We typically perform $2\times 10^6$ sweeps at each
value of $\beta$ at each lattice size, and we typically calculate correlators every
$25$ sweeps.

We calculate masses $M$ and energies $E$ from the exponential decay of appropriate
correlation functions. For example, the energy  $E_{gs}$ of the lightest state with the
quantum numbers of the operator $\phi$ is given by
\begin{equation}
C(t=an_t) = \langle \phi^\dagger(t) \phi(0) \rangle 
= \sum_n |\langle vac|\phi^\dagger|n \rangle|^2 e^{-E_n t} 
\stackrel{t\to\infty}{\propto} e^{-E_{gs}an_t} .
\label{eqn_M}
\end{equation}
These masses and energies are obtained in lattice units, i.e. as $aE_{gs}$, and are
relative to the vacuum once $L_t$ is large enough, so that the vacuum energy is zero.
If $\phi$ has the quantum
numbers of the vacuum then the lightest state of interest is the first excited state
above the vacuum. In this case it can be convenient to use the vacuum-subtracted operator
$\phi - \langle \phi \rangle$, which will remove the contribution of the vacuum
in eqn(\ref{eqn_M}), so that the lightest non-trivial state appears as the leading
term in the expansion of states.

The accuracy of such a calculation of $aE_{gs}$ is constrained by the fact that the
statistical errors are roughly independent of $t$ (for pure gauge theories) while
the desired `signal' is decreasing  exponentially with $t$. So if the overlap of
the desired state onto our basis is small, the relevant correlator will disappear
into the statistical `noise' before we get to large enough $t$ for the correlator
to be dominated by the simple exponential from which we can extract the
ground state energy. Similarly the correlator of a 
more massive state will disappear into the `noise' at smaller $t=an_t$
and this may make ambiguous the judgement of whether it is dominated by
a single exponential. All this may provide an important source of
systematic error.

The glueball masses calculated using eqn(\ref{eqn_M}) are in lattice units, i.e. $aM$,
and to remove this factor of $a$ we can take a ratio of masses or energies. For this
purpose we simultaneously calculate the confining string tension, $a^2\sigma$, and form
the dimensionless ratio $aM/a\surd\sigma = M/\surd\sigma$ from which the lattice
scale has disappeared. The leading lattice corrections to such a ratio are $O(a^2)$
so we extrapolate to the continuum limit using
\begin{equation}
\left.\frac{M}{\surd\sigma}\right|_{a}
=
\left.\frac{M}{\surd\sigma}\right|_{a=0}
+ c_1 a^2\sigma + c_2 (a^2\sigma)^2+ ...
\label{eqn_MKcont}
\end{equation}
We will usually only need the leading correction. In reality the coefficients $c_i$
are power series in $g^2$ but this dependence on $a$ is too weak (logarithmic) to be
visible in our data, so we follow common practice in treating the $c_i$ as constants.

The values of $\beta$ we use are listed in Table~\ref{table_param}. The lattice sizes
next to these $\beta$ values are the ones we use for our string tension and glueball
mass calculations. The lattices listed on the right are used for calculating topology
and also, in some cases, to check for finite volume corrections. We also show the
average plaquette at each value of $\beta$ as well as the string tension (see
Section~\ref{section_sigma}) and the mass gap  (see Section~\ref{section_glueballs}).

\section{String tension}
\label{section_sigma}

We calculate the string tension by calculating the ground state energy $E(l)$
of a flux tube of length $l$ that closes on itself by winding once around a
spatial torus of size $l$. We use eqn(\ref{eqn_M}) where the operator $\phi$
is the product of $SU(3)$ link matrices taken around a non-contractible closed
path that winds once around the spatial torus. The simplest such operator is the
spatial Polyakov loop
\begin{equation}
\phi(n_t) = l_p(n_t) =  \sum_{n_y,n_z} \mathrm{Tr} 
\left\{\prod^{L_x}_{n_x=1} U_x(n_x,n_y,n_z,n_t)\right\} 
\label{eqn_poly}
\end{equation}
where we take the product of the link matrices in the $x$-direction 
around the $x$-torus of length $l=aL_x$. Here
$(x,y,z,t)=(an_x,an_y,an_z,an_t)$, and the sum over $n_y$ and $n_z$ produces
an operator and hence a state with zero transverse momentum, $p_y = p_z = 0$.
Since  $\phi$ is clearly invariant under translations in $x$, the state
also possesses $p_x=0$. We do the same for loops over the $y$ and $z$ spatial
tori and average over all three. We do the same replacing the link matrices
by blocked link matrices, which is the essential step in obtaining a useful
overlap onto the ground state. This provides a vector space of operators
on which we perform a variational calculation to extract the operator that is
our best approximation to the ground state wavefunctional. From the correlator
of this operator we extract a value for $E(l)$.

We extract the string tension $\sigma$ from $E(l)$ using the `Nambu-Goto' formula
\begin{equation}
E(l)
\stackrel{NG}{=}
\sigma l \left(1-\frac{2\pi}{3\sigma l^2}\right)^{1/2}\,,
\label{eqn_gsNG}
\end{equation}
which arises from the light-cone quantisation of the bosonic string
and is known to provide an excellent approximation to the lattice
calculations
\cite{AABBMT-string}
for reasons that have now become well understood. (See for example
\cite{string_theory1,string_theory2}.)

To extract the ground state energy of the flux tube using eqn(\ref{eqn_M})
we need to be confident that we have gone to large enough $t=n_t$ that
the correlator is dominated by a single exponential within errors, and moreover
that these errors are small enough for this observation to be significant.
A useful way to test for a single exponential is to extract an effective
energy
\begin{equation}
aE_{eff}(n_t) = -\ln \frac{C(n_t)}{C(n_t-1)}.
\label{eqn_Eeff}
\end{equation}
If a single exponential dominates $C(n_t)$ for $n_t\geq \tilde{n}_t$, then $E_{eff}(n_t)$
should become independent of $n_t$ for $n_t\geq \tilde{n}_t$, and in that case
$E_{eff}(\tilde{n}_t)$ provides an estimate of the energy $E(l)$ .
(In practice we extract our $E$ from a fit over a range of $n_t$.)
That is to say, we search for a `plateau' in the values of $E_{eff}(n_t)$ 
against $n_t$. An important point is that as long as our operators are
entirely spacelike, the reflection positivity of our action ensures that
$E_{eff}(n_t) \geq E(l)$ for any $n_t$, up to statistical errors, and
indeed that the value of $E_{eff}(n_t)$ must be monotonically decreasing with
increasing $n_t$.

\begin{figure}[htb]
\begin	{center}
\leavevmode
\input	{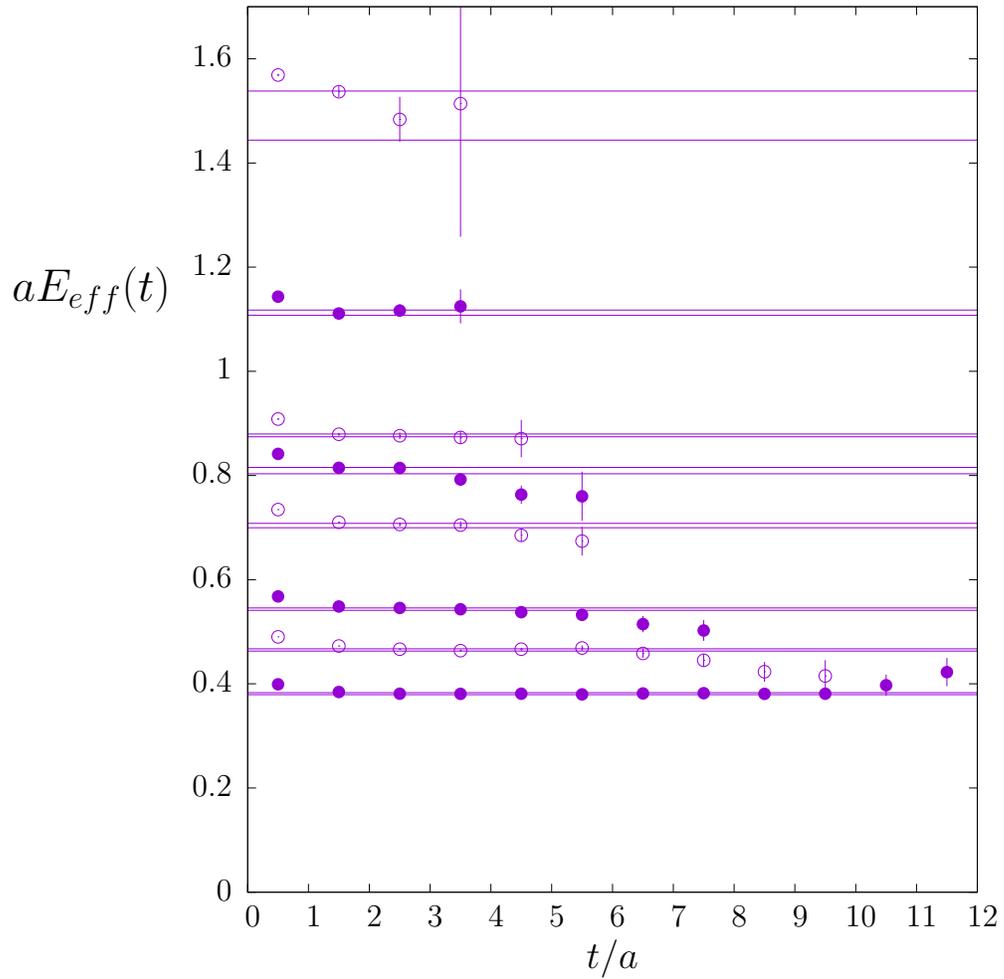}
\end	{center}
\caption{Effective energy of the ground state of the winding flux tube from
  which we extract the string tension. For $\beta = 6.50, 6.338,6.235,6.0625,5.99,5.80,5.6924$
  in ascending order. Pairs of lines represent $\pm 1\sigma$ error bands for the best estimates
  of the energies.}
\label{fig_Eeffl}
\end{figure}

In Fig.~\ref{fig_Eeffl} we plot the values of $aE_{eff}(t)$ for each of our $\beta$
values together with the $\pm 1\sigma$ error band for our best estimate of the
energy $E(l)$. At our smallest value of $\beta$, $a$ and hence $aE$ are large
and the exponential disappears into the statistical errors before we see an
unambiguous effective energy plateau -- hence the very wide error band around
our estimate of $E(l)$. At other values of $\beta$ there is little ambiguity and
the error bands are correspondingly narrow. We also note that the asymptotic
exponential typically begins at $t=a$ at smaller $\beta$ and at $t=2a$ at larger $\beta$.
Since we expect that our space of multiply blocked operators will have an overlap
onto the ground state that is weakly dependent on $a(\beta)$, this is as expected.
It is also the reason that for our lowest value of $\beta$, where no effective energy
plateau apears, we are confident in choosing a value of $E$ that encompasses, within
errors, the values of $aE_{eff}(t)$ for $t=a$ and $t=2a$.

\section{Glueball spectrum}
\label{section_glueballs} 

We calculate glueball masses from Euclidean correlators as in eqn(\ref{eqn_M}).
Since only states $|n\rangle$ such that $\langle vac|\phi^\dagger|n \rangle \neq 0$
will contribute to the correlator, we can restrict the contributing states to have
particular quantum numbers by choosing operators $\phi$ with those quantum numbers.
Since the error to signal ratio of such a correlator increases exponentially in
$t=an_t$, we can only hope to identify an effective mass plateau using
eqn(\ref{eqn_Eeff}) if the overlap of the state onto the operator $\phi$ is
large enough. So in order to avoid missing intermediate states in the spectrum
one needs a sufficiently large basis of operators.

In this paper we employ a basis of 27 different operators. The operators are
contractible closed loops of links, and we take their traces. We include all
loops of length 4 and length 6 as well as fifteen of length 8 and eight of
length 10, with all their rotations and parity inverses. In addition to
constructing the loops out of elementary links we also construct them out of
link matrices that have been `blocked'
\cite{MT-block1,MT-block2}
using the algorithm described for example in
\cite{BLMTUW-2004}.
The length of a blocked link is $2^{n_b}a$, where $n_b$ is the blocking level,
and we use all blocked link matrices up to a maximum level that depends on
the lattice size and is 6 for our largest $38^4$ lattice which corresponds
to blocked links of length $32a$. Our loops are of length $4,6,8,10$ in units of
the blocked links being employed in their construction.

As we increase $\beta$, and hence decrease $a(\beta)$, we increase the spatial
lattice size $L_s$ so that the lattice volume is approximately constant in physical
units. We choose  for our physical units our calculated values of the string tension, 
$\sigma$, so that we keep $a\sqrt\sigma L_s$ roughly constant. Since our blocked
operators cover all length scales on such a lattice, we expect that their overlap onto
the desired glueball states will only change weakly with $\beta$ and this indeed
appears to be the case.

We calculate all the cross-correlators,
$C_{ij}(t=an_t) = \langle \phi_i^\dagger(t) \phi_j(0) \rangle$,
of all the operators $\phi_i,\phi_j$ with given quantum numbers (see below) and
this means that we can calculate the correlator of any linear combination in the
vector space generated by the $\{\phi_i\}$ and so we can 
perform a variational calculation to find the linear combination
$\Phi_0$ that minimises the energy (and hence maximises $\exp\{-Et\}$).
We then look for an effective mass plateau in the correlator of $\Phi_0$ to
provide our best estimate of the ground state energy. We now take 
the subspace of $\{\phi_i\}$ that is orthogonal to $\Phi_0$ and
repeat the variational procedure  to find the linear combination
$\Phi_1$ that minimises the energy in this subspace and hence provides,
from its correlator, our best estimate of the mass of the first excited
state. And so on. Note that one needs to ensure that the $\phi_i$ one keeps are linearly
independent so as to avoid singularities in the standard variational
procedure.

The number of our operators $\phi_i$ in a given quantum number sector can be
very large (up to nearly 800) and we sometimes do not use all of our
operators where doing so would make the analysis very unwieldy. In a few
quantum number sectors the number is not large, but in any case the
minimum is 30 operators, which should be large enough for the lightest
couple of states, which is all we need in those cases.

In this section we start by describing the rotational quantum numbers
of our lattice states and how these relate to the continuum spins.
We then address the important issue of corrections due to the finite
volume of our lattice. Finally we provide our results for the
glueball spectrum with some demonstrations of how reliable are
our mass estimates.

\subsection{Quantum numbers}
\label{subsection_reps}

In the continuum theory the relevant quantum numbers of glueballs are
their spin $J$, parity $P$ and their sign under charge conjugation, $C$.
(They also have momentum but since we are interested in the masses we
choose to calculate states with zero momentum.)

On a cubic lattice the rotational symmetry is restricted to the octahedral subgroup
of the full rotation group and there are only five irreducible representations
which we shall label generically as $R$. These five representations are often
labelled as $A_1,A_2,E,T_1,T_2$. The dimensions of these are $1,1,2,3,3$
respectively. The states that we calculate will belong to one of these
representations and we will generically label them by $R^{PC}$ or more
specifically as $A_1^{PC}$ etc. As $a(\beta)\to 0$ the glueball size diverges
in lattice units and we recover the full rotational invariance where the
states fall into the $2J+1$ multiplets labelled by $J$. Thus the lattice states
can also be labelled by the continuum spin $J$ to which they tend continuously
as $a(\beta)\to 0$. (Ignoring ambiguities due to level crossing, which should
be appropriate once we are close enough to the continuum limit.) Of course
at non-zero $a$ the energy degeneracy will be broken except for that part of
it that is enforced by the dimensionality of the lattice representations.

The way that the states labelled by the spin $J$ of the full rotation group
are partitioned amongst the representations of the subgroup of the symmetry
rotations of the cube is well-known (see for example Section 9.4 and Table 9.4 of
\cite{Hamermesh})
and we reproduce the result for the lowest spins in Table~\ref{table_J_R}.
One can readily check that for each $J$ the total number of states in the
corresponding cubic representations, taking into account their degeneracies,
is indeed $2J+1$. Where these states fall into different cubic representations
they will not be degenerate, but one expects that the breaking of the
degeneracy will be small once $a(\beta)$ is small. Thus one can try to identify
the spin $J$ of a lattice state by seeing if there are a set of near-degenerate
states in the corresponding cubic represntations listed in Table~\ref{table_J_R}.
In practice this works well for the lightest states, but breaks down for heavier
states where the spectrum of states becomes sufficiently dense that an
apparent near-degeneracy (within errors) is increasingly likely to be accidental.
This method has been used to good effect, within its limitations, in for example
\cite{MP-2005},
and it is the method we will employ in the present paper. There also exists
a more powerful method which consists in ascertaining the approximate
rotational properties of the states using an appropriate variation of a
Fourier transform. This approach has been used successfully in
\cite{HM-2005,HMMT-spin}
and more recently in
\cite{PCSDMT-2019}
but requires a dedicated calculation of a kind that is beyong the scope of the
present paper.

\subsection{Finite volume corrections}
\label{subsection_volume} 

The finite volume dependence of the glueball spectrum has two main components.
The first is due to self-energy corrections that are modified by the periodic
boundary conditions
\cite{Luscher-V}.
For example when a glueball emits a virtual glueball,
one of the loop propagators can wind around the spatial torus before being
reabsorbed. These corrections are typically exponentially suppressed  in
$m_Gl_s$ where $m_G$ is the mass gap and $l_s$ is the spatial periodicity,
and since in our typical calculations $m_Gl_s \sim 10$ the resulting mass
shift will be invisible within our statistical errors.

The second finite volume influence on the glueball spectrum consists of extra
states whose energy $\to \infty$ as $l_s\to\infty$, so that they disappear in
the thermodynamic limit, but can have the same quantum numbers as glueballs and so
can mix with the latter and be mistaken for them in a finite volume.
These states are composed of flux tubes that wind around a spatial torus, and are
generically called torelons. A single flux tube has non-trivial quantum numbers under
a certain centre group transformation under which glueballs tranform trivially,
and hence has zero overlap with contractible glueball operators and so is irrelevant
here. (Although it is of course relevant to calculating the string tension as in
Section~\ref{section_sigma}.) The simplest non-trivial such finite volume state will
consist of a torelon with a conjugate torelon  closed around the same spatial torus.
If the ground state energy of the torelon is $E_T$, then the minimum energy of such
a state will be $E_{T\bar{T}} \sim 2E_T \sim 2\sigma l_s$ and it will in principle appear
as an extra state in our glueball correlator. Such a `ditorelon' state may be a bound
state or a scattering state. There will be a set of scattering states where each of the
two torelons has an equal and opposite momentum (since our glueball operators have
zero total momentum). One can estimate their energies using the allowed lattice
momenta, e.g. $ap_x=2\pi n/L_x$, although the actual momenta may have substantial
shifts from these values due to the interactions. (It is only the total momentum
of the state that has these quantised values.) Since the ditorelon states
correspond to double trace operators while our glueball operators are single
trace, we expect the overlap of the ditorelons onto our glueball operator
basis to have some `large-$N$' suppression which may mean that some of these
finite volume states may become invisible in our actual calculations.
However it has long been known that the lightest of these ditorelon states will
certainly appear as the lightest glueball states in the $A_1^{++}$ and $E^{++}$
representations once the volume is small enough. (See e.g.
\cite{CMGTMT-1988}.)
Since this simplest ditorelon operator is real, invariant under reflections and
also invariant under rotations of $\pi$, it will only contribute to the $A_1^{++}$
and $E^{++}$ representations. Nonetheless this poses a potential problem since the
glueball states that are lightest and hence of greatest phenomenological interest
are the $J^{PC}=0^{++},2^{++}$ ground states and these fall into the $A_1^{++}$
and $E^{++}$ representations respectively. So it will be  important to make sure
that the spatial volume is large enough for the ditorelon not to influence these
mass estimates. For the $J^{PC}=2^{++}$ state $2$ of the $2J+1=5$ components are in
the $E^{++}$ and $3$ are in the $T_2^{++}$ and since the ditorelon does not contribute
to the latter this provides a useful crosscheck. However the $0^{++}$ only appears
in the $A_1^{++}$. Since the ditorelon mass is $\propto l_s$ one can make
the spatial volume large enough for the ditorelon not to influence these
important mass estimates. Then by comparing calculations on different volumes
we can identify the location of any such ditorelon states in the glueball spectra.
If we keep $l_s$ roughly constant in units of $\sigma$ as we vary $\beta$ we
can keep such a state in the same location in the ordered spectrum and so
obtain reliable continuum limits of the other `true' glueballs. 

In addition to the ditorelon one can form a tritorelon composed of three winding
flux tubes. In $SU(3)$ this has trivial centre symmetry quantum numbers and so can
mix with local glueball states. The operator has an imaginary part and so it
can contribute to some $C=-$ states. However its energy is $\sim 3\sigma l_s$
and so it will not influence the lightest states given that we will work on
lattices where the ditorelons are much heavier than the lightest states.
Moreover this is a triple trace operator whose overlap onto the single trace
operators we use in our glueball calculations is likely to be small enough
that these states make no impact on our spectrum. 

We also note that the above ditorelon is the simplest of a larger family of ditorelon
states. A winding flux tube has a spectrum of excitations
\cite{AABBMT-string}
and one can have a ditorelon consisting of an excited torelon as well as a
ground state (conjugate) torelon, or indeed both torelons being excited. Such
states may contribute to various $R^{PC}$ representations but will be heavier and
will be less easy to identify since their volume dependence is much weaker
at intermediate volumes. It is however not clear that they will have a large
enough overlap onto our glueball operator basis to be relevant and we
do not attempt to identify them explicitly in the present work.

To establish some explicit control over these finite volume ambiguities
we have performed calculations on
three different volumes at $\beta=6.235$. This corresponds to one of our smaller 
values of $a(\beta)$ and that enables us to obtain reasonably reliable
mass estimates for our heavier
states as well as our lightest states. In addition to the $26^326$ lattice,
which corresponds to the standard physical volume that we will use in our glueball
calculations below, we also perform calculations on a smaller $18^326$ lattice
and a larger $34^326$ lattice. In this comparison we will normally extract the
masses from the same range of $n_t$ in the correlators on these three lattices so
as to minimise enhancing the statistical fluctuations betweeen the calculations.
This means that some of these masses may differ slightly from those presented
in the following section.

We begin with the $P,C=+,+$ sector of states, our mass estimates being listed in
Table~\ref{table_Vcheckpp_b6.235}. The states are ordered by the effective mass
$aM_{eff}(t)$ of the corresponding correlator at $t=a$, which will normally, but
not always, correspond to being ordered in the value
of the mass taken from the effective mass `plateau'. To make
the comparison clearer we have left gaps where a state on a smaller lattice
has no partners on the larger lattices. At the bottom of the table is listed
the energy of a pair of non-interacting torelons on each lattice; presumably
the mass of a ditorelon would be close to this. In the $A_1^{++}$ sector
of the $18^326$ lattice we clearly see a state at $aM \simeq 0.70$ which does
not appear in the spectra of the larger lattices. This has a mass close to twice
that of the torelon and is undoubtedly a ditorelon. Similarly it is clear that
the spectrum on the $26^326$ lattice has an extra state: between the very well
defined first excited state at $aM \simeq 0.86$ and the well defined states at
$aM \simeq 1.23$ and $aM \simeq 1.33$ there are 4 states on the $26^326$ lattice
but only 3 states on the $34^326$ lattice. Which of these states is the extra
one is ambiguous since all 4 states have quite similar masses. We note that
the third of these four states on the $26^326$ lattice has a lower mass,
$aM \simeq 1.054$, than the preceeding two; this is because it has a smaller overlap
so that there is a larger gap between the value of the mass and the value of
$aM_{eff}(t=a)$. We have chosen this to be the extra state, mainly because
we expect the ditorelon to have a reduced overlap onto the single trace
operators of our basis. However from a purely matching point of view, we
could equally well have chosen the next higher state at $aM \simeq 1.19$.
In both cases this extra state has a mass close to twice that of that lattice's
torelon, as expected. In the $E$ representation on the $18^326$ lattice it is
the ground state with $aM \simeq 0.66$ that is definitely an extra state and hence
the candidate for a ditorelon, and on the $26^326$ lattice it is the excited
state with $aM \simeq 1.206$ that appears to be the extra state and hence a
ditorelon. If we now compare the other states from the $34^326$ and $26^326$
lattices we see that they agree well with each other: there are no
differences that are greater than 2 standard deviations, except $2.6\sigma$ for
the $A_2$ ground state. More surprisingly, the same is true for most (but not
all) states on the small $18^326$ lattice.

We perform a similar comparison for the other values of $P$ and $C$. In
Table~\ref{table_Vcheckmp_b6.235} we do so for the $P,C=-,+$ sector of states.
We see that all the masses on the $26^326$ and $34^326$ lattices are within
$2\sigma$ of each other, with most differences much less than that. So that
it appears that the $26^326$ lattice is large enough within our statistics.
This is not the case for the $18^326$ lattice which shows large differences
for a number of states. The conclusion is the same for the $P,C=+,-$ and
$P,C=-,-$ sectors of states, as we seefrom Table~\ref{table_Vcheckpm_b6.235}
and Table~\ref{table_Vcheckmm_b6.235} respectively.

\begin{figure}[htb]
\begin	{center}
\leavevmode
\input	{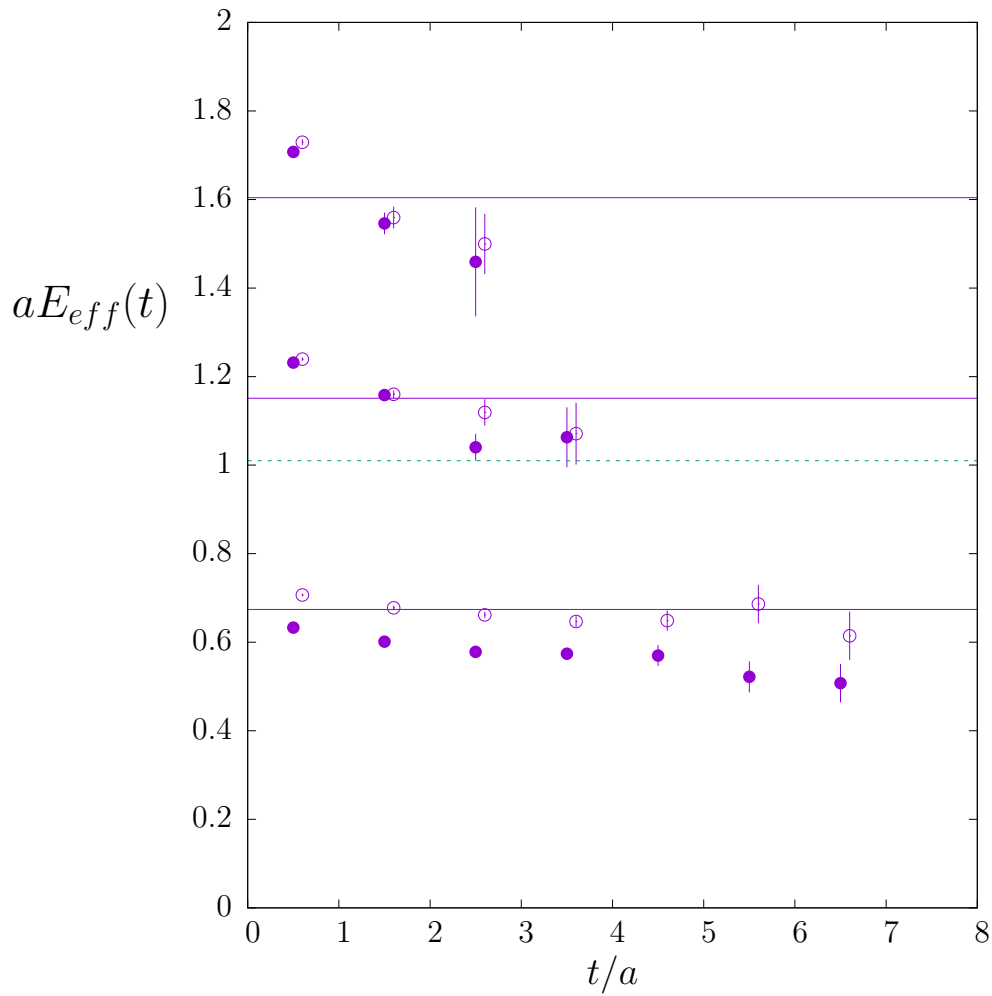}
\end	{center}
\caption{Effective energies of the lightest 3 states from our ditorelon basis of operators
  for the $A1^{++}$, $\bullet$, and $E^{++}$, $\circ$, representations. On a $18^326$ lattice
  at $\beta=6.235$. Horizontal lines are estimated energies of lightest ditorelon states
  neglecting interactions between the torelons; dashed line for the lightest tritorelon
  state.}
\label{fig_Eeffll_l18}
\end{figure}

While we can provide a plausible identification of the ditorelon contribution
to the spectrum, as we have done above, it would be useful to complement that
with an alternative calculation. One way to do this is to
include operators that are explicitly ditorelon. We shall now provide a first
limited step in that direction at $\beta=6.235$ where we have performed our
above finite volume comparison. We take operators that are a product of a
zero momentum Polyakov loop times its conjugate. (Note that in so far as the
torelons interact with each other this mometum is not conserved so this
operator should couple to ditorelons with non-zero relative momenta.)
We include only the three
largest blocking levels since, as we have checked, it is these that provide
essentially the whole of the flux tube ground state wave function at this value
of $\beta$. These ditorelon operators should contribute primarily to the $A1^{++}$
and $E^{++}$ representations. Taking into account the fact that the two torelon
operators in the ditorelon can have different blocking levels this means that
we have 6 distinct double trace operators in the $A1 ^{++}$ representation and
12 in the $E^{++}$ (since it has dimension two). Note that to contribute
efficiently to other representations we would need to include Polyakov loop
operators with non-trivial quantum numbers, employing suitable transverse deformations,
and with non-zero momenta (equal and opposite so as to give zero total momentum).
This would also give a more complete basis for the  $A1^{++}$ and $E^{++}$
representations, but would go well beyond the scope of the present work. In the
same exploratory spirit of this calculation we include only 12 different loops
in our glueball basis, rather than the 27 used in most of our other calculations,
and we have lower statistics (by a factor of two) and we restrict ourselves to
the three largest blocking levels. That is to say, we have 36 and 108 single trace
glueball operators for $A1^{++}$ and $E^{++}$ respectively. We perform the calculations
on our smaller $18^326$ lattice as well as on our standard $26^326$ lattice.
We begin by showing in Fig.~\ref{fig_Eeffll_l18} the  effective mass plots
for the lightest three $A1^{++}$ and $E^{++}$ states that we obtain on the $18^326$
lattice using only our ditorelon operators. We compare these to the lightest 
three states obtained with two free torelons on the same lattice. (We also
indicate the lightest free tritorelon energy, although its relevance is not
clear given that it should have a reduced overlap onto our ditorelon basis.)
In interpreting this plot one must bear in mind that these ditorelon operators
should have a non-trivial projection onto glueball states, and where these
are lighter they will drive the behaviour of the effective masses at larger $t$.
We begin by noting that the $E^{++}$ ground state has the same energy as the free
ditorelon suggesting a negligible interaction energy. And it is the lightest $E^{++}$
state overall. By contrast the lightest $A1^{++}$ ditorelon shows a nice plateau
at intermediate $t$ but one which is about $15\%$ below the free ditorelon
energy, indicating a significant interaction energy in this representation:
presumably a genuine bound state. At larger $t$ the effective energy appears to
drift lower, presumably towards the mass of the lightest $A1^{++}$ glueball.
Finally we remark that while the excited ditorelon
states begin roughly within striking range of the corresponding free ditorelon
energies, they appear to be drifting rapidly lower, presumably driven by their
overlaps onto the lighter glueball states, and there is no evidence that
they harbour any bound states. We turn next to Fig.~\ref{fig_EeffGllA1++_l18}
where we show effective mass plots for our lightest four $A1^{++}$ states
on the $18^326$ lattice. We show the results using our basis of 36 single trace
operators and also when this basis is augmented by the 6 ditorelon operators.
We see that the ground state and the third excited state are unaffected by
the ditorelon operators; presumably because they are far enough from
the energy of the free ditorelon. In contrast, the overlaps of the first
and second excited states onto the operator basis are both significantly
increased when the ditorelon operators are included. This is consistent
with our earlier conclusion that the first excited state is the extra
ditorelon state when compared to the spectra on larger volumes, but
is not so specific. In Fig.~\ref{fig_EeffGllE++_l18} we repeat the plot for
the four lightest $E^{++}$ states, and the picture is very similar. The
overlap of the two lightest states is increased by the inclusion of the
ditorelon operators, but the higher states, which are much further in energy
from the free ditorelon energy, are unaffected. Again while this is consistent
with our earlier conclusion that the lightest $E^{++}$ state is in fact
a ditorelon, it is more ambiguous. The ambiguity is due to the fact that
the ditorelon state, if any, already appears in the spectrum obtained
with our single trace glueball operators. That is to say there is a strong
mixing between the single and double trace operators, at least on this smaller
volume. The corresponding spectra obtained on the larger $26^326$ lattice
are shown in Fig.~\ref{fig_EeffGllA1++_l26} and Fig.~\ref{fig_EeffGllE++_l26}. 
The picture here is quite different in two main ways. The first and trivial
difference is that the individual flux tubes are longer and more massive, so
that the ditorelon is much more massive and appears amongst the higher
excited states. This should, in principle, make its identification more
difficult. The second and more interesting difference is that the inclusion
of the ditorelon operators appears to lead to an extra state in both the
$A1^{++}$ and $E^{++}$ representations, i.e. the second excited state when using
the ditorelon extended operator basis, and these states are close in
energy to that of a free ditorelon. There thus appears to be little
ambiguity in identifying these as ditorelon states. Since we use the same
single trace basis on the $18^326$ and $26^326$ lattices the fact that
these states appear with that basis on the former lattice but not on the
latter suggests that the overlap of the ditorelon onto the conventional
glueball basis decreases as its length increases. This means that
the value of $E_{eff}(t=a)$ will increase relative to the ditorelon
energy, so that it will appear higher in the spectrum and with larger
errors. If we make the single trace operator basis much larger, then the
overlap should increase, perhaps sufficiently for the ditorelon states
to reappear as low-lying states in that spectrum. This is indeed what
appears to happen: our finite volume comparison between the  $26^326$
and  $34^326$ lattices used a much larger basis and this appears to
resurrect the ditorelon states in the conventional glueball spectrum.
We can conclude that even though these ditorelon calculations are intended
to be exploratory, they already reveal some usefulness in complementing
our earlier comparison of spectra across different volumes.

We conclude that the important finite volume effects on a lattice
that has the physical spatial volume of a $26^3$ lattice at $\beta=6.235$
are the extra ditorelon states in the $A_1^{++}$ and $E^{++}$ representation.
As we see in Table~\ref{table_Vcheckpp_b6.235} this is probably the 5'th 
$A_1^{++}$ state and the 5'th  $E^{++}$ state. We will work on
lattices at other values of $\beta$ which are chosen so as to be roughly
the same physical size (in units of the string tension) and so provided
there is no strong lattice correction to the level ordering, we can assume
that the location of the ditorelons in the spectrum of these extra states is the
same. When we come to perform continuum extrapolations of our lattice masses
we shall limit ourselves to the lightest 4 $A_1^{++}$ states and the lightest
4 $E^{++}$ states, which means that if our identification of the ditorelon
states in Table~\ref{table_Vcheckpp_b6.235} is correct, then we should not
have any contamination from these ditorelon states in our final results.
This is not guaranteed of course: there are several states close together in
the relevant energy range which may enhance mixing and changes in level
ordering on slightly different volumes. Nonetheless, by the same token
this means that our energy estimates of the genuine glueball states
should be quite accurate.

\subsection{Glueball masses}
\label{subsection_masses} 

We calculate a number of the lightest glueball masses for each set of $R^{PC}$
quantum numbers. Since our mass estimates becomes less reliable as the 
the mass becomes larger, our first task is to provide some evidence for their
reliability. This we do by providing effective mass plots for all the states
to which we will eventually assign a cotinuum spin $J$, i.e those states that
are of greatest phenomenological interest. We do so for our $\beta=6.50$
calculation on a $38^4$ lattice, because it has the smallest lattice spacing
and so is the closest to the continuum limit and has the finest resolution 
of the glueball correlators.

\begin{figure}[htb]
\begin	{center}
\leavevmode
\input	{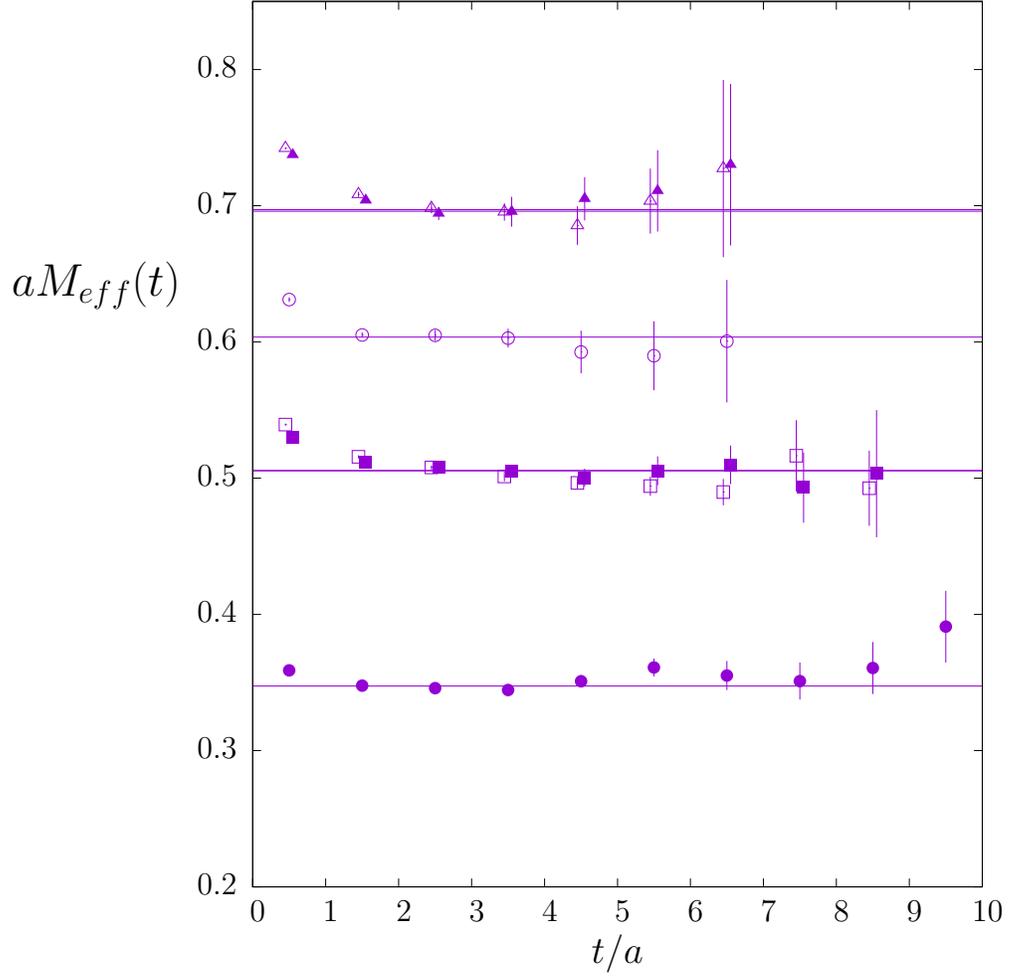}
\end	{center}
\caption{Effective masses for the ground ($\bullet$) and first excited ($\circ$)
  $J^{PC}=0^{++}$ states. Also the ground and first excited $J^{PC}=2^{++}$ states:
  2 of the 5 components from the $E$ representation ($\blacksquare,\blacktriangle$)
  and the other 3 from the $T_2$ representation ($\square,\vartriangle$). Some points
  shifted for clarity. Horizontal lines are our best mass estimates.
  On the $38^4$ lattice at $\beta=6.50$.}
\label{fig_Meff02pp_b6.5}
\end{figure}

We begin, in Fig.~\ref{fig_Meff02pp_b6.5}, with the lightest and first
excited $J^{PC}=0^{++},2^{++}$ glueballs. The  $0^{++}$ ground state is the
lightest of all the glueballs and constitutes the theory's mass gap.
We show also, as horizontal lines, our best estimates of the various masses.
We see that we have
well defined effective mass plateaux in all cases and that these begin from
either $t=a$ or from $t=2a$. In the case of the $0^{++}$ ground state we see a
noticeable increase in the $aM_{eff}(t)$ at larger values of $t$, but since we
know that the exact  value of $aM_{eff}(t)$  must be monotonically decreasing
as $t$ increases this rise must be a statistical fluctuation. In general the
effecive masses do show fluctuations up or down at large $t$ which is
presumably due to the incompleteness of our straightforward error analyses.
In addition for excited states the incompleteness of our variational basis
means that our best operator for an excited state may contain components
of lower mass states. When these components are small, which we try to achieve
with our operator basis being large, one will see the desired effective mass
plateau over an intermediate $t$ range followed by a slow drift at larger $t$
to a secondary plateau (and so on). So sometimes a downward drift at large $t$
will be real rather than a fluctuation. In any case our interest is in the
initial effective mass plateau. Returning to Fig.~\ref{fig_Meff02pp_b6.5} what we
see is that in the cases shown the mass estimates are robust. It is also clear that
the $T_2$ and $E$ states that make up the five components of the $J=2$ state
are degenerate within errors as we would expect once $a(\beta)$ is small enough
and the spatial volume is large enough.

Note that since the states of $T_1$ and $T_2$ come in degenerate triples and those
of $E$ in doublets, we shall average the corresponding correlators to produce single
values for the masses. That is to say, whenever in this paper we refer to a
state of one these representations we are actually referring to an average
of the 3 or 2 states that would be exactly degenerate with infinite statistics.  
This is what we have done in Fig.~\ref{fig_Meff02pp_b6.5} for the $E$ and $T_2$
masses. So the former state actually represents two states and the latter
three giving a total of five as needed for $J=2$.

\begin{figure}[htb]
\begin	{center}
\leavevmode
\input	{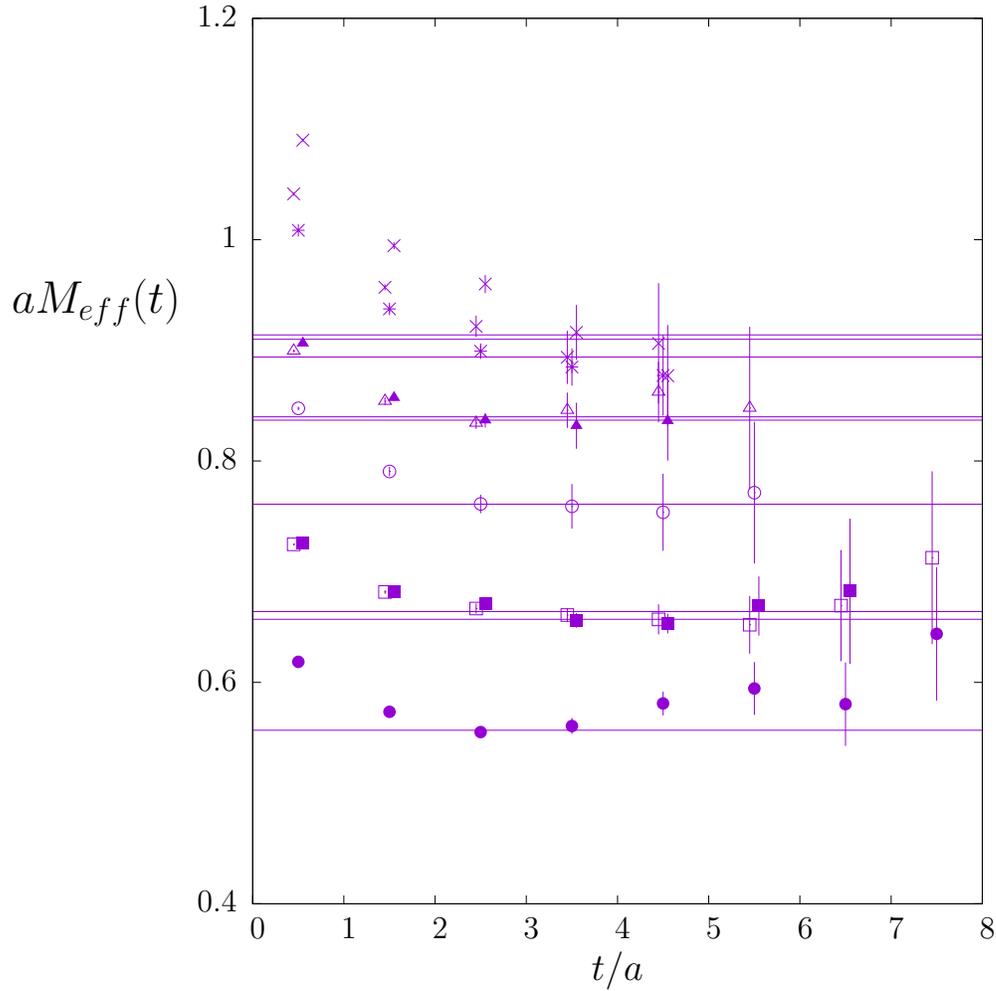}
\end	{center}
\caption{Effective masses for the ground ($\bullet$) and first excited ($\circ$)
  $J^{PC}=0^{-+}$ states. Also the ground and first excited $J^{PC}=2^{-+}$ states:
  2 of the 5 components from the $E$ representation ($\blacksquare,\blacktriangle$)
  and the other 3 from the $T_2$ representation ($\square,\vartriangle$). And the
  lightest three $J^{PC}=1^{-+}$ states ($\star,\times,+$), each triply degenerate.
  Some points shifted for clarity. Horizontal lines are our best mass estimates.
  On the $38^4$ lattice at $\beta=6.50$.}
\label{fig_Meff012mp_b6.5}
\end{figure}

\begin{figure}[htb]
\begin	{center}
\leavevmode
\input	{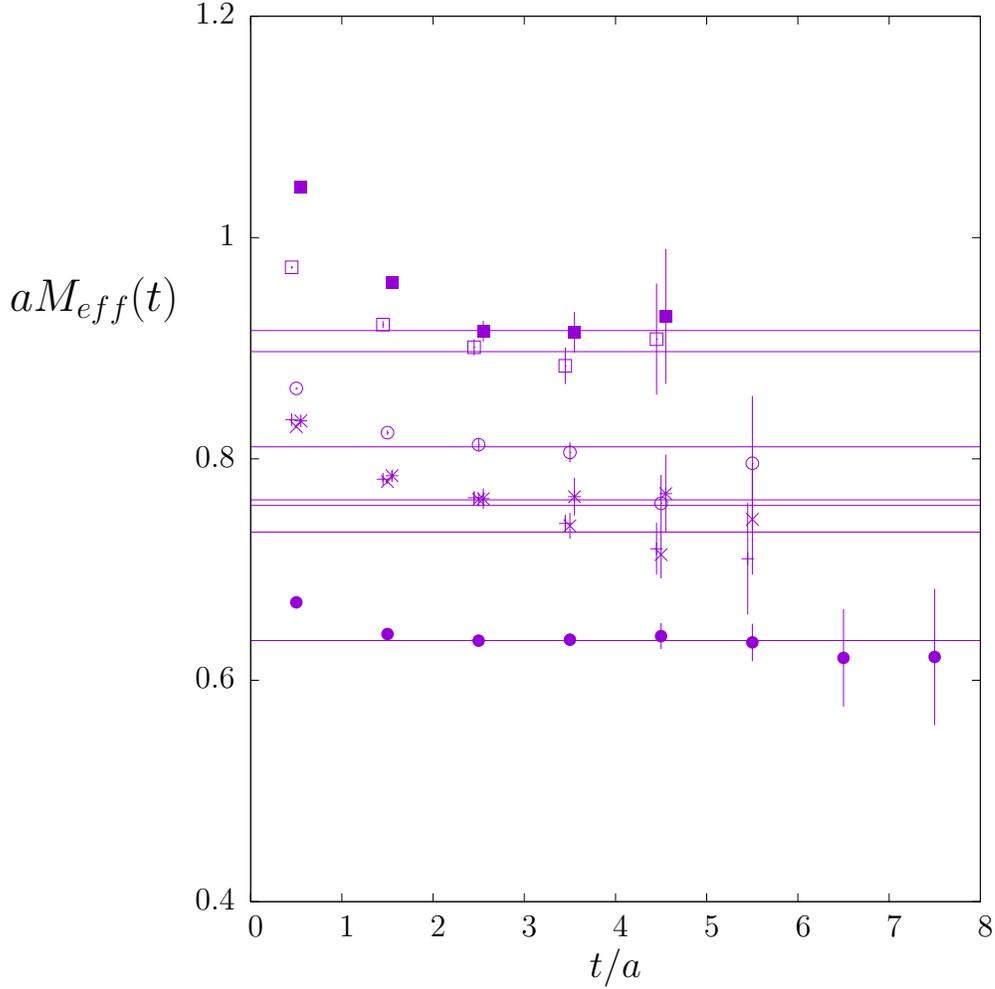}
\end	{center}
\caption{Effective masses for the ground ($\bullet$) and for the first excited ($\circ$)
  $J^{PC}=1^{+-}$ states. Also the  $J^{PC}=2^{+-}$ ground state: 2 of the 5 components
  from the ground state $E$ ($\blacksquare$) and the other 3 from the first
  excited $T_2$ state ($\square$). And the  $J^{PC}=3^{+-}$ ground state:
  one component from the $A_2$ ground state ($\star$), three from the $T_2$
  ground state ($+$), and three from the first excited $T_1$ state ($\times$).
  Some points shifted for clarity. Horizontal lines are our best mass estimates.
  On the $38^4$ lattice at $\beta=6.50$.}
\label{fig_Meff123pm_b6.5}
\end{figure}

\begin{figure}[htb]
\begin	{center}
\leavevmode
\input	{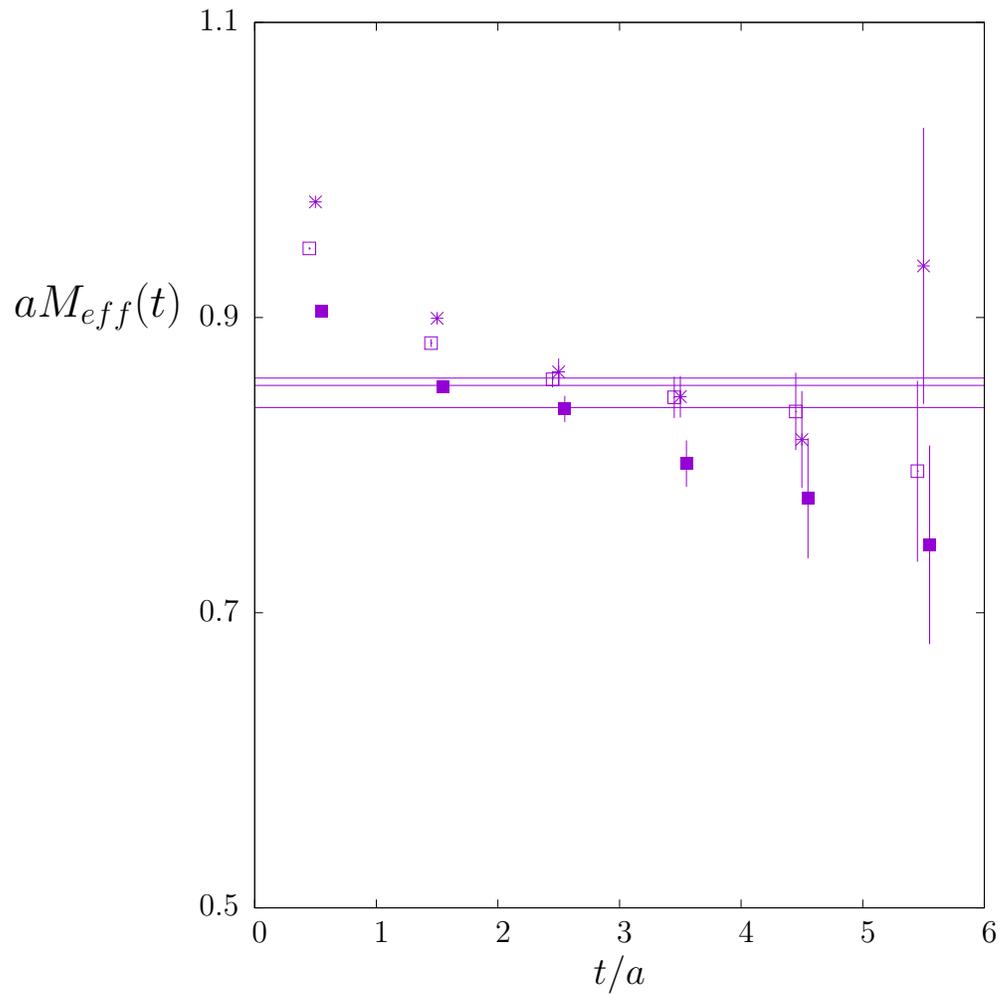}
\end	{center}
\caption{Effective masses for the $J^{PC}=2^{--}$ ground state;
  2 of the 5 components from the $E$ representation ($\blacksquare$)
  and the other 3 from the $T_2$ representation ($\square$). Also the
  triply degenerate $J^{PC}=1^{--}$ ground state ($\star$).
  Some points shifted for clarity. Horizontal lines are our best mass estimates.
  On the $38^4$ lattice at $\beta=6.50$.}
\label{fig_Meff12mm_b6.5}
\end{figure}

In Fig.~\ref{fig_Meff012mp_b6.5} we display the effective masses of some $P,C=-,+$ states.
The lightest and first excited $0^{-+}$ have well defined plateaux starting at around
$t=2a$. As do the ground and first excited $2^{-+}$, with a near-degeneracy between
the $E$ and $T_2$ components. The lightest three $T_1^{-+}$ triplets are heavier and
have worse overlaps: for two of the states there is evidence for a plateau
starting near $t=2a$ but less evidence for the third. We assign all three to $J=1$
(see Section~\ref{subsection_spins}) but with modest confidence for some.
In Fig.~\ref{fig_Meff123pm_b6.5} we display the effective masses of some $P,C=+,-$ states.
The lightest and first excited $1^{+-}$ have well defined plateaux starting at around
$t=2a$. The $2^{+-}$ are reasonably convincing. The near degeneracies needed to
identify the $J=3$ state are convincing, and the effective mass plateau is quite
plausible. 
Finally in Fig.~\ref{fig_Meff12mm_b6.5} we display the $1^{--}$ and $2^{--}$
effective masses and it is clear that here we can only make an estimate by assuming
that, as in the previous states, the plateau starts from $t=2a$.

The message here is that our lighter states are robust but that as the mass becomes
higher the extraction of a mass will typically become more ambiguous. In addition we
have seen that for most states the effective mass plateau begins at $t=2a$, or
even $t=a$, in this calculation at $\beta=6.50$. Since our operator basis is designed
so that the correlators scale approximately with the physical mass scale, this implies
that for $\beta\leq 6.0$ the effective mass plateau will usually begin at $t=a$.
Since at our smaller $\beta$ values the lattice spacing $a$ and hence the
lattice mass $aM$ are considerably larger, an effective mass plateau is
harder to establish except for the lightest states, and in these cases the fact that
we can assume the plateau typically begins from $t=a$ becomes useful.

We list our resulting mass estimates in Tables~\ref{table_M_b5.80_b5.6924}
to  \ref{table_M_b6.50}.

\section{Continuum limit}
\label{section_continuum} 

We take the masses in  Tables~\ref{table_M_b5.80_b5.6924} to  \ref{table_M_b6.50}
and express them  as $M/\surd\sigma$ in units of the string tension that we have
already calculated. It is this dimensionless ratio that we extrapolate to the continuum
limit in Section~\ref{subsection_massratios} using eqn(\ref{eqn_MKcont}).
However while expressing the continuum masses in terms of the string tension is
theoretically elegant it is not ideal if we are thinking of phenomenological
applications where we want to express masses in GeV units. For this purpose
one typically  expresses masses in units of the $r_0$ scale
\cite{Sommer-r0}
and hence into `GeV' units, as discussed in Section~\ref{subsection_units}.
Of course this still involves some ambiguity because the $SU(3)$ gauge theory
and QCD are two different theories and there can be no single rescaling that will
match their physics.
In addition to expressing the masses in physical units we will obviously want to identify the
continuum spins of as many states as possible, which we do in Section~\ref{subsection_spins}.

\subsection{Continuum  mass ratios}
\label{subsection_massratios}

We use eqn(\ref{eqn_MKcont}) to extrapolate our mass ratios, $M/\surd\sigma$, to the
continuum limit. We choose to use just the leading $O(a^2\sigma)$ correction
and where necessary we drop masses at the largest values of $\beta$ in order
to obtain acceptable fits. We list in Table~\ref{table_MK_R} the resulting continuum
extrapolations for states with various quantum numbers. Some of the
entries in Table~\ref{table_MK_R} are starred, indicating particularly poor fits
where the $\chi^2$ per degree of freedom is greater than 2.5. How good the fits
are is cleary an important issue. This involves not only the $\chi^2$ per degree
of freedom but also the range fitted. So  to provide evidence for the reliability
of these continuum extrapolations we shall show explicitly a number of them below.
Since the most important states are those whose continuum spin we are able to identify,
these are the ones we shall show.

Note that by deliberately limiting ourselves in Table~\ref{table_MK_R} to the lightest
four $A_1^{++}$ and $E^{++}$ states we expect to exclude from our spectra the
finite-volume ditorelon states that were discussed in Section~\ref{subsection_volume}.

\begin{figure}[htb]
\begin	{center}
\leavevmode
\input	{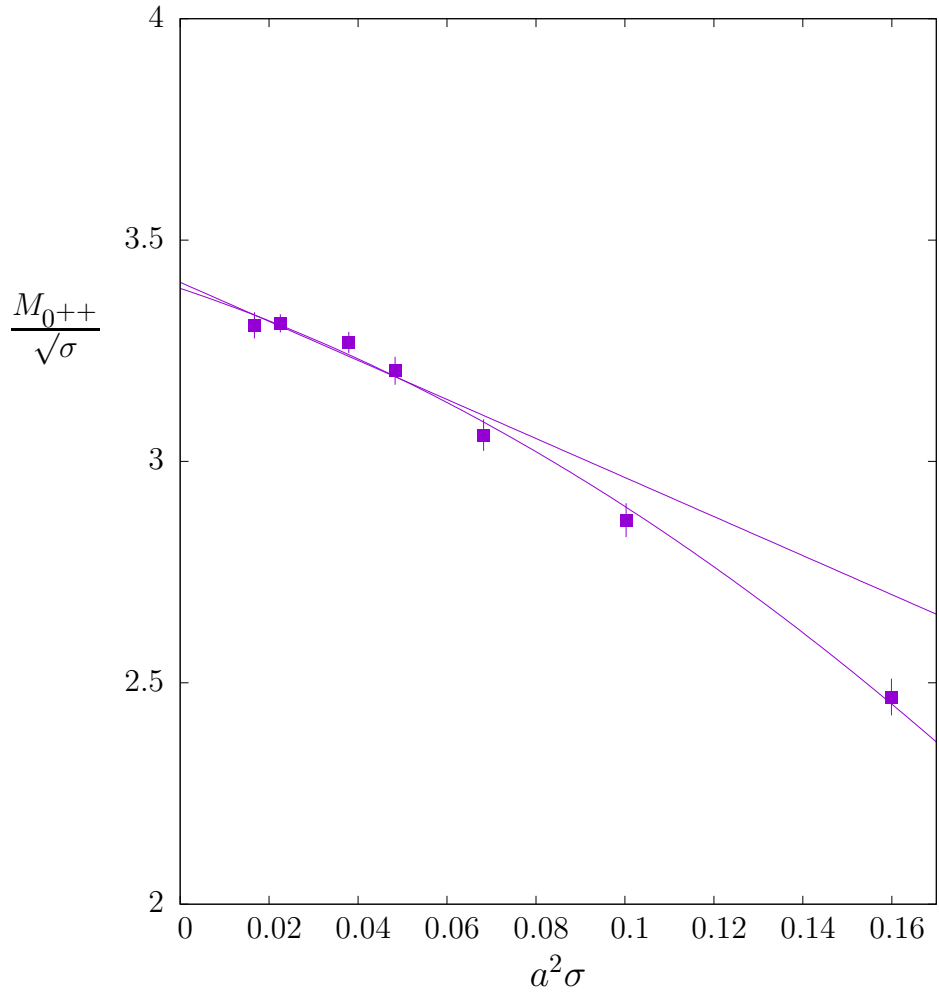}
\end	{center}
\caption{The lightest $J^{PC}=0^{++}$ glueball with linear and quadratic
extrapolations to the continuum limit.}
\label{fig_MJ0ppgsK_cont}
\end{figure}

We begin by showing in Fig.~\ref{fig_MJ0ppgsK_cont} two extrapolations of the very
lightest glueball mass, the $A_1^{++}$ ground state which has $J=0$. One extrapolation
keeps just the leading $O(a^2\sigma)$ correction, while the other includes an additional
$O(a^4\sigma^2)$ correction. As is well known this state has a deep dip around
$\beta \sim 5.5$ where one has a crossover between the regimes of weak and strong bare
coupling. This dip is a reason for attempting to improve the linear fit with a
higher order correction. However, since our calculations go to quite large $\beta$
the continuum limits obtained with these two extrapolations turn out
to be identical within errors:
$M_{0^{++}}/\surd \sigma = 3.405(21)$ and $3.391(23)$ respectively. We shall employ
the first value, primarily because it makes no difference, but also because the
strong to weak coupling transition involves a change in the functional form of
the expansion in $\beta$, and there is no guarantee that simply including a
higher order term derived in weak-coupling is the right way to parametrise it.

\begin{figure}[htb]
\begin	{center}
\leavevmode
\input	{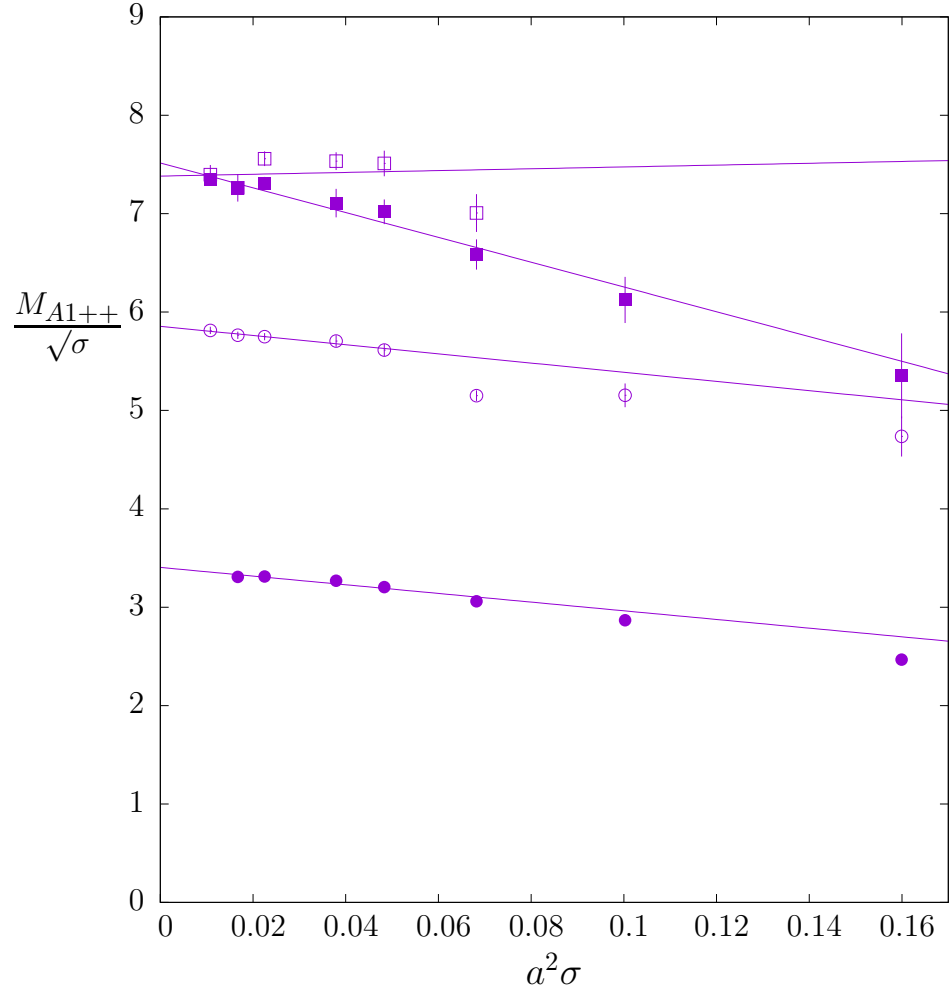}
\end	{center}
\caption{Continuum extrapolations of the lightest two $J^{PC}=0^{++}$ glueballs,
  $\bullet$ and $\circ$, as well as the next two heavier $A1^{++}$ glueballs.}
\label{fig_MA1ppK_cont}
\end{figure}

In Fig.~\ref{fig_MA1ppK_cont} we display the continuum extrapolations of the
lightest four $A_1^{++}$ states. The lightest two have $J=0$; and the very lightest
has already been shown in Fig.~\ref{fig_MJ0ppgsK_cont}. The third state
probably belongs to $J=4$ (see below) and we have no $J$ assignment for the
fourth state. As we see, the continuum limits appear to be quite unambiguous in all cases.

In Fig.~\ref{fig_MJ2ppK_cont} we move on to the lightest and first excited $E^{++}$
and $T_2^{++}$ states. Recall that since the states in the $E$ come in degenerate pairs
(up to fluctuations), we average the correlators to produce a single mass.
Similarly for the triply degenerate $T_2$ (and $T_1$). So, for example, the lightest
$T_2$ mass corresponds to the lightest three states and the first excited $T_2$ mass
corresponds to the next three states. This labelling of the states is used throughout
the paper. Returning to Fig.~\ref{fig_MJ2ppK_cont} we see that for these states,
which we shall see below correspond to $J=2$, there is remarkably little lattice
dependence over our whole range of $\beta$, leading to continuum extrapolations
that are unambiguous.

Some heavier states with $P=+$ and $C=+$ are displayed in Fig.~\ref{fig_MJ3ppK_cont}
and Fig.~\ref{fig_MJ4ppK_cont}. Here one sometimes sees a scatter of points indicating
worse fits, and this is presumably due to the increasing uncertainty in extracting the
masses of heavier states. Nonetheless the extrapolations are plausible within the
errors quoted in Table~\ref{table_MK_R}.

\begin{figure}[htb]
\begin	{center}
\leavevmode
\input	{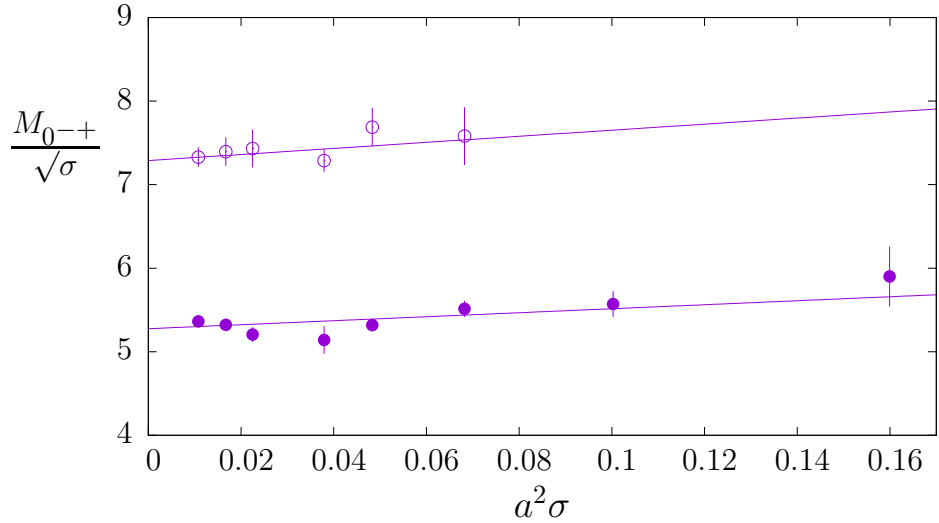}
\end	{center}
\caption{Continuum extrapolation of the two lightest $J^{PC}=0^{-+}$ glueballs
  as obtained from the lightest two states of the $A1^{-+}$ representation.}
\label{fig_MJ0mpK_cont}
\end{figure}

\begin{figure}[htb]
\begin	{center}
\leavevmode
\input	{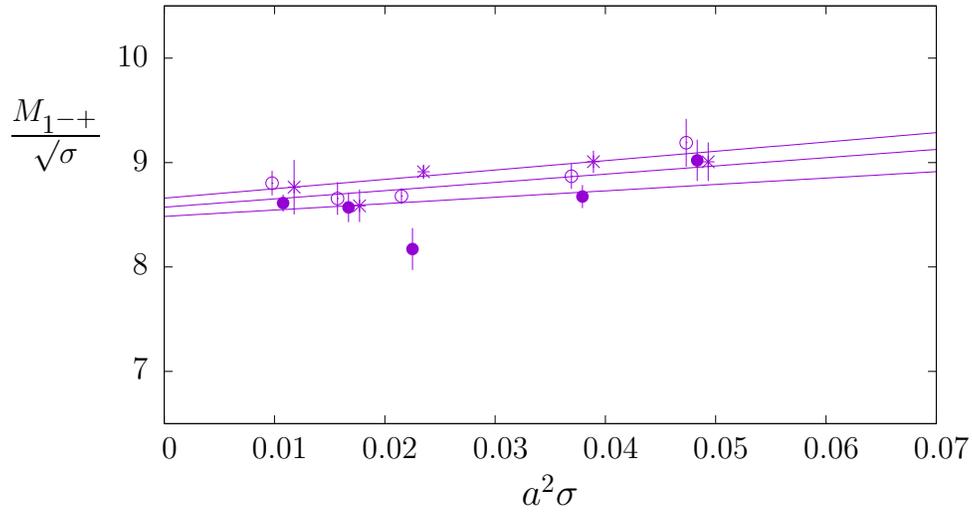}
\end	{center}
\caption{Continuum extrapolation of the masses of the lightest three $J^{PC}=1^{-+}$ glueballs
  which appear to be (nearly) degenerate. They 
  correspond to the lightest three $T1^{-+}$ states, each of which is triply degenerate
  (corresponding to the three components of each of these  three $J=1$ states).}
\label{fig_MJ1mpK_cont}
\end{figure}

Moving now to states with $P=-$ and $C=+$. The lightest two $A_1^{-+}$ states
plotted in Fig.~\ref{fig_MJ0mpK_cont} show small $O(a^2)$ corrections and are
simple to extrapolate to the continuum. The lightest two $E^{-+}$ and $T_2^{-+}$
states, plotted in Fig.~\ref{fig_MJ2mpK_cont}, also show small corrections
and hence easy extrapolations. In Fig.~\ref{fig_MJ1mpK_cont} we plot the lightest
three $T_1^{-+}$ states (each triply degenerate) which have very similar masses.

\begin{figure}[htb]
\begin	{center}
\leavevmode
\input	{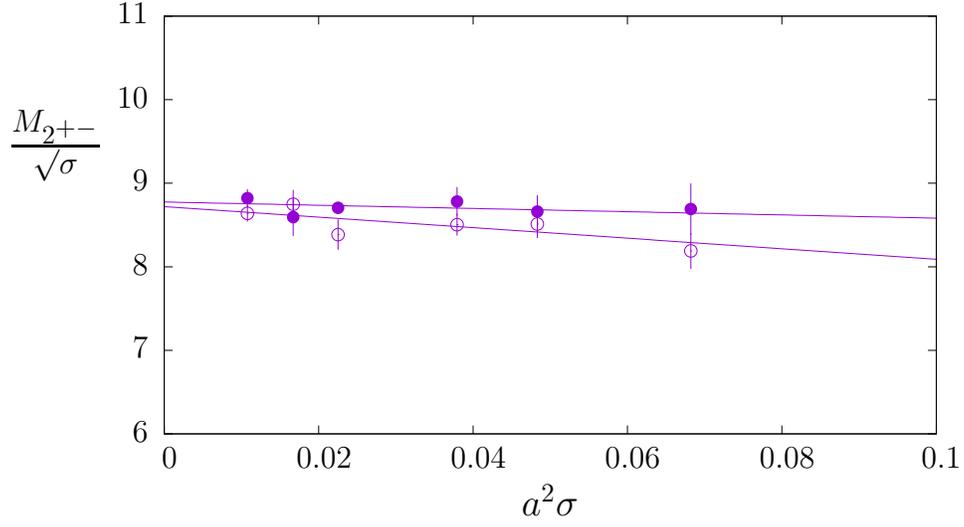}
\end	{center}
\caption{Continuum extrapolation of the masses of the lightest $J^{PC}=2^{+-}$ glueball,
  composed of the lightest (doubly degenerate) $E^{+-}$ state, $\bullet$, and the
  first excited (triply degenerate) $T2^{+-}$ state, $\circ$.}
\label{fig_MJ2pmK_cont}
\end{figure}

\begin{figure}[htb]
\begin	{center}
\leavevmode
\input	{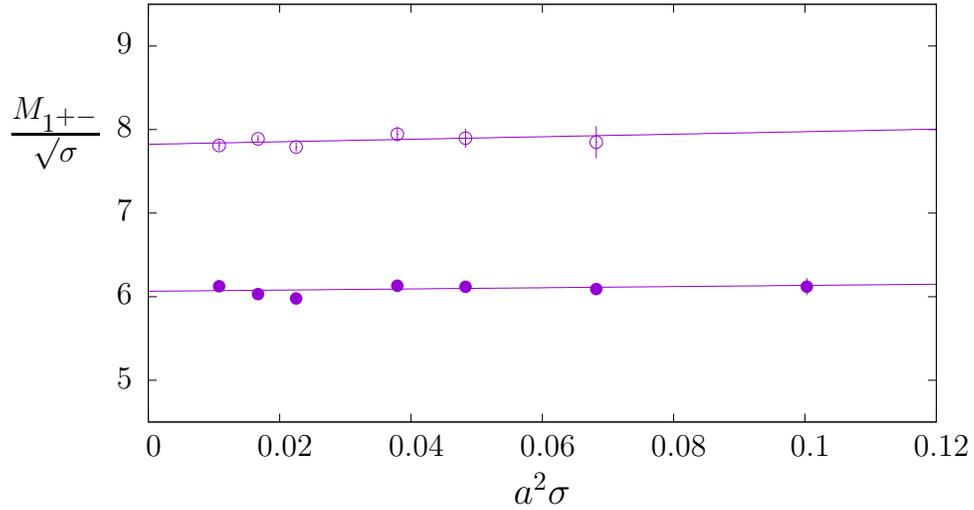}
\end	{center}
\caption{Continuum extrapolation of the masses of the lightest
  and first excited $J^{PC}=1^{+-}$ glueball states corresponding
  to the ground and second excited (triply degenerate) $T1^{+-}$ states.}
\label{fig_MJ1pmK_cont}
\end{figure}

\begin{figure}[htb]
\begin	{center}
\leavevmode
\input	{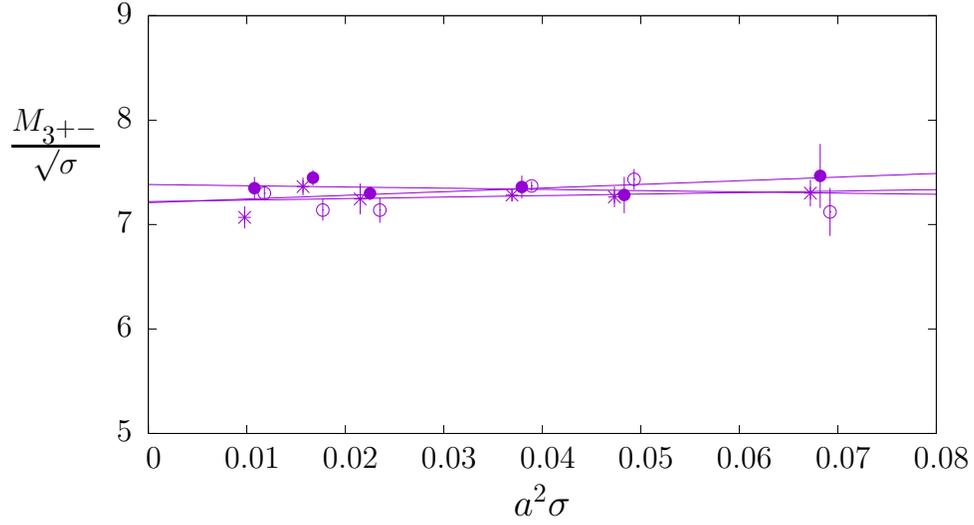}
\end	{center}
\caption{Continuum extrapolation of the lightest $J^{PC}=3^{+-}$ glueball
  as obtained from the lightest $A2^{+-}$ state ($\bullet$), the first
  excited $T1^{+-}$ state ($\circ$ and triply degenerate) and the lightest
  $T2^{+-}$ states ($\star$ and triply degenerate). Hence the seven components
  of the $J=3$ ground state.}
\label{fig_MJ3pmK_cont}
\end{figure}

\begin{figure}[htb]
\begin	{center}
\leavevmode
\input	{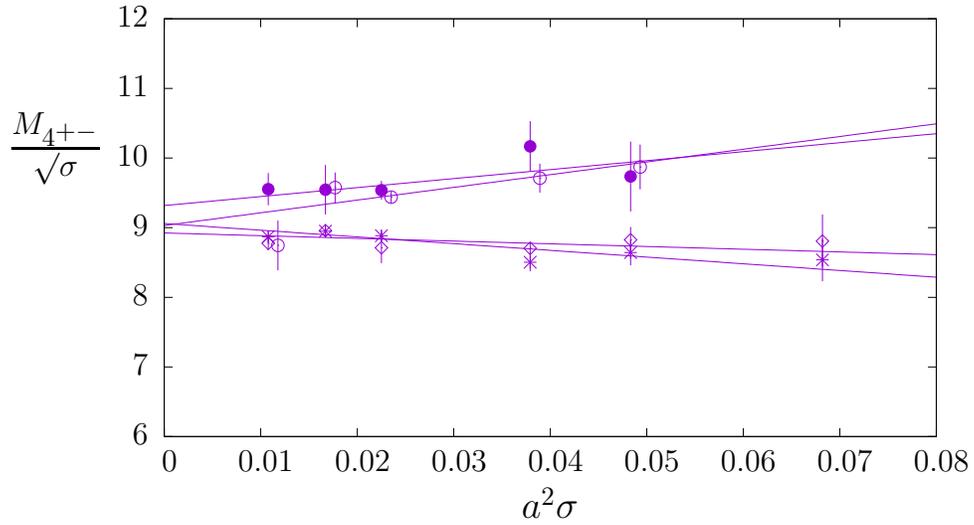}
\end	{center}
\caption{Continuum extrapolation of the lightest $J^{PC}=4^{+-}$ glueball
  as obtained from the lightest $A1^{+-}$ ($\bullet$), the doubly degenerate
  first excited $E^{+-}$ ($\circ$), the triply degenerate third excited $T1^{+-}$
  ($\lozenge$),  and the triply degenerate second excited $T2^{+-}$ ($\star$).
  Hence the nine components of this $J=4$ state. This assignment is uncertain.}
\label{fig_MJ4pmK_cont}
\end{figure}

\begin{figure}[htb]
\begin	{center}
\leavevmode
\input	{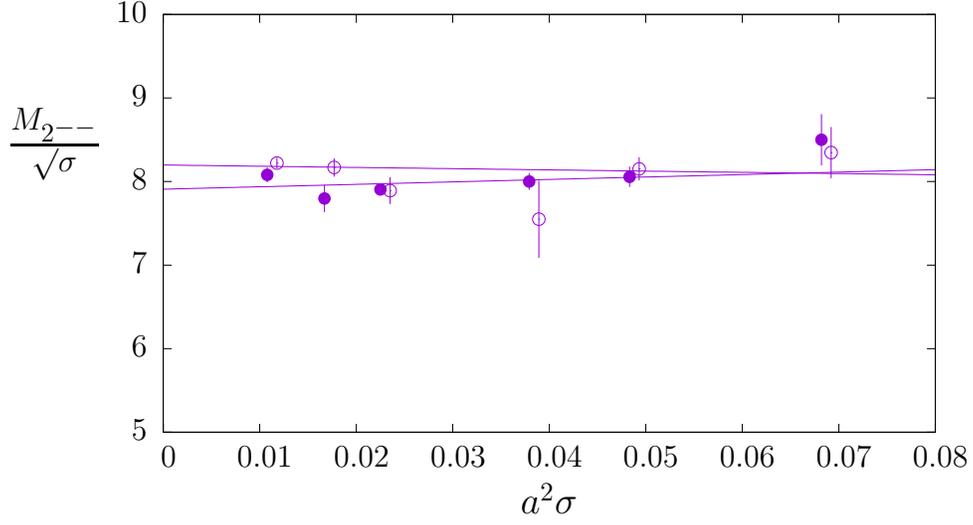}
\end	{center}
\caption{Continuum extrapolation of the lightest $J^{PC}=2^{--}$ glueball,
  as obtained from the lightest $E^{--}$ state ($\bullet$, doubly degenerate)
  and the lightest $T2^{--}$ state ($\circ$, triply degenerate).
  Hence the five components of the two $J=2$ state.}
\label{fig_MJ2mmK_cont}
\end{figure}

\begin{figure}[htb]
\begin	{center}
\leavevmode
\input	{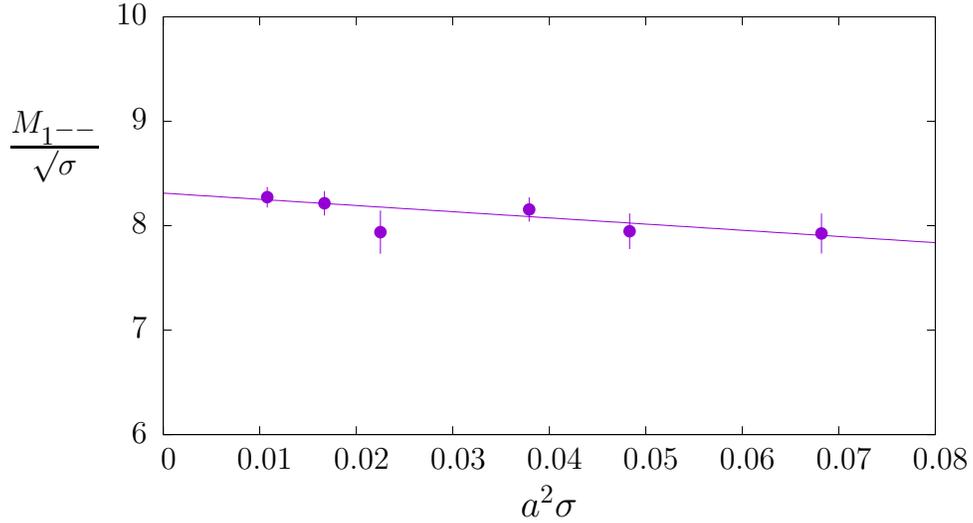}
\end	{center}
\caption{Continuum extrapolation of the mass of the lightest
  $J^{PC}=1^{--}$ glueball state corresponding
  to the (triply degenerate) $T1^{+-}$ ground state.}
\label{fig_MJ1mmK_cont}
\end{figure}

Turning to the  $C=-$ states, we plot in Figs~\ref{fig_MJ2pmK_cont},~\ref{fig_MJ1pmK_cont},
~\ref{fig_MJ3pmK_cont} and ~\ref{fig_MJ4pmK_cont} a number of states with $P=+$.
Those in the first three figures have well defined extrapolations, but the heaviest
states in the last figure are increasingly uncertain. Some states with $P=-$
are shown in Figs~\ref{fig_MJ2mmK_cont},~\ref{fig_MJ1mmK_cont} and while the
extrapolations of the $E^{--}$ and $T_1^{--}$ ground states are unambiguous, that
of the $T_2^{--}$ ground state is only plausible.

\subsection{Physical units}
\label{subsection_units} 

We have used the string tension as the energy scale for our continuum extrapolation of the
glueball masses both because it is a quantity that is of fundamental dynamical significance
in the theory and because one can calculate it very accurately in our lattice calculation,
as described in Section~\ref{section_sigma}. However for phenomenological applications
one would like to present the spectrum in `GeV' units. Strictly speaking this is of course
not possible: the pure gauge theory and QCD, with its various quark masses at their physical
values, are two different theories. Nonetheless phenomena such as the OZI rule and lattice
calculations showing that $SU(3)$ is `close to' $SU(\infty)$, both for the pure gauge theory
\cite{SU3-SUN}
and also when quarks are added 
\cite{QCD-N},
motivates attempting a translation into  GeV units, something which has traditionally been
done in past lattice calculations.

One possibility is to use the QCD value for the string tension $\sigma$. Since the  QCD
string will contain virtual $q\bar{q}$ pairs -- which eventually cause it to break -- this is
not unambiguous. For a rough estimate one might look to the long-distance linearly rising piece
of the potential describing heavy quarkonia, but since the dynamics of these heavy states is
primarily determined by the interactions at short distances this is not very constraining.
The traditional approach is to take the slope of the observed Regge trajectories and to
use a simple asymptotic rotating string model, and this gives something like
$\surd\sigma \sim 440\, MeV$. How good this interpretation is for the relatively low $J$
states that in practice determine these Regge trajectories is of course uncertain.

An alternative scale that has been widely used is to determine the distance $r$ at which the
slope of the static (heavy) quark potential, the force $F(r)$, has some value. For example
$r^2F(r)|_{r=r_0} = 1.65$ determines the scale $r_0$
\cite{Sommer-r0}
and there are arguments that this should be relatively insensitive to the inclusion of quarks. An
analysis of various lattice QCD calculations with at least 3 light quarks leads to the estimate
\cite{RS-2014}
\begin{equation}
                 r_0 = 0.472(5)fm =  \frac{1}{418(5) MeV}.
\label{eqn_r0MeV}
\end{equation}
We have not attempted to calculate $r_0/a$ in our calculations but there are calculations
covering our range of $\beta$ in
\cite{HWalpha-1998,SNRS-2001}
which provide interpolating formulae for $r_0/a|_\beta$. We have used these formulae
to interpolate to our values of $\beta$ and have extrapolated the resulting values of
$r_0\surd\sigma$ to the continuum limit. We find good fits which agree within errors
and give us
\begin{equation}
                \surd\sigma = \frac{1.160(6)}{r_0} = 485(6) MeV
\label{eqn_KMeV}
\end{equation}
using the value for $r_0$ in eqn(\ref{eqn_r0MeV}). Using this scale we can convert
the continuum mass ratios in Table~\ref{table_MK_R} to GeV units and this results
in the masses listed in Table~\ref{table_MGeV_R}.

\subsection{Continuum glueball spins}
\label{subsection_spins} 

In the continuum theory a glueball transforms according to an irreducible representation of
the (proper) rotation group $SO(3)$ and so will possess some continuum spin $J$, and it will be part
of a set of $2J+1$ degenerate states. All such states will in principle be included in the set of
states we have obtained by extrapolating our lattice calculations, albeit with the obvious
caveats arising from the limitations of our basis, our statistics etc. However our calculated
states are all labelled by the irreducible representations $A_1,A_2,E,T_1,T_2$ of the
octahedral subgroup $O$
of $SO(3)$, so we need to map back from these to the multiplets corresponding to a given $J$.
The mapping can be readily derived (see Section 9.4 of
\cite{Hamermesh}),
and is summarised for the lowest values of $J$ in Table~\ref{table_J_R}. (There is a simple
relation allowing one to calculate the mapping for any $J$ once one has the mappings for
$J\leq 5$, as explained in Section 9.4 of
\cite{Hamermesh}).
Taking into account the triple degeneracy of
the $T_1$ and $T_2$ representations, and the double degeneracy of $E$, we can see that
the sum of states is indeed $2J+1$ in each case, as it should be. We have, optimistically
included the mapping up to $J=8$ although, as we will see, we are not able to calculate
masses beyond $J=4$.

In a calculation with infinite statistics identifying states of spin $J$ from our continuum
calculations would be unambiguous: we would simply search for $2J+1$ degenerate states
located in the towers of states in each of the appropriate cubic representations listed
in Table~\ref{table_J_R}. In practice, with finite statistical and systematic errors and
with the energy gaps between neighbouring states in each tower decreasing with increasing
energy, this strategy has severe limitations. An alternative approach would be to
exploit the fact that even on a cubic lattice the full $SO(3)$ rotational invariance
becomes a better approximation on physical length scales as $a\to 0$. Once one calculates
the variationally selected glueball wave-functional at each $\beta$ one can analyse its
approximate rotational properties so as to determine the probable continuum spin and
see if it approaches the corresponding continuum angular wave-function as $a(\beta)$
decreases. A pioneering calculation of this type
\cite{HM-2005,HMMT-spin}
in $D=3+1$ gauge theories appears promising. Moreover in $D=2+1$, where the rotation group is
Abelian so that there is no $J$-dependent degeneracy to exploit, this technique
has proved quite powerful in recent work
\cite{PCSDMT-2019}
that builds upon the earlier exploratory studies
\cite{HM-2005}.

In the present paper we attempt to identify states of various spin $J$ using the decomposition
in Table~\ref{table_J_R} applied to our results in Table~\ref{table_MK_R}. The arguments will
be as follows, leading to the identifications presented in Table~\ref{table_M_J_R}. We will
also find it useful to refer to the  spectra obtained at various values of $a$,
on the assumption that lattice corrections are small, as appears to be the case for most states.

\begin{figure}[htb]
\begin	{center}
\leavevmode
\input	{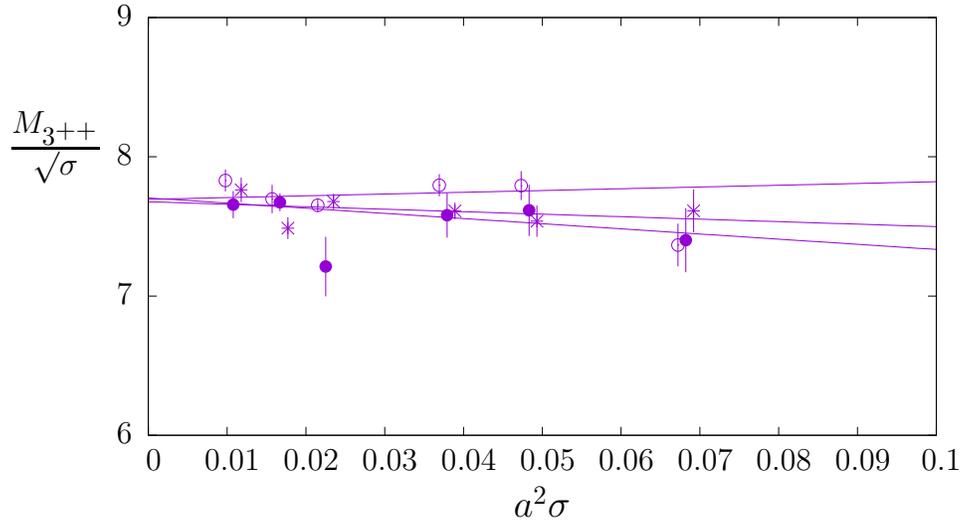}
\end	{center}
\caption{Continuum extrapolation of the lightest $J^{PC}=3^{++}$ glueball.
  Obtained from the lightest $A2^{++}$ state ($\bullet$), the lightest $T1^{++}$ state 
  ($\circ$,  triply degenerate) and the third excited $T2^{++}$ states ($\star$,
  triply degenerate): hence the seven components of the $J=3$ ground state.}
\label{fig_MJ3ppK_cont}
\end{figure}

We begin with the $P,C=+,+$ sector of states.
First we note that the lightest two masses in $R^{PC}=A_1^{++}$ are very much lighter
(as always, compared to errors) than any states in other $R^{++}$. So these are, unambiguously,
the ground and first excited $J^{PC}=0^{++}$ states. The lightest two $E^{++}$ states (each
doubly degenerate) are nearly degenerate with the  lightest two $T_2^{++}$ states (each triply
degenerate) but not with any other states (by a large margin). Moreover this near-degeneracy
is the case at all values of $\beta$ as we see in Fig.~\ref{fig_MJ2ppK_cont}.
So these are unambiguously, the ground and first excited $J^{PC}=2^{++}$ states. 
One might wonder if the lightest $T_1^{++}$ state in Table~\ref{table_MK_R} is $J=1$? We note however
that it is essentially degenerate with the first excited $T_1^{++}$ and also with the lightest
$A_2^{++}$ and with either the second ot third excited  $T_1^{++}$ state, which are nearly degenerate.
In addition the second excited $A_1^{++}$ and  $E^{++}$ have similar masses. This suggests that
we have a $3^{++}$ ground state and a  $4^{++}$ ground state as in Table~\ref{table_M_J_R},
which happen to have very similar masses. If we look at the states as a function of $\beta$,
with the $3^{++}$ in Fig.~\ref{fig_MJ3ppK_cont} and the $4^{++}$ in Fig.~\ref{fig_MJ4ppK_cont},
then the assignments look plausible, particularly in the former case. 
However it is clear that these choices are no more than plausible, because all the states
shown in Fig.~\ref{fig_MJ3ppK_cont} and Fig.~\ref{fig_MJ4ppK_cont} are very close in mass.
An alternative interpretation could be that the lightest $T_1^{++}$ is $J=1$ and that the remaining
states listed above make up the $J=6^{++}$ ground state. However this would mean that the $J=6$
ground state is lighter than the $J=4$ ground state, which appears less likely to us.
Hence our choice. However this discussion illustrates the limitations of our approach. 

We now turn to the $P,C=-,+$ sector of states. The lightest two $A_1^{-+}$ states have no partners in
other representations and so are unambiguously the ground and first excited $0^{-+}$ states. The
lightest two  $E^{-+}$ states are nearly degenerate with the  lightest two $T_2^{-+}$ states but not with
any other states, so it is unambiguous that they are the ground and first excited $2^{-+}$ states.
Again this is reinforced by noting the near-degeneracy at all values of $\beta$ as shown
in Fig.~\ref{fig_MJ2ppK_cont}. The lightest three $T_1^{-+}$ states (each triply degenerate) have no
obvious partners in other representations, so we have labelled all of them as being $1^{-+}$
states. We do so with some unease: why the apparent degeneracy of these three states? (See
also Fig.~\ref{fig_MJ1mpK_cont}.) Perhaps
only one is truly $1^{-+}$ and the other two are nearly degenerate because they appear together
as components of a higher spin state? Unfortunately we have no plausible partners in other
representations to realise this suggestion.

In the  $P,C=+,-$ sector of states the lightest glueball is the $T_1^{+-}$ and it has no partners elsewhere:
hence it is unambigously the $1^{+-}$ ground state. The second excited  $T_1^{+-}$  has no
partners elsewhere so we are confident that it is the first excited $1^{+-}$ glueball. Both
their continuum limits are robust as we see from Fig.~\ref{fig_MJ1pmK_cont}. The $A_2^{+-}$ ground state
is nearly degenerate with the $T_2^{+-}$ ground state and with the first excited $T_1^{+-}$, with no
nearby states elsewhere. And, as we see in Fig.~\ref{fig_MJ3pmK_cont}, this is so at all $\beta$.
Hence we are confident that this is the $3^{+-}$ ground state.
There appears to be little ambiguity in identifying the $3^{+-}$ ground state with
the $A_2^{+-}$ and $T_2^{+-}$ ground states and the first excited  $T_1^{+-}$, since these states
are not very heavy and it is clear that there are no partner states elsewhere.
We further note that the ground state $E^{+-}$ and the first excited $T_2^{+-}$ are nearly degenerate
and so we assign them to the $2^{+-}$ ground state. This is also the case throughout
our range of $\beta$ as we see in Fig.~\ref{fig_MJ2pmK_cont}. However
the mass and errors are large, and there are other states with masses about
$2\sigma$ away, so the assignment has some uncertainty. However looking at the corresponding
spectra at $\beta=6.50, 6.338, 6.235$ does increases the evidence for no partner states elsewhere.
Finally we can speculate that the $A_1^{+-}$ ground state partners the first excited $E^{+-}$, the
second excited  $T_2^{+-}$ and the third excited  $T_1^{+-}$ to form the  $4^{+-}$ ground state.
This claim is slightly weakened by what we observe in the spectra at smaller $\beta$, as shown
in Fig.~\ref{fig_MJ4pmK_cont} and is somewhat flimsy.

Finally, in the $P,C=-,-$ sector the states are all heavy making it difficult to calculate much of the
spectrum with sufficient accuracy to be able to search for near-degeneracies. The lightest three
states are the $E^{--}$, $T_1^{--}$ and $T_2^{--}$ ground states and there are no potential partners to
these. Given the possibilies listed in Table~\ref{table_J_R}, there is no ambiguity in assigning
the $E^{--}$ and $T_2^{--}$ ground states to give the $2^{--}$ ground state, 
and the $T_1^{--}$ ground state to be the  $1^{--}$ ground state.

Having made these identifications we average the masses of the states in Table~\ref{table_MK_R}
that correspond to a state of given $J^{PC}$, taking some account of any splittings between the
masses in the different cubic representations. This leads to the masses listed in
Table~\ref{table_MK_J}, for a number of continuum states with various values of $J^{PC}$.
Where we have some hesitation about the spin assignment, as discussed above, we qualify
the value with a single star. Two stars indicate a speculation.

Our predictions for the glueball masses are given in terms of the string tension.
If one wishes one may express the masses in units of the mass gap, the mass of the
lightest $0^{++}$ glueball, simply by multiplying the various mass ratios by the inverse
of $m_{0^{++}}/\surd\sigma$. Since many of these states may be of phenomenological
interest it is useful to provide an estimate in physical GeV units and this
we do in Table~\ref{table_MGeV_J} using the value of $\surd\sigma$ estimated in
Section~\ref{subsection_units}.

\section{Comparisons}
\label{section_comparisons} 

It is clearly useful to compare our calculations to earlier ones. Since the number of these
is too large to attempt a comprehensive comparison, we shall choose a few representative 
calculations (within which one may find references to other calculations).

One of the earlier realistic attempts to calculate the ground states of all the representations of the
cubic group was in
\cite{CMMT-1989},
and this was extended to a smaller value of $a(\beta)$ in
\cite{CM-UKQCD-1993}.
The values for the masses of the states of greatest phenomenological interest, i.e. those that
have masses $ < 3 {\rm GeV}$, are consistent with ours (within two standard deviations) albeit with
statistical errors at least ten times larger than ours. The point here is that the approximate
masses of these states, and their level ordering, have been established for a very long time.

When one reduces the statistical errors, one also needs to reduce the systematic errors so that
they do not become the dominant error. An important example is the need to improve the extrapolation
to the continuum limit: in  
\cite{CMMT-1989}.
the mass ratios were consistent within errors with being independent of $\beta$, so that a
constant extrapolation was justified, but with smaller errors a variation with $\beta$
becomes visible and one needs to include at least an $O(a^2\sigma)$ correction.
One also needs to go to smaller values of  $a(\beta)$ to better control the continuum
extrapolation, as for example in
\cite{CM-UKQCD-1993}.

Another systematic error arises in the calculation of heavier glueballs. One needs the mass in
lattice units, $aM$, to be not too large if one is to observe some kind of effective
mass plateau in the correlation function. One way to achieve this is to go to small enough
$a(\beta)$. (At the same time reaping benefits in the continuum extrapolation.) An
alternative is to note that it is the temporal lattice spacing $a_t$ that appears in
the correlator, so if one works on lattices where the spatial, $a_s$, and temporal lattice spacings
differ with  $a_t \ll a_s$ one can achieve modest values for $a_tM$ even if $M$ is large,
while working on $L_s^3L_t$ lattices with $L_s \ll L_t$ which are of a manageable size.
This technique proved invaluable in some of the earliest calculations of the lightest
scalar and tensor glueballs 
\cite{KIGSMT-1983}
in providing evidence for effective mass plateaux. It has been used subsequently to good
effect in a comprehensive calculation of the mass spectrum that includes many heavy glueball states 
\cite{MP-1999,MP-2005}.
One weakness with such calculations is that they typically involve larger spatial
lattice spacings $a_s$ than those on symmetric lattices, and hence risk larger sytematic
errors in the continuum extrapolation where the leading correction is $O(a_s^2)$.
So, for example, the smallest value of $a_s$ in
\cite{MP-1999,MP-2005}
is $a_sM_g \simeq 3a_tM_g \simeq 0.84$, where $M_g$ is the mass gap, whereas the calculations
in the present paper go down to $a_sM_g \simeq 0.35$. A second issue is that it is difficult
to estimate the corrections to the inputted tree level value of $a_s/a_t$, and the introduction of
a physical scale such as $r_0$ or $\surd\sigma$ will depend on this ratio. While the resulting
uncertainty may be only at the (few) percent level, this becomes significant in precise calculations.

A further important systematic error arises once one is more ambitious and wishes to calculate some
significant number of excited states in each $R^{PC}$ sector. The problem arises from the finite
size of one's operator basis: one will completely miss states which do not have a substantial overlap
onto the basis. Such missing states can lead to problems in a comparison with theoretical predictions
and in the identification of the near degenerate multiplets associated with a given continuum spin $J$.
Here the best one can do is to make the operator basis as large as practical, as we have
done in the present work.

If one is interested in the spin $J$ that a lattice state extrapolates to, the identification
of the $2J+1$ nearly degenerate states in the appropriate cubic representations becomes
rapidly problematic as the mass increases. Typically the density of states increases with
increasing mass, and with finite errors that also increase with the mass,
any apparent degeneracies are likely to be accidental.
An alternative approach to determining $J$ is to use sets of loop operators that are approximate
rotations of each other by angles that are some fraction of $\pi/2$. Since the important
operators have an extension that is constant in physical units, the approximate rotation
invariance becomes better as $a(\beta)$ deacreases, and one can resolve higher spins.
This method has been used to good effect in
\cite{HM-2005}
where the spins of an extensive set of states in various $R^{PC}$ representations were
determined. 

We now turn to a comparison with the specific results of
\cite{MP-2005,HM-2005}.
In our calculations we have a large basis of 27 loops, with each typically constructed out
of 5 or 6 different levels of blocked links, which is a considerably larger basis of operators
than that used in the earlier calculations we have discussed above. This has enabled us to
calculate more excited states within each quantum number sector. Having these extra
states also helps in the identification of their continuum $J$ using the $2J+1$ degeneracy
partitioned amongst the lattice states as in Table~\ref{table_J_R}. In practice we
are able to so identify more states than in
\cite{MP-2005}.
However apart from that we do broadly agree: for the 12 states listed in Table XXI of
\cite{MP-2005}
we agree with the spin assignment and mass (within errors) of 10 and only have serious doubts
about 2 of the heaviest states, the $J^{PC}=0^{+-}$ and the $3^{--}$. The former state
is based on the very heavy $A1^{+-}$ state which is quite possibly nearly degenerate
with excited $E^{+-}$, $T1^{+-}$ and $T2^{+-}$ states making it $4^{+-}$ rather than $0^{+-}$.
The claimed $3^{--}$ is based on the $A2^{--}$ ground state and we do not see an excited
$T1^{+-}$ with a mass that is plausibly degenerate. But only better calculations will be
able to tell. As for the lightest phenomologically interesting states, our masses agree
remarkably well except for a significant discrepancy for the $0^{++}$ ground state.
(Which is slightly increased by the fact that the scale used in 
\cite{MP-2005}
is $r_0^{-1}=410MeV$ while we use a later determination, $r_0^{-1}=418MeV$.)
Since this state has large lattice spacing corrections, we put this down to our
better control of the $O(a_s^2)$ corrections in taking the continuum limit.

The very different method used in
\cite{HM-2005}
for determining the spin $J$ of lattice states is more powerful and allows more
states to be categorised than we are able to do. This despite our much higher statistics
and larger range of $\beta$ which lead to much smaller statistical errors as well
as a better control of the continuum limit. Within these large errors we are
in agreement for most spins and masses, as we see by comparing our Table~\ref{table_MK_R}
to Tables 7.10,7.11 in 
\cite{HM-2005}.
There are some differences: in the $PC=++$ sector our candidate for the $J=4$ ground state
seems too light; in the $PC=-+$ sector two of our three nearly degenerate $T_1$ states might
be part of the $J=5$ ground state according to 
\cite{HM-2005}.
We note that with this method one could identify states with spin up to $J=6$ in the
now rather old calculations of
\cite{HM-2005},
something we could not hope to do despite our statistical errors being a factor
of 3 or 4 smaller and our range of $a(\beta)$ being significantly larger.

\section{Topological fluctuations}
\label{section_topology}

An interesting property of Euclidean non-Abelian gauge fields is that they possess non-trivial
topological properties, characterised by a topological charge $Q$ which is integer-valued 
in a space-time volume with periodic boundary conditions. This charge can be expressed
as the integral over Euclidean space-time of a topological charge density, $Q(x)$, where
\begin{equation}
Q(x)= {\frac{1}{32\pi^2}} \epsilon_{\mu\nu\rho\sigma}
Tr\{F_{\mu\nu}(x)F_{\rho\sigma}(x)\}.
\label{eqn_Q_cont}
\end{equation}
Recalling that the plaquette matrix $U_{\mu\nu}(x) = 1 + a^2 F_{\mu\nu}(x) + ....$
on smooth fields, one can write a lattice topological charge density $Q_L(x)$ as
\begin{equation}
Q_L(x) \equiv {\frac{1}{32\pi^2}} \epsilon_{\mu\nu\rho\sigma}
Tr\{U_{\mu\nu}(x)U_{\rho\sigma}(x)\}
= a^4 Q(x) +O(a^6).        
\label{eqn_Q_lat}
\end{equation}
While this definition is adequate for sufficiently smooth fields, it lacks the explicit
reflection properties of the continuum operator in eqn(\ref{eqn_Q_cont}), 
since all the plaquettes $U_{\mu\nu}(x)$ are defined as forward going in terms of our
coordinate basis. So it needs to be extended to include
terms with backward going plaquettes as well, if it is to be used on the `rough'
Monte Carlo generated fields, and so we employ the version
of this operator that is antisymmetrised with respect to forward and backward directions
\cite{DiVecchia-FFD}.

The fluctuations of realistic lattice gauge fields affect the observed value of $Q_L(x)$
in two important ways. Firstly we note that the
fluctuations of $Q_L(x)$ are related to expectation value of the composite operator
$Q_L^2(x)$ whose operator product expansion contains the unit operator
\cite{DiVecchia-FFD}.
So these fluctuations are powerlike in $\beta$ while the average of $Q_L(x)$ is $O(a^4)$ and
hence exponentially suppressed in $\beta$. Thus as $\beta$ increases the fluctuations
around $Q_L(x)$ and $Q_L=\sum_x Q_L(x)$ diverge compared to the physically interesting
mean values. In addition the lattice composite operator also receives a multiplicative
renormalisation $Z(\beta)$ such that $Z(\beta\to\infty)=1$ but which at accessible
values of $\beta$ strongly suppresses the charge
\cite{Pisa_ZQ}.

In practice all this means that one cannot extract the topological charge of a typical lattice
gauge field by directly calculating $Q_L=\sum_x Q_L(x)$ on that gauge field. However we note that
the fluctuations obscuring the value of $Q$ are ultraviolet, while the physically relevant
topological charge is on physical length scales. Thus if we perform a very limited local smoothening
of the fields to suppress the ultraviolet fluctuations, this should not affect physics
on long distance scales, and the value of $Q_L=\sum_x Q_L(x)$ calculated on these smoothened
fields should provide a reliable estimate of $Q$. Moreover, recalling that the total
topological charge of a gauge field is unchanged under smooth deformations, we can expect
that even under a moderately large amount of continued smoothening the value of  $Q_L$ will not
change, even though $Q_L(x)$ itself does gradually change. One convenient way to smoothen the
gauge fields is to locally minimise the action. Such a `cooling' of the original `hot'
lattice gauge field
\cite{MT-cool}
involves sweeping through the lattice one link at a time, precisely
like the Monte Carlo except that one chooses the link matrix that minimises the total
action of the plaquettes containing that link matrix. This is a standard technique
that one can find described in more detail in, for example,
\cite{DSMT-Q}. An alternative and attractive smoothing method with purturbatively proven
renormalisation properties is the gradient flow
\cite{Luscher:2010iy,Luscher:2011bx,Luscher:2013vga}
which has been shown to be numerically equivalent to cooling
\cite{Bonati:2014tqa,Alexandrou:2015yba,Alexandrou:2017hqw}.
Cooling appears to perform nearly two orders of magnitude faster than the gradient flow and
since we are aiming for large statistics we adopted cooling.
After the first couple of cooling sweeps the fields are already quite smooth, as we shall see below.
Since we are minimising the action and since in the continuum the minimum action field with a given
$Q$ is a multi-instanton field, we expect that under systematic cooling the lattice field will be driven
to become some multi-instanton field, which one can see by calculating the distribution $ Q_L(x)$ on such
a field. Of course, because of the discretisation of space-time the topological properties of lattice fields
are only approximately like those of a continuum field. One can deform a large instanton by gradually
shrinking its non-trivial core and on a lattice this core can shrink to within a hypercube. At this
point what was an instanton has been transformed into a gauge singularity and the value of $Q$
will now differ from its original value by one unit. (Equally, one can gradually grow an instanton
out of a hypercube.) This process can occur during cooling but it can equally occur during
the course of our Monte Carlo. In the latter case, it is these changes in $Q$ that allow us
to sample all possible values of $Q$ and hence maintain the ergodicity in $Q$ of our Markov process.
As $\beta\uparrow$ the distance between physical and ultraviolet scales grows and these
changes in $Q$ become increasingly suppressed -- in fact more strongly than the usual critical
slowing down (see below). 

One of our reasons for studying the topology of our $SU(3)$ gauge fields is to establish that
in our range of $\beta$ we have not in fact lost ergodicity in the topology $Q$ of the fields.
In the process we shall provide some insights into how reliable is the cooling method for
establishing the topological charge of the lattice gauge fields. As a side-product we will
also provide an accurate calculation of the topological susceptibility, which has been
much studied because of its appearance in sum rules for the lightest pseudoscalars in QCD.
For a more detailed discussion of these issues we refer to, for example,
\cite{DSMT-Q}.

\subsection{Topology on the lattice}
\label{subsection_Q_lattice} 

Since we calculate the topological charge $Q$ by cooling the lattice gauge fields,
it is relevant to ask how (un)ambiguous is the identification of $Q$ on those
cooled fields. (Note that from now on we will use the labels $Q$ and $Q_L$
interchangeably unless specified otherwise.)
The answer will depend on the value of $\beta$ and since the intrinsic
ambiguity in defining topology is a lattice artifact we might expect that any ambiguity
is greatest at the smallest values of $\beta$ where the lattice spacing is largest.
We therefore look at our smallest value of $\beta$, $\beta=5.6924$, an intermediate
value, $\beta=6.0625$, and our largest value, $\beta=6.50$. We note that the final
values of $Q$ that we shall quote are those that are obtained after 20 cooling sweeps.

\begin{figure}[htb]
\begin	{center}
\leavevmode
\input	{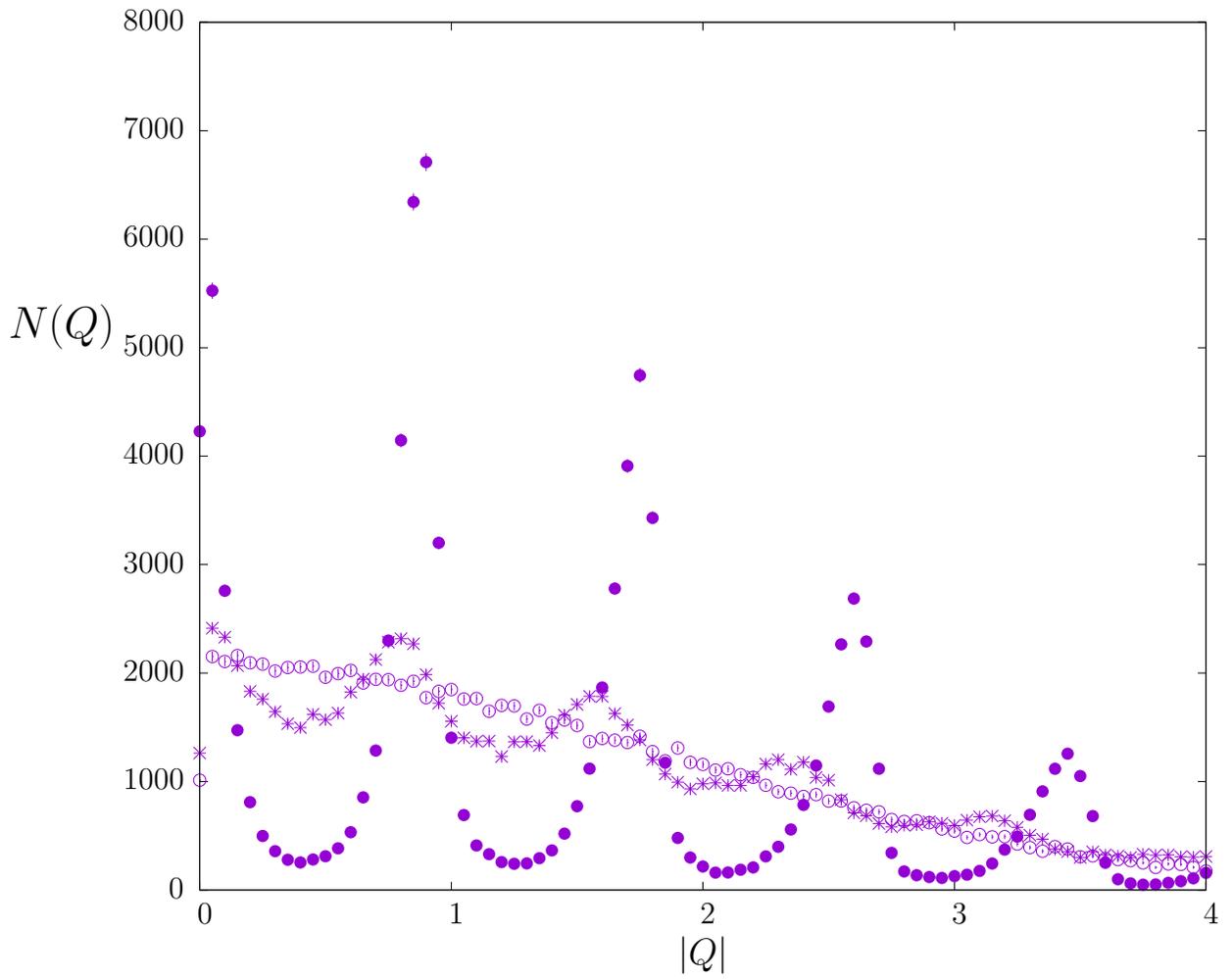}
\end	{center}
\caption{Histogram of lattice $Q$ after 4 cooling sweeps, $\circ$, 8 cooling sweeps, $\star$,
  and after 20 cooling sweeps, $\bullet$ on a $8^316$ lattice at $\beta=5.6924$.}
\label{fig_Q_p_b5.6924}
\end{figure}

We begin with $\beta=5.6924$, which we expect to be our `worst' case. In Fig.~\ref{fig_Q_p_b5.6924}
we show the distributions of $Q$ obtained after 4, 8 and 20 cooling sweeps on a total of
96000 $8^3 16$ lattice gauge fields. We see that after 20 cooling sweeps any ambiguity
in identifying $Q$ -- arising from the overlap of the tails between the peaks -- is at
most at the percent level. The peaks are shifted away from integer values due to the fact
that the typical instanton size, $\rho/a$ in units of the lattice spacing, is not 
large at this $\beta$. (One can calibrate the shifts using classical instanton
solutions discretised on a lattice.) However, while the value of $Q$ is well defined after
20 cooling sweeps, it is clearly quite ambiguous after 8 cooling sweeps, and totally 
invisible after 4 cooling sweeps. 

\begin{figure}[htb]
\begin	{center}
\leavevmode
\input	{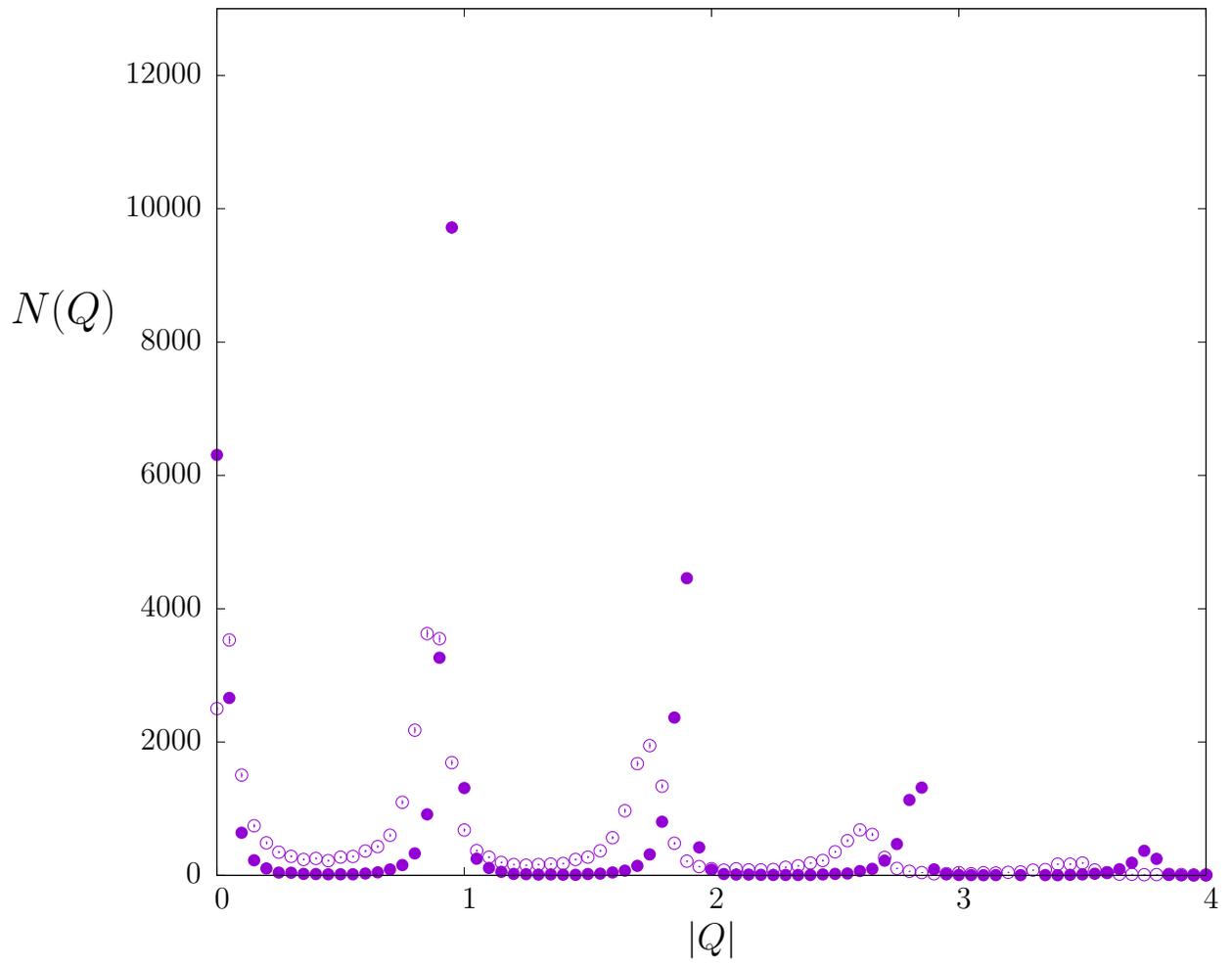}
\end	{center}
\caption{Histogram of lattice $Q$ after 4 cooling sweeps, $\circ$,
  and after 20 cooling sweeps, $\bullet$ on a $14^320$ lattice at $\beta=6.0625$.}
\label{fig_Q_p_b6.0625}
\end{figure}

We turn now to $\beta=6.0625$ where we calculate $Q$ on a total of 40000 $14^3 20$
lattice gauge fields. In Fig.~\ref{fig_Q_p_b6.0625} we show the distributions of $Q$
obtained after 4 and 20 cooling sweeps. It is clear that after 20 cooling sweeps
any ambiguity in identifying the value of $Q$ is totally negligible. More interestingly,
the value after only 4 cooling sweeps is reasonably well defined with only a minor ambiguity.
We also note that the peaks have moved closer to integer values, as should be the case
since we expect the instanton sizes to be larger in lattice units at this larger $\beta$.

Finally we show in  Fig.~\ref{fig_Q_p_b6.50} the distributions of $Q$
obtained after 4 and 20 cooling sweeps on 44000  $26^3 38$  lattice gauge fields
generated at  $\beta=6.50$. Now any ambiguity is completely negligible even after
only 4 cooling sweeps. 

\begin{figure}[htb]
\begin	{center}
\leavevmode
\input	{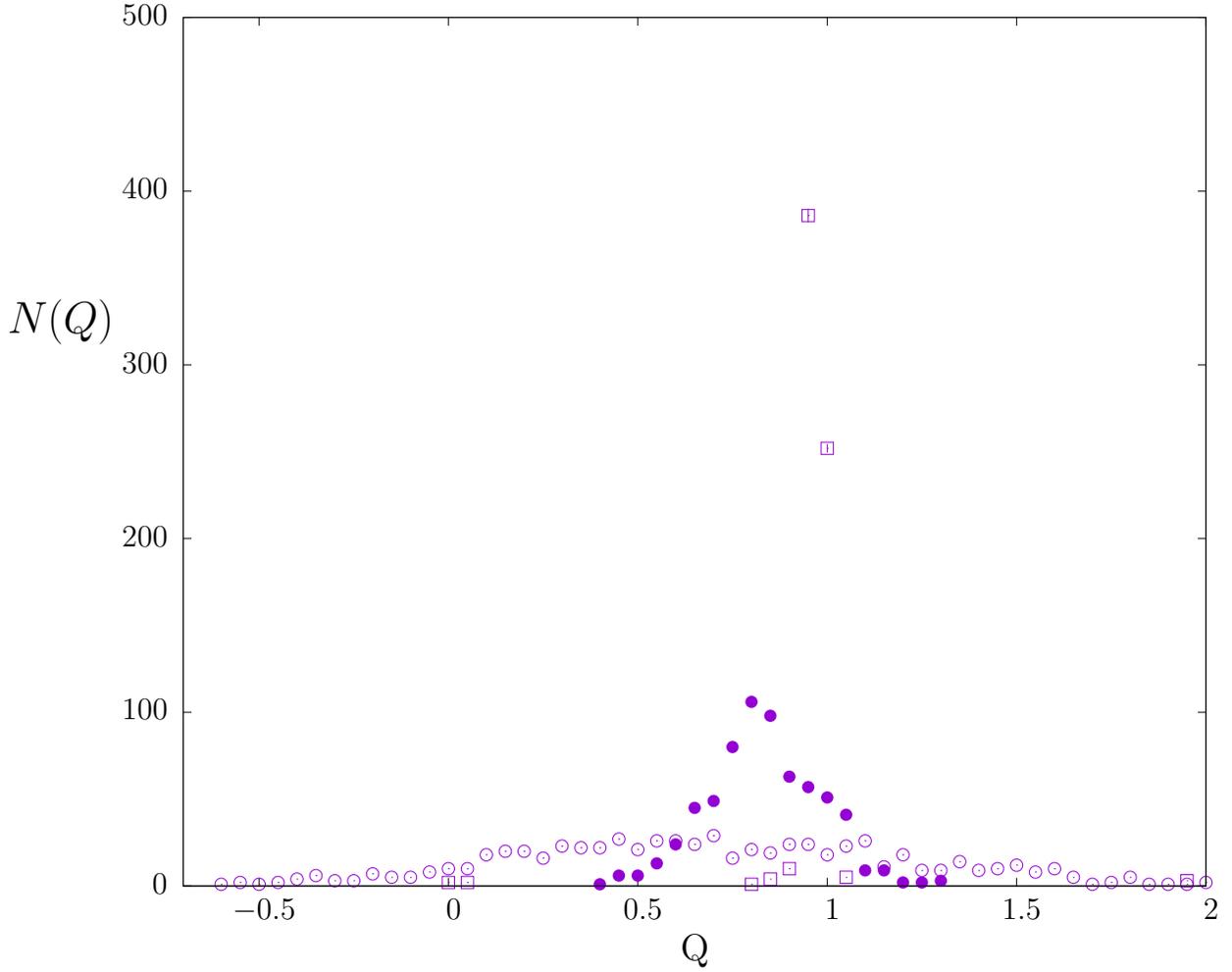}
\end	{center}
\caption{Histogram of $Q$ for fields with $Q\in [-0.40,1.35]$ after 2 cooling
  sweeps, $\bullet$. For the same fields a histogram for $Q$ after 1 cooling
  sweep, $\circ$, and after 20 cooling sweeps, $\Box$. Points corresponding to $N(Q)=0$ are suppressed.}
\label{fig_Qrevlong_Q1c2_b6.50}
\end{figure}

We see from the above that after 20 cooling sweeps the value of $Q$ is very well defined
over our whole range of $\beta$. Now, although it is plausible that the value of  $Q$
after 20 cools reflects the value of $Q$ of the original `hot' field configuration, it is
worth asking how well that expectation holds up. The first obvious comment is that the smaller the
number of cooling sweeps, the more convincing is the calculated value of $Q$. This is because the
cooling is a local process to the same extent as the Monte Carlo heat bath, so a few cooling sweeps
will deform the original fields only on correspondingly small distance scales. However one can ask
if the fields with a certain value of $Q$ after a very few cooling sweeps have the same value
of $Q$ after 20 cooling sweeps. To address this kind of question better we have taken
a sample of 2750 fields generated at $\beta=6.50$ and measured the value of $Q$
for 1,2,3,4,5,10,15,....,50 cooling sweeps. In Fig.~\ref{fig_Qlong_p_b6.50} we plot
the values of $Q$ after 2 and after 20 cooling sweeps. We see that even after only
2 cooling sweeps the values of $Q$ are peaked around near-integer values with little overlap
so that we can assign each value of $Q$ to an integer, $Q_I$, with a high level of confidence.
And since 2 cooling sweeps can only deform the fields over the shortest distance scales
we can be confident that this topological charge provides a very accurate representation
of the topological charge of the original lattice gauge fields.
The interesting question now is whether the fields with a given $Q_I$ after 2 cooling sweeps
maintain that same value when we continue cooling the lattice gauge fields
to, say, 20 cooling sweeps. The answer is yes to a very good approximation.
As an example we show in Fig.~\ref{fig_Qrevlong_Q1c2_b6.50} the values of $Q$ obtained
after 2 cooling sweeps which lie between the minima that define the peak near $Q=1$, as
in  Fig.~\ref{fig_Qlong_p_b6.50}, and which we therefore label by $Q_I=1$.
We follow these same fields to 20 cooling sweeps and plot their values of $Q$ in the figure.
They nearly all have $Q\simeq 1$. In fact of the 665 lattice fields that we assigned to
$Q_I=1$ at 2 cooling sweeps, 658 had $Q\simeq 1$ after 20 cooling sweeps with 4 at $Q\simeq 0$
and 3 at $Q\simeq 2$. Moreover, we observe at intermediate cooling sweeps that the
7 fields that end up with $Q_I \neq 1$ migrate gradually from $Q_I = 1$ to $Q_I \neq 1$
through intermediate values of $Q$ in a way that is consistent with a narrow (anti)instanton
shrinking out of the lattice with increased cooling.

We have given the fields with $Q_I=1$ as an example. In Table~\ref{table_QI2_QI50} we
show what happens to those fields possessing other values of $Q_I$ at 2 cooling sweeps
when one continues to cool these fields to a total of 50 cooling sweeps. We
indicate the range of $Q$ that we assign to each integer $Q_I$. That such a assignment
should possess little ambiguity should be evident from the histogram of $Q$ after 2 cooling
sweeps in Fig.~\ref{fig_Qlong_p_b6.50}. We observe that the value of $Q_I$ is almost
completely unchanged between 2 and 50 cooling sweeps in all cases.

One might ask why we start with fields after 2 cooling sweeps rather than after just
1 cooling sweep. The answer is indicated in Fig.~\ref{fig_Qrevlong_Q1c2_b6.50} where we
take the fields that we assign to $Q_I=1$ after 2 cooling sweeps and we display
the values of $Q$ for those same fields after only one cooling sweep. While we
do indeed observe a distribution of values that has a mean not too far from what
one might expect, the width of the distribution is so large that it will clearly
overlap heavily with fields that belong to neighbouring values of $Q_I$ (and which are not
shown). So after one cooling sweep we cannot make a plausible estimate of $Q_I$ unless
we run more highly cooled fields backwards as we have done here. This is even more so
if we consider the fields prior to any cooling at all. 
 
What this study demonstrates is that calculating $Q$ after 20 cooling sweeps provides
a reliable method for calculating the true integer-valued topological charge of our
Monte Carlo generated lattice fields -- at least once $\beta$ is not too small.  

The shift of the value of $Q$ as calculated after 20 cooling sweeps from the integer value
that it would have in the continuum limit is primarily due to the finite size in lattice
units of the typical instanton in the cooled fields. That is to say, it is a lattice artifact.
So if we assign the corresponding integer $Q_I$ to a given field -- an assignment which
we have seen is essentially unambiguous over our range of $\beta$ -- we can think of
it as partially compensating for the lattice artifacts. Of course since different lattice artifacts
may mutually cancel, there is no guarantee that this will hasten the overall approach to the
continuum limit. Nonetheless this integer-valued topological charge $Q_I$ has at least as much
validity as the measured charge $Q$ and we shall present results for both.

The possibility of assigning an integer $Q_I$ to a lattice field depends on there being
a clear multi-peak structure in the histogram of values of $Q$ with a very small overlap
between neighboring peaks. As we see from  Fig.~\ref{fig_Q_p_b6.0625} and
Fig.~\ref{fig_Q_p_b6.50} this is the case not only after 20 cooling sweeps but
even after only 4 cooling sweeps for these values of $\beta$. So we also calculate
an integer $Q_I$ for all cooling sweeps where a reliable assignment is possible.
(Variations in how one does this will make a negligible  difference as long as
one more-or-less assigns fields between neighbouring minima in the histogram of $Q$
to the same value of $Q_I$.)
Examples of such an assignment for $\beta=6.0625$ and $\beta=6.50$ are shown in
Table~\ref{table_Q_QI} where we compare the values of $\langle Q^2 \rangle$
and $\langle Q_I^2 \rangle$ against the number of cooling sweeps. As we see,
a property of $\langle Q_I^2 \rangle$ is that it is essentially independent of the
number of cooling sweeps, unlike the value of $\langle Q^2 \rangle$. (Note that
the statistical errors shown are almost completely correlated when cooling,
which is why we display them separately at the bottom of the Table.)
This is a consequence of the fact that, as we have seen above, to a very good
approximation the topological charge does not change with cooling. Beyond a
very few cooling sweeps the topological charge only changes through an instanton
shrinking out of the lattice (or the unlikely reverse), a process which becomes
rapidly much rarer as $\beta$ increases for any given number of cooling sweeps.

Note that the fact that $\langle Q_I^2 \rangle$ does not vary with the number of cooling
sweeps means that we can evaluate it at some fixed number of cooling sweeps for any
of our values of $\beta$. We shall choose to use $20$ cooling sweeps unless
otherwise stated.

\begin{figure}[htb]
\begin	{center}
\leavevmode
\input	{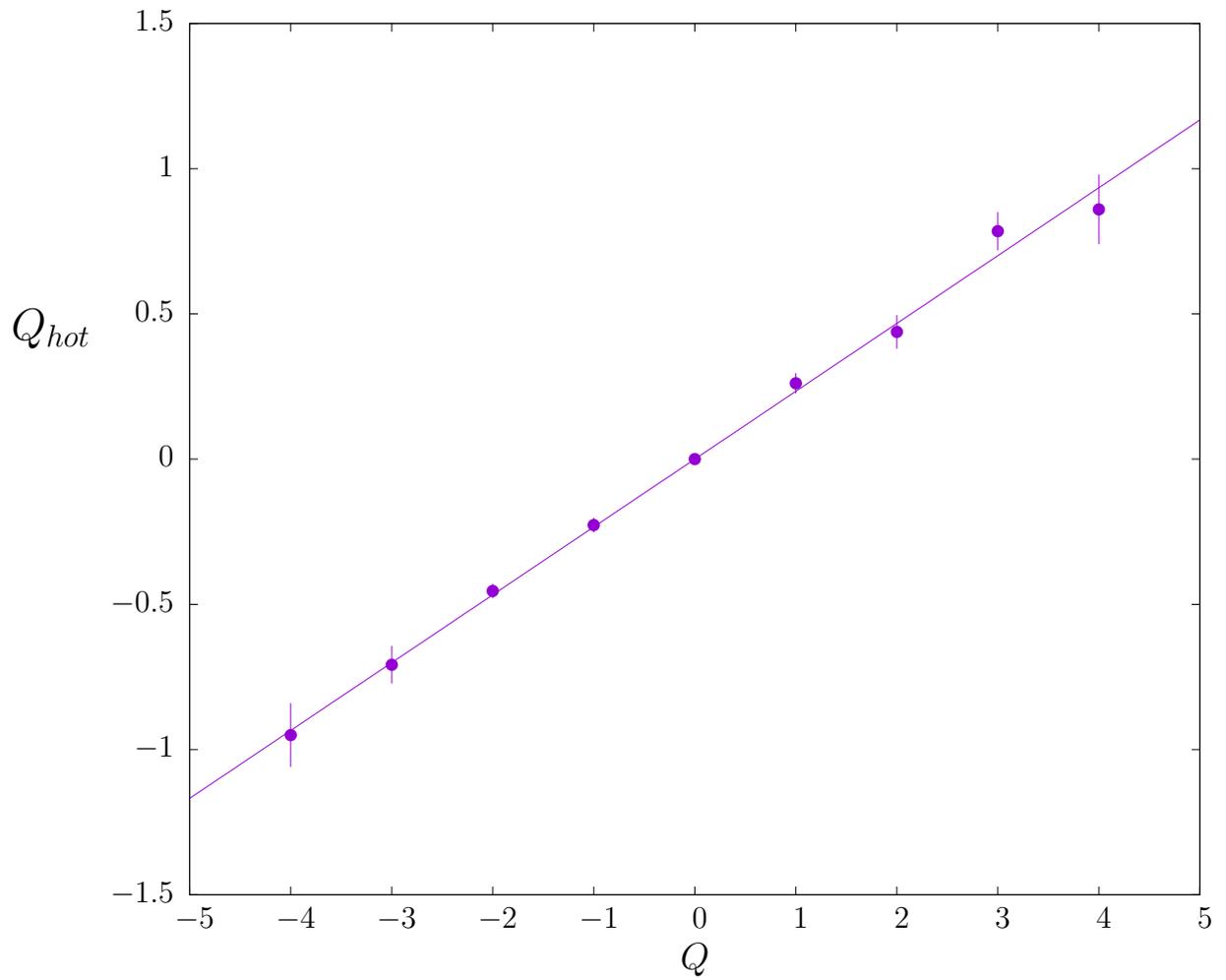}
\end	{center}
\caption{Average topological charge on lattice fields which
  have an integer-valued  charge $Q$ after 20 cools; on $26^338$ lattices at $\beta=6.5$.}
\label{fig_QhQ20_b6.5}
\end{figure}

As an aside, it is interesting to see how the average value of the lattice
topological charge $Q_{hot}$ on the lattice fields prior to cooling, is related
to the true (cooled) $Q_I$. We show the result of the comparison in
Fig.~\ref{fig_QhQ20_b6.5}. We see that the relation is a simple multiplicative
constant to a good approximation, i.e. $Q_{hot}=Z(\beta) Q_I$. Here we
have indicated that we expect this constant to vary with the coupling $\beta$.
Such a linear dependence is expected if one performs a perturbative
calculation of $Z(\beta)$
which at 1 loop leads to $Z_{1-loop}(\beta)=1-5.451/\beta$ in $SU(3)$
\cite{Pisa_ZQ}.
Note that the correction is $O(1/\beta)$ in our range of $\beta$ which
means that we expect  $Q_{hot} \ll Q_I$ and this is certainly what we
observe in Fig.~\ref{fig_QhQ20_b6.5}. 
We have repeated this analysis for our other values of $\beta$ and
we list the values obtained for $Z(\beta)$ in Table~\ref{table_Z},
together with the 1 loop perturbative predictions. We note that
the latter are, given the large size of the leading correction, surprisingly
close to our values.

\begin{figure}[htb]
\begin	{center}
\leavevmode
\input	{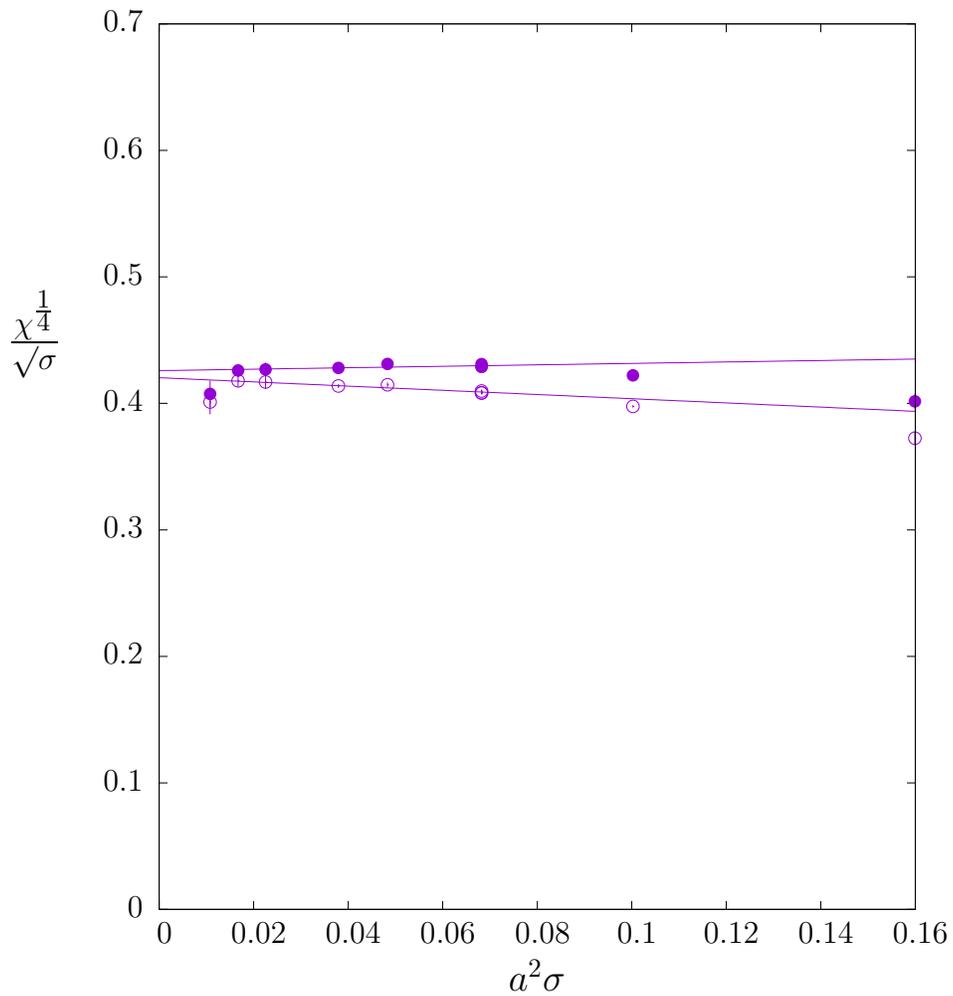}
\end	{center}
\caption{Topological susceptibility in units of the string tension.
  Continuum extropolation using raw $Q$, $\circ$, and integer
  valued $Q$, $\bullet$, both after 20 cooling sweeps.}
\label{fig_KhiK_nc20}
\end{figure}

As we remarked earlier, the fluctuations of the topological charge in the
pure gauge theory are of particular interest since they are related, through
the $U_A(1)$ anomaly to the masses of the lightest pseudoscalars, up to
$1/N_{colour}$ corrections
\cite{susc_eta1,susc_eta2}.
In Table~\ref{table_QQkhi} we list our values for both $\langle Q^2 \rangle$
and $\langle Q_I^2 \rangle$ at our values of $\beta$. The interesting quantities
are these values divided by the lattice volumes, i.e. the topological susceptibility
$a^4\chi_t=\langle Q^2 \rangle / L_s^3L_t$, where $L_s^3L_t$ is the lattice volume,
and the susceptibility expressed in physical units, e.g. $\chi^{1/4}_t/\surd\sigma$ where
$a^2\sigma$ is the string tension at that $\beta$. These values are also shown
in Table~\ref{table_QQkhi}. We extrapolate these values to the continuum limit with a
leading  $O(a^2\sigma)$ correction obtaining the continuum values given in the Table.
The best fits are displayed in Fig.~\ref{fig_KhiK_nc20}. Not surprisingly
the fits deviate from our measurements at the two coarsest lattice spacings.
At the largest $\beta$ value we are beginning to lose ergodicity in topology
(see below) so that our errors may be underestimates, but the overall fits
are still statistically acceptable. In Table~\ref{table_QQkhi}
we show the result of fitting to two different ranges of $\beta$. The results
are consistent within errors as are the continuum limits using $Q$ and $Q_I$.
If we take the most concervative value, based on the integer valued charge
in the range $5.99 \leq \beta \leq 6.50$, we obtain the continuum limit
\begin{equation}
  \frac{\chi^{\frac{1}{4}}_t}{\surd\sigma} = 0.4187 \pm 0.0053
\label{eqn_chit_K}
\end{equation}
To convert the scale from $\surd\sigma$ to $r_0$ one can use the conversion
factor given in Table~\ref{table_MK_J} giving
\begin{equation}
  r_0\chi^{\frac{1}{4}}_t  = 0.4857 \pm 0.0067.
\label{eqn_chit_r0}
\end{equation}
We note that this agrees, within errors, with the value $r_0\chi^{\frac{1}{4}}_t=0.4928(62)$
obtained in
\cite{DelDebbio05}
where the calculation of $Q$ is performed by the very different method of calculating the
number of exact fermionic zero-modes using the Neuberger overlap-Dirac operator
\cite{Neuberger1,Neuberger2}.
One can now use eqn(\ref{eqn_r0MeV}) to convert to `MeV' units. If we do so we obtain
\begin{equation}
   \chi^{\frac{1}{4}}_t = 203 \pm 4 MeV
\label{eqn_chit_MeV}
\end{equation}
which agrees with the value  $\chi^{\frac{1}{4}}_t = 208(6) MeV$ obtained in
\cite{MLFP10}
using the gradient (Wilson) flow technique. (Note that this value has been
rescaled from that in the paper using the more recent value of $r_0$
given in eqn(\ref{eqn_r0MeV}). Note also that true systematic error in obtaining
these physical `MeV' units is no doubt underestimated.)

Our above calculation assumes that finite volume corrections to the topological
susceptibility are negligible when working on the lattice volumes shown in
Table~\ref{table_QQkhi}. To check that this is so we have calculated the susceptibility
on different lattice sizes at $\beta=5.8941$, $\beta=5.99$ and $\beta=6.235$. When expressed
in units of the string tension the lattice volumes in Table~\ref{table_QQkhi}
are spanned by the volumes $10^316$ and $12^316$ at $\beta=5.8941$, by
$14^320$ at $\beta=5.99$ and by $18^326$ at $\beta=6.235$. So we have
performed calculations of the integer valued topological charge on somewhat larger
lattices at these three $\beta$ values to see if there are any pronounced changes
with the volume. The results are shown in Table~\ref{table_khiV} and we see that
there are no significant finite volume corrections within our small errors.

\subsection{Critical slowing down}
\label{subsection_Q_freeze} 

Since our Monte Carlo exploration of the phase space of all fields proceeds through local steps, by
changing one link matrix at a time, the value of $Q$ only changes when the core of a topological charge
shrinks down to the scale of a lattice hypercube where it becomes nothing more than a gauge singularity.
(Or the reverse process.) Once the lattice spacing $a$ is small enough, a topological charge whose
core is a few lattice spacings across will be embedded in a background gauge field that is
relatively smooth, and one can calculate its weight in the path integral using standard
semiclassical methods. To be specific the weight of an instanton of size $\rho$ is
\begin{equation}
D(\rho) d\rho \propto {\frac{d\rho}{\rho}}{\frac{1}{\rho^4}}
e^{-{\frac{8\pi^2}{g^2(\rho)}}}\,,
\label{eqn_Drho1}
\end{equation}          
where we omit factors varying weakly with $\rho$. We note the scale-invariant integration measure,
a factor to account for the fact that a ball of volume $\rho^4$ can be placed in $1/\rho^4$ different ways
in a unit volume; and a factor arising from the classical instanton action, $S_I=8\pi^2/g^2(\rho)$,
where $g^2(\rho)$ is the running coupling on the scale $\rho$. When we insert the asymptotically free form
of the coupling, we obtain
\begin{equation}
D(\rho)  \propto \Bigl({\frac{\rho}{\xi}}\Bigr)^6\,,
\label{eqn_Drho2}
\end{equation}
where $\xi$ is a physical length scale of the theory, such as $r_0$, or $\surd\sigma$, or $\Lambda_{\rm MOM}$.
This formula should be reliable as long as $\rho \ll \xi$. On the lattice it will also break down once
$\rho \sim a$ in a way that depends on the particular lattice action, but it should be accurate once
$\rho \gtrsim few \times a$. In order for $Q$ to change on the lattice, an instanton will have to pass
through sizes $\rho \sim few \times a$ and as we see from eqn(\ref{eqn_Drho}) such fields are suppressed
by a factor $\sim (a/\xi)^6$. 
That is to say, our sequence of Monte Carlo generated lattice fields will be locked into given values of $Q$
once we are at large enough $\beta$ and we will have lost ergodicity in $Q$. Whether any such loss of
ergodicity in $Q$ would have a significant impact on the physics that we are interested
in is not clear. (In QCD the situation is quite different due to the associated zero-modes of the Dirac
operator.) Nonetheless it is worth monitoring this potential loss of ergodicity, as we shall now do.

\begin{figure}[htb]
\begin	{center}
\leavevmode
\input	{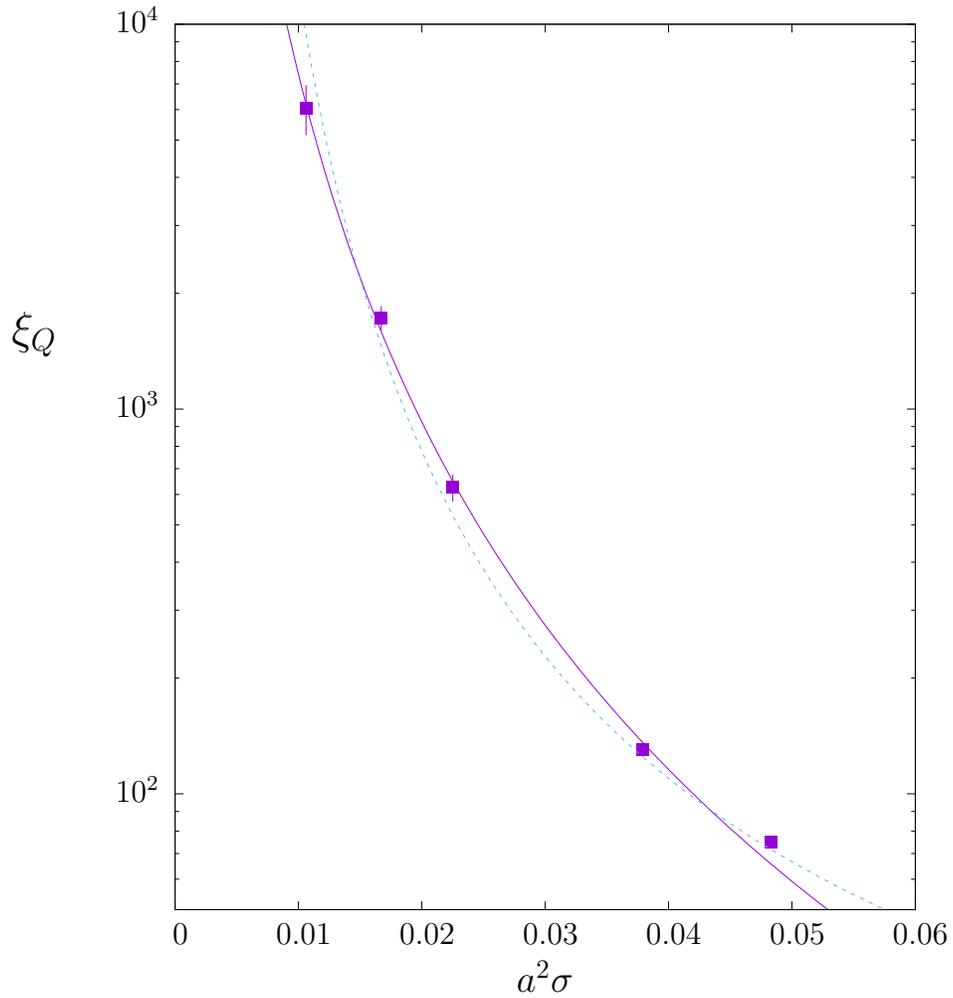}
\end	{center}
\caption{Correlation length $\xi_Q$  for the SU(3) lattice topological charge
  defined by the relation  $\langle Q(is)Q(is+\xi_Q)\rangle/\langle Q^2\rangle = e^{-1}$
  where $is$ and $is+\xi_Q$ label the number of MC sweeps. Solid line
  is $\xi_Q \propto 1/(a\surd\sigma)^{6}$, dashed line is $\xi_Q \propto \exp\{c/a\surd\sigma\}$.
  Calculations are for $\beta \in [5.99,6.50]$.}
\label{fig_Q20cor_su3}
\end{figure}

We focus on two quantities. The first is to calculate directly the number of Monte Carlo sweeps, $\xi_Q$,
that it takes to `decorrelate' the value of $Q$, as defined by 
\begin{equation}
  \frac{\langle Q(is)Q(is+\xi_Q)\rangle}{\langle Q^2\rangle} = e^{-1}\,,
\label{eqn_Drho}
\end{equation}
where $is$ and $is+\xi_Q$ label the number of MC sweeps and $is$ is averaged over.
This is of course not a physical quantity: it
depends on our particular implementation of the Monte Carlo algorithm. However by comparing the
value of $\xi_Q(\beta)$ to the total number of sweeps at a given $\beta$ we can judge how ergodic
is our calculation. The values of $\xi_Q(\beta)$ that we obtain in our calculations are plotted
in Fig.~\ref{fig_Q20cor_su3}. (We restrict our calculation to $\beta \ge 5.99$ since we only
calculate $Q$ every 25 sweeps and so our determination of $\xi_Q$ can only be reliable once
$\xi_Q\gg 25$.) Given that our calculations are typically based on $\sim 10^6$ sweeps
it is clear that we do not have a serious breakdown of ergodicity at any $\beta$ although
it is also clear that at $\beta=6.50$ we are at the borderline given that our sequences of
field configurations are generated with 40000 prior thermalisation sweeps beginning from
a $Q=0$ `frozen' starting configuration. It is interesting to note that, as shown in
Fig.~\ref{fig_Q20cor_su3}, the dependence of  $\xi_Q(\beta)$ on  $a(\beta)$ can be fitted
by $\xi_Q\propto \{a\surd\sigma\}^{-6}$. This is what one would naively expect using
eqn(\ref{eqn_Drho2}) with $\rho\sim a$, about which value one expects a shrinking instanton
to become a dislocation and then a $Q=0$ gauge singularity. (Or the reverse.) However it is
also possible that a tunnelling process is involved in the topology change, as discussed in
\cite{QcorA,QcorB},
in which case one would expect an exponential dependence on the free energy. Since
the relevant length scale is $\sim 1/a$ and the typical dynamical energy scale is
$\sim \surd\sigma$, this would suggest a correlation length
$\xi_Q\propto \exp\{const/a\surd\sigma\}$. Fitting such a form to our calculated values we
obtain the dashed curve in Fig.~\ref{fig_Q20cor_su3}. This is an acceptable fit given
the quality of the data; the overshoot of the $\beta=6.50$ value cannot be taken too
seriously given the large error on that value (and, no doubt, the even larger relative
error on the value of that error estimate). Nonetheless, while our calculated values of
$\xi_Q$ are not precise enough to discriminate between these different pictures of
topology change, they serve their main purpose which is to confirm that the `freezing'
of the topological charge is not severe enough to raise doubts concerning our mass
and string tension calculations.

The second quantity we calculate is the number of sweeps, $\tau_{|\Delta Q|=1}$, that is takes to
change $Q$ by one unit. We list our calculated values in Table~\ref{table_tau}. Clearly the
probability of producing or annihilating an instanton as described above will be
proportional to the space-time volume, so it is useful to give its value for
a standard physical volume which we choose to be $V=(3/\surd\sigma)^4$, which is
very close to the volumes we actually used for calculating $Q$. The resulting values,
labelled $\tilde{\tau}_{|\Delta Q|=1}$, are also listed in Table~\ref{fig_Q20cor_su3}
and, not surprisingly,they are close to the values of $\tau_{|\Delta Q|=1}$.
Now, since these changes in $Q$ are randomly $\pm 1$ it takes roughly $Q^2$ such changes
to disorder a charge of $Q$. On our lattices we have $\langle Q^2\rangle \sim 3$ so the number
of sweeps to disorder the lattice charge is $\sim\langle Q^2\rangle \tilde{\tau}_{|\Delta Q|=1}
\sim 3\tilde{\tau}_{|\Delta Q|=1}$. This comes to about $\sim 4\times 10^3$ for $\beta=6.50$,
which is not very far from our other measure, displayed in  Fig.~\ref{fig_Q20cor_su3}.
We note that since $\langle Q^2\rangle \propto V$ and since $\tau_{|\Delta Q|=1}\propto 1/V$,
we expect this measure of decorrelation to be approximately volume independent.
In any case we confirm that our calculations in this paper are reasonably ergodic in $Q$.

\section{Possible improvements}
\label{section_improvements} 

The calculations in this paper cover a large enough range of $a(\beta)$ for the continuum
extrapolations to be under reasonable control. And our large basis of loop operators means
that it is unlikely that we are missing any glueball states in the part of the mass spectrum
that we have presented. In addition we have checked that even at our largest values of $\beta$
we maintain reasonable ergodicity as far as the topology of the gauge fields is concerned.  

In this section we briefly reflect upon the main ways in which our calculations can be improved.

While our calculation of the lightest states is adequate, the calculation of heavier states
is less so. Here using lattices with an asymmetric discretisation, $a_t \ll a_s$ can be
very useful. However since the glueball masses are calculated as $a_tM$ while the
physical scale, whether provided by $\sigma$ or $r_0$, also involves the value of $a_s$,
we need to know the value of the ratio $\xi=a_s/a_t$ to form dimensionless ratios
such as $M_{0^{++}}/\surd\sigma$ or $M_{0^{++}}r_0$.
To overcome any issues with the shift in the value of the  asymmetry $\xi=a_s/a_t$
from its inputted tree-level value, one can perform two calculations in parallel. The
first is with $a_s=a_t$ and is much like the one presented in this paper but one that can
be done with
a much smaller basis of loop operators. It would be designed to calculate the lightest
$0^{++}$, $2^{++}$ and $0^{-+}$ states as well as the string tension. (And perhaps $r_0$
and the spectrum of string excitations.) This would give us accurate values of the continuum
limit of ratios such as $M_{0^{++}}/\surd\sigma$ and $M_{2^{++}}/\surd\sigma$. Our second
calculation would be with some well chosen value of $\xi=a_s/a_t > 1$ with a large basis
of loop operators and would be designed to provide accurate values of the heavy as well as the
light glueball masses. This latter calculation would give the continuum limit of, for example,
$M_G/M_{0^{++}}$ for the glueball $G$. Then one can, for example, multiply it by the value of
$M_{0^{++}}/\surd\sigma$  obtained in the first calculation to obtain $M_G/\surd\sigma$.

The main reason to calculate the heavier states is to provide a target for theoretical and
model calculations. However with only 5 rotational representations on the lattice, each will
typically end up having a densely packed set of excited states making the challenge
ambiguous. So it is (almost) essential to be able to identify the spin, $J$, towards which
each lattice state tends as $a\to 0$, and also its $2J+1$ (near) degenerate partners.
Doing so by simply identifying nearly degenerate states partitioned appropriately amongst
the  5 rotational representations of the lattice is not an option once the states
become dense enough. This means incorporating something like the approach in
\cite{HM-2005}
into the glueball calculations. Doing so could in addition have the benefit of identifying
the lightest high spin glueballs and hence the leading glueball Regge trajectory and
so test the old idea that this trajectory is the Pomeron.

Amongst the more massive states will be multiglueball scattering states and we need to know
if any of our supposed single glueball states are in fact of this type. (This is closely
related to the issue of the decay widths of heavier glueballs.) While one expects
there to be some suppression of the overlap between our single trace operators and the
multi-trace operators that are the natural wave-functionals for multiglueball states,
this is something that needs to be examined explicitly. A way to do this is to include
some appropriate multi-trace operators in our basis. For example we can take our
variationally selected lightest $0^{++}$ glueball operator, give it a non-zero
momentum, then take a product of two such operators with equal and opposite momenta
to give a zero-momentum state. Do so for various momenta and possibly different
glueballs. States in the resulting spectrum that have a small overlap onto such
multi-glueball states can be identified with more confidence as being single glueballs.
For a first step in this direction see for example Section 7.2.3 of 
\cite{HM-2005}
where small overlaps are found indicating that the multi-glueball states are likely to
be invisible to current glueball calculations. However this needs to be studied more
systematically and should form part of an improved calculation.

The above discussion also extends to the question of identifying ditorelon states.
For the lightest such states their large variation with the lattice volume makes them
easy to detect. However when such states are heavier and are located in the dense part
of the spectrum, identifying finite volume shifts becomes ambiguous and a direct
study of overlaps, as for multiglueball states, becomes useful and should be one
of the incorporated improvements.

While do not claim the above improvements to be exhaustive, they would represent a
major qualitative improvement to this and earlier glueball spectrum calculations.

\section{Discussion}
\label{section_discussion} 

In this paper we provide a new calculation of the glueball spectrum of $SU(3)$ gauge theories
which is designed to improve upon some limitations of earlier calculations. We have a very large
basis of glueball operators which gives us confidence that we are not missing any states in
the important low-lying part of the spectrum. We perform high statistics calculations out
to lattice spacings as small as $a(\beta=6.50)\simeq 0.042fm$, which both improves our
control of our extrapolations to the continuum limit and gives us a finer resolution of the
glueball correlators, which is particularly useful for the heavier glueballs. We perform a
high statistics calculation to check that our spatial volume is large enough and identify
the most important multi-torelon states so that we can exclude them from the spectrum.
We perform a high statistics study of the topological fluctuations of the gauge fields,
with particular care in establishing the reliability of our measure of topology, so as
to establish the ergodicity of our Monte Carlo sequences throughout our calculations.
Given our cubic lattice, our glueball states fall into representations of the octahedral
subgroup  of the full rotation group, but using the observed near-degeneracies of various
states we are able to assign a continuum spin to more states than previous calculations.
At the same time, such an extensive calculation highlights important areas where improvements
are needed: a parallel calculation with a lattice asymmetry allowing a much finer
discretisation of the correlators
\cite{KIGSMT-1983,MP-2005}
would be useful for the massive states; to systematically establish the continuum spins
of the glueball states one needs to incorporate methods such as those used in
\cite{HM-2005};
to be sure one that one can distinguish multiglueball scattering states in the spectrum
one needs to include the relevant operators into the calculations; and the same applies
in controlling the contribution of the finite volume multi-torelon states.

Our results for the glueball spectrum of the continuum theory are listed in
Tables~\ref{table_MK_R}, \ref{table_MGeV_R} and \ref{table_MGeV_J}. Table~\ref{table_MK_R}
lists the masses in units of the confining string tension, and labels them
by their parity $P$, charge conjugation $C$, and the irreducible representation
of the rotation group of our cubic lattice. Table~\ref{table_MGeV_R} translates
these masses into `GeV' units as explained in Section~\ref{subsection_units}.
In Table~\ref{table_MGeV_J} we list those glueballs whose continuum spins $J$
we can identify.

If we assume that the phenomenologically relevant states are those with masses
$\lesssim 3 GeV$ then our calculations in Table~\ref{table_MGeV_J} confirm what has been
established for a long time: 
these are the $0^{++}$, $2^{++}$, $0^{-+}$, $2^{-+}$, $1^{+-}$ ground states and the
first excited $0^{++}$. Of course our calculations are much more accurate than those
early calculations, and improve significantly upon the more recent ones as well,
but at this stage the greatest phenomenological uncertainty lies in the translation of
the masses into GeV units and that is something that a calculation in the pure gauge
theory cannot deal with by itself; here we need glueball calculations in full QCD
to be further pursued.

We have emphasised that an important role for the glueball spectrum lies in providing a benchmark
challenge for theoretical and model approaches that attempt to understand the non-perturbative
physics of the $SU(3)$ gauge theory and hence of QCD, usually in some approximation.
For this purpose our excited states in
Table~\ref{table_MGeV_J} are useful, as indeed is the more extensive set of states listed
in Table~\ref{table_MK_R} since a calculation in the continuum theory can always project
the states into the representations of the octahedral subgroup of the full rotation group.

\section*{Acknowledgements}

We thank Massimo D'Elia for discussions.
AA has been financially supported by the European Union's Horizon 2020 research and innovation
programme ``Tips in SCQFT'' under the Marie Sk\l odowska-Curie grant agreement No. 791122.
MT acknowledges support by Oxford Theoretical Physics and All Souls College. The numerical
computations were carried out on the computing cluster in Oxford Theoretical Physics.


%
%
%

\clearpage

\begin{table}[htb]
\centering
\begin{tabular}{|cc|ccc|c|} \hline
$\beta$ & lattice & $\tfrac{1}{3}\text{ReTr}\langle U_p\rangle$ & 
$a\surd\sigma$ & $am_G$ & lattices \\ \hline
5.6924 & $10^316$ & 0.5475112(71) & 0.3999(58)  & 0.987(9)   & $8^316$  \\
5.80   & $12^316$ & 0.5676412(36) & 0.31666(66) & 0.908(12)  & $10^316$ \\
5.8941 & $14^316$ & 0.5810697(18) & 0.26118(37) & 0.7991(92) & $12^316$,$10^316$ \\
5.99   & $18^4$ &  0.5925636(11) & 0.21982(77) & 0.7045(65) & $14^320$ \\
6.0625 & $20^4$ &  0.6003336(10) & 0.19472(54) & 0.6365(43) & $14^320$ \\
6.235  & $26^4$ &  0.6167723(6) & 0.15003(30) & 0.4969(29) & $34^326$,$18^326$ \\
6.3380 & $30^4$ &  0.6255952(4) & 0.12928(27) & 0.4276(37) & $22^330$ \\
6.50   & $38^4$ &  0.6383531(3) & 0.10383(24) & 0.3474(22) & $26^338$ \\ \hline
\end{tabular}
\caption{Parameters of the $SU(3)$ calculations as well as the string tension, $\sigma$,
  and the  mass gap, $m_G$. On the right are shown other lattice sizes used for finite volume
  studies and for calculations of the topological charge.}
\label{table_param}
\end{table}

\begin{table}[htb]
\centering
\begin{tabular}{|ccc|} \hline
\multicolumn{3}{|c|}{continuum $J \sim$ cubic $R$} \\ \hline
$J$    &        &  cubic $R$  \\  \hline
 0   & $\sim$ & $A1$   \\
 1   & $\sim$ & $T1$      \\
 2   & $\sim$ & $E+T2$     \\
 3   & $\sim$ & $A2+T1+T2$     \\
 4   & $\sim$ & $A1+E+T1+T2$     \\
 5   & $\sim$ & $E+2T1+T2$      \\
 6   & $\sim$ & $A1+A2+E+T1+2T2$     \\
 7   & $\sim$ & $A2+E+2T1+2T2$     \\
 8   & $\sim$ & $A1+2E+2T1+2T2$     \\ \hline
\end{tabular}
\caption{Projection of continuum spin $J$ states onto the cubic representations $R$.}
\label{table_J_R}
\end{table}

\begin{table}[htb]
\centering
\begin{tabular}{|c|ccc|} \hline
\multicolumn{4}{|c|}{$aM$ , $\beta=6.235$ , PC=++} \\ \hline
$R$   & $34^326$ & $26^326$ &  $18^326$  \\ \hline
A1  &  0.4965(41)  & 0.4969(29)  & 0.4773(53)   \\ 
    &              &             & 0.7002(73)   \\ 
    &  0.8566(70)  & 0.8626(52)  & 0.842(18)   \\ 
    &  1.130(15)   & 1.096(10)   & 1.050(9)   \\ 
    &  1.129(14)   & 1.134(11)   &    \\ 
    &              & 1.054(27)   &    \\ 
    &  1.144(13)   & 1.193(12)   &    \\ 
    &  1.238(16)   & 1.226(17)   &    \\ 
    &  1.329(22)   & 1.322(18)   &    \\   \hline
A2  &  1.166(12)   & 1.127(9)   &  1.100(11) \\ 
    &  1.31(9)     & 1.25(6)    &  1.13(6)  \\ 
    &  1.456(22)   & 1.429(16)  &  1.455(22)  \\   \hline
E   &             &             &  0.6627(50)  \\ 
    &  0.7354(26) & 0.7342(45)  &  0.742(9)  \\
    &  1.013(9)   & 1.013(6)    &  0.974(19)  \\ 
    &  1.088(17)  & 1.067(25)   &  1.054(23)  \\ 
    &  1.127(12)  & 1.122(8)    &  1.054(26)  \\ 
    &             & 1.206(11)   &    \\ 
    &  1.242(15)  & 1.243(15)   &    \\ 
    &  1.376(25)  & 1.332(17)   &    \\   \hline
T1  &  1.111(19)  &  1.148(7)   &  1.160(24)  \\ 
    &  1.116(23)  &  1.137(23)  &  1.194(9)  \\ 
    &  1.392(12)  &  1.407(12)  &  1.38(5)  \\
    &  1.423(13)  &  1.392(51)  &    \\   \hline
T2  & 0.7439(35) & 0.7374(31) &  0.728(7) \\ 
    & 1.021(6)   & 1.025(6)   &  1.036(7)  \\ 
    & 1.106(19)  & 1.096(18)  &  1.100(21)  \\ 
    & 1.158(7)   & 1.152(8)   &  1.166(7)  \\   \hline
2aE(l) & 1.479(12)   & 1.087(5)   &  0.674(5)  \\   \hline
\end{tabular}
\caption{Finite volume check on glueball masses with $P=+$ and $C=+$
  at $\beta=6.235$ on the lattices shown. The masses are extracted from
  the same initial time and are not always the same as our best
  mass estimates listed in other tables. Intemediate gaps appear where there is
  a clear extra state on a smaller lattice. Last line gives twice the
  energy of a winding flux tube.}
\label{table_Vcheckpp_b6.235}
\end{table}

\begin{table}[htb]
\centering
\begin{tabular}{|c|ccc|} \hline
\multicolumn{4}{|c|}{$aM$ , $\beta=6.235$ , PC=-+} \\ \hline
$R$   & $34^326$ & $26^326$ &  $18^326$  \\ \hline
A1  &  0.806(13)  & 0.781(13)   &  0.772(12)  \\ 
    &  1.08(5)    & 1.12(4)     &  0.924(23)  \\ 
    &  1.408(25)  & 1.437(20)   &  1.216(13)  \\ 
    &  1.507(25)  & 1.485(22)   &  1.456(19)  \\   \hline
A2  &  1.420(24)  & 1.404(21)   &  1.519(21)  \\ 
    &  1.62(5)    & 1.60(3)     &  1.84(4)  \\   \hline
E   &  0.970(10)  & 0.958(5)    &  0.963(5)  \\ 
    &  1.224(12)  & 1.249(8)    &  1.257(10)  \\ 
    &  1.394(18)  & 1.422(15)   &  1.423(15)   \\   \hline
T1  &  1.277(52)  & 1.226(30)   &  1.258(33) \\ 
    &  1.303(11)  & 1.302(11)   &  1.322(10)  \\ 
    &  1.335(11)  & 1.337(10)   &  1.398(11)  \\
    &  1.43(7)    & 1.46(7)     &  1.59(8)  \\   \hline
T2  &  0.958(7)   & 0.962(5)    &  0.971(6)  \\ 
    &  1.234(9)   & 1.219(6)    &  1.251(8)  \\ 
    &  1,29(4)    & 1.23(4)     &  1.256(32)  \\   \hline
\end{tabular}
\caption{Finite volume check on glueball masses with $P=-$ and $C=+$
  at $\beta=6.235$ on the lattices shown. Where possible the masses are extracted from
  the same initial time and so are not always the same as our best
  mass estimates listed in other tables.} 
\label{table_Vcheckmp_b6.235}
\end{table}

\begin{table}[htb]
\centering
\begin{tabular}{|c|ccc|} \hline
\multicolumn{4}{|c|}{$aM$ , $\beta=6.235$ , PC=+-} \\ \hline
$R$   & $34^326$ & $26^326$ &  $18^326$  \\ \hline
A1  &  1.397(22)  & 1.431(20)  &  1.486(30)  \\   \hline
A2  &  1.101(13)  & 1.095(9)   &  1.117(10)  \\ 
    &  1.304(20)  &  1.307(18) &  1.37(2)  \\ 
    &  1.395(33)  &  1.395(19) &  1.62(3)  \\   \hline
E   &  1.324(14)  &  1.306(10) &  1.383(12)  \\ 
    &  1.397(18)  &  1.416(13) &  1.445(15)  \\ 
    &  1.453(20)  &  1.455(16) &  1.539(17)  \\   \hline
T1  &  0.910(10)  &  0.897(10) &  0.928(11)  \\ 
    &  1.063(21)  &  1.071(18) &  1.067(16)   \\ 
    &  1.175(8)   &  1.169(8)  &  1.156(7)  \\
    &  1.301(24)  &  1.307(33) &  1.26(3)  \\   \hline
T2  &  1.107(8)   &  1.109(7)  &  1.104(9)  \\ 
    &  1.290(12)  &  1.258(27) &  1.285(37)  \\ 
    &  1.337(9)   &  1.333(12) &  1.331(9)   \\   \hline
\end{tabular}
\caption{Finite volume check on glueball masses with $P=+$ and $C=-$
  at $\beta=6.235$ on the lattices shown. Where possible the masses are extracted
  from the same initial time and so are not always the same as our best
  mass estimates listed in other tables.} 
\label{table_Vcheckpm_b6.235}
\end{table}

\begin{table}[htb]
\centering
\begin{tabular}{|c|ccc|} \hline
\multicolumn{4}{|c|}{$aM$ , $\beta=6.235$ , PC=--} \\ \hline
$R$   & $34^326$ & $26^326$ &  $18^326$  \\ \hline
A1  &  1.507(18) &  1.512(21) & 1.523(24)   \\ 
    &  1.50(2)   &  1.59(3)   & 1.53(3)   \\   \hline
A2  &  1.313(17) & 1.340(18) & 1.357(21)   \\ 
    &  1.59(3)   & 1.57(3)   & 1.60(3) \\   \hline
E   &  1.203(10) & 1.186(10) & 1.213(8)   \\ 
    &  1.34(9)   & 1.38(6)   & 1.48(7)   \\ 
    &  1.39(10)  & 1.41(8)   & 1.566(20)   \\   \hline
T1  &  1.191(31)  & 1.163(24)  & 1.233(22)   \\ 
    &  1.283(40)  & 1.305(31)  & 1.351(11)   \\ 
    &  1.479(16)  & 1.490(12)  & 1.458(12)   \\   \hline
T2  &  1.184(24)  & 1.200(28)  & 1.221(9)   \\ 
    &  1.268(42)  & 1.283(25)  & 1.31(4)   \\ 
    &  1.41(7)    & 1.50(6)    & 1.25(3)   \\   \hline
\end{tabular}
\caption{Finite volume check on glueball masses with $P=-$ and $C=-$
  at $\beta=6.235$ on the lattices shown. Where possible the masses are extracted
  from  the same initial time and so are not always the same as our best
  mass estimates listed in other tables.} 
\label{table_Vcheckmm_b6.235}
\end{table}

\begin{table}[htb]
\centering
\begin{tabular}{|c|cc|} \hline
\multicolumn{3}{|c|}{$aM$} \\ \hline
$R^{PC}$   & $\beta=5.6924$ & $\beta=5.80$ \\ \hline
$A_1^{++}$ & 0.987(9) &  0.908(12)  \\ 
          & 1.89(8)  &  1.632(38)  \\ 
          &          &  1.94(8)  \\   \hline
$E^{++}$  & 1.95(5)  &  1.550(17)  \\ 
          &          &  2.09(7)  \\   \hline
$T_2^{++}$ & 2.02(5)  & 1.586(23)   \\   \hline
$A_1^{-+}$ & 2.36(14) & 1.764(49)   \\   \hline
$E^{-+}$   &          & 2.09(7)   \\  \hline
$T_2^{-+}$ &          & 2.03(5)   \\  \hline
$T_1^{+-}$ &          & 1.938(32)   \\  \hline \hline
$aE_l$    & 1.491(47) & 1.1126(50)   \\  \hline
\end{tabular}
\caption{Glueball masses at $\beta=5.6924$ and $\beta=5.80$ on a $10^316$ and $12^316$ lattices
  respectively, for all octahedral representations $R$, and for both values of parity, P,
  and charge conjugation, $C$. Also the energy $E_l$ of a winding flux tube.}
\label{table_M_b5.80_b5.6924}
\end{table}

\begin{table}[htb]
\centering
\begin{tabular}{|c|c|c|c|c|} \hline
\multicolumn{5}{|c|}{$aM$ , $\beta=5.8941$} \\ \hline
  $R$   & P=+,C=+ & P=-,C=+ &  P=+,C=-   &  P=-,C=-   \\ \hline
$A_1$ & 0.799(10)  & 1.440(25) & 2.67(25)  &  2.74(51)  \\ 
      & 1.345(14)  & 1.98(9)   &    &    \\
      & 1.72(4)   &    &    &    \\
      & 1.83(5)   &    &    &    \\  \hline
$A_2$ & 1.933(60)   & 2.62(25)  &  1.95(8)  & 2.10(12   \\  \hline
$E$   & 1.280(15)   & 1.661(33) &  2.27(8)  & 2.22(8)   \\
      & 1.747(24)   & 2.07(8)   &    &    \\
      & 1.814(36)   &    &    &    \\ 
      & 1.976(48)   &    &    &    \\  \hline
$T1$  & 1.924(40)   & 2.20(6)   & 1.591(18) &  2.07(5)  \\
      & 1.976(56)   & 2.17(8)   & 1.86(6)   &  2.22(12)  \\
      & 2.35(11)    & 2.19(10)  & 2.05(5)   &    \\   \hline
$T2$  & 1.276(11)   & 1.669(29) & 1.907(23) &  2.18(8)  \\
      & 1.796(28)   & 2.05(5)   & 2.14(6)   &    \\
      & 1.913(24)   &           &    &    \\ 
      & 1.988(40)   &           &    &    \\  \hline 
\multicolumn{5}{|c|}{$aE_l = 0.8770(27)$} \\ \hline
\end{tabular}
\caption{Glueball masses at $\beta=5.8941$  on a $14^316$ lattice
  for all octahedral representations, $R$, and for both values of parity, P,
  and charge conjugation, $C$. Also the energy $E_l$ of a winding flux tube.}
\label{table_M_B5.8941}
\end{table}

\begin{table}[htb]
\centering
\begin{tabular}{|c|c|c|c|c|} \hline
\multicolumn{5}{|c|}{$aM$ , $\beta=5.99$} \\ \hline
  $R$   & P=+,C=+ & P=-,C=+ &  P=+,C=-   &  P=-,C=-   \\ \hline
$A_1$ & 0.7045(65)  &  1.169(15) & 2.14(11)   & 2.37(13)   \\ 
      & 1.234(13)   &  1.69(5)  &    &    \\
      & 1.543(27)   &  1.93(8)  &    &    \\ 
      & 1.651(28)   &  1.82(8)  &    &    \\ 
      & 1.729(41)   &           &    &    \\  \hline 
$A_2$ & 1.674(40)   &  1.94(12) & 1.601(38) &  2.00(7)  \\
      & 1.79(6)     &  2.21(14) & 1.88(5)   &   \\
      &             &           & 2.04(7)   &    \\  \hline
$E$   & 1.082(8)    &  1.442(20)  & 1.904(42) &  1.771(26)  \\
      & 1.490(16)   &  1.790(35)  & 2.17(7)   &  2.02(6)  \\
      & 1.611(27)   &  2.12(6)    &           &  2.21(10)  \\ 
      & 1.627(28)   &             &           &     \\  \hline
$T1$  & 1.713(22)   & 1.983(43)   & 1.345(13)   &  1.747(37)  \\
      & 1.715(25)   & 2.02(5)     & 1.634(21)   &  2.09(5)  \\
      & 1.97(6)     & 1.98(4)     & 1.736(24)   &  2.15(7)  \\ 
      & 2.02(5)     & 2.16(5)     & 1.94(4)     &    \\  \hline
$T2$  & 1.083(6)    & 1.411(12)   & 1.597(21)   &  1.792(30)  \\
      & 1.459(16)   & 1.767(25)   & 1.871(36)   &  1.935(47)  \\
      & 1.669(21)   & 1.849(38)   & 1.90(5)     &  2.103(53)  \\ 
      & 1.657(24)   & 1.934(37)   &    &    \\ 
      & 1.756(24)   & 1.994(46)   &    &    \\  \hline
\multicolumn{5}{|c|}{$aE_l = 0.8095(61)$} \\ \hline
\end{tabular}
\caption{Glueball masses at $\beta=5.99$  on a $18^4$ lattice
  for all octahedral representations, $R$, and for both values of parity, P,
  and charge conjugation, $C$. Also the energy $E_l$ of a winding flux tube.}
\label{table_M_b5.99}
\end{table}

\begin{table}[htb]
\centering
\begin{tabular}{|c|c|c|c|c|} \hline
\multicolumn{5}{|c|}{$aM$ , $\beta=6.0625$} \\ \hline
  $R$   & P=+,C=+ & P=-,C=+ &  P=+,C=-   &  P=-,C=-   \\ \hline
$A_1$ & 0.6365(43)  & 1.001(32)  & 1.98(7)   & 1.93(8)   \\ 
      & 1.111(11)   & 1.419(26)  & 2.13(8)   & 1.99(9)   \\
      & 1.384(28)   & 1.83(7)    & 2.05(7)   & 2.15(12)   \\ 
      & 1.467(17)   & 1.75(6)    &    &    \\ 
      & 1.482(31)   &            &    &    \\ 
      & 1.54(3)     &            &    &    \\  \hline
$A_2$ & 1.476(31)   & 1.85(7)  & 1.433(21) & 1.667(38)   \\
      & 1.649(29)   & 1.94(8)  & 1.68(4)   & 1.888(78)   \\ 
      & 1.81(6)     &          & 1.85(6)   & 2.05(9)    \\  \hline
$E$   & 0.9623(61)  & 1.213(33) & 1.710(33)  & 1.558(19)   \\
      & 1.298(8)    & 1.60(3)   & 1.891(40)  & 1.824(32)   \\
      & 1.432(15)   & 1.81(5)   & 1.845(42)  & 1.98(5)   \\ 
      & 1.465(11)   &           &    &    \\ 
      & 1.514(19)   &           &    &    \\  \hline
$T1$  & 1.518(15)  & 1.689(21)  & 1.194(8)  & 1.588(22)   \\
      & 1.54(2)    & 1.727(23)  & 1.435(8)  & 1.81(3)   \\
      & 1.74(3)    & 1.754(20)  & 1.547(16) & 1.86(4)   \\ 
      &            & 1.91(4)    & 1.695(17) &    \\  \hline
$T2$  & 0.948(13)  & 1.270(15)  & 1.418(11) &  1.47(9)  \\
      & 1.335(10)  & 1.575(16)  & 1.701(29) &  1.58(14)  \\
      & 1.487(13)  & 1.56(9)    & 1.656(25) &  1.81(3)  \\ 
      & 1.482(11)  &            & 1.83(3)   &  1.98(5)  \\ 
      & 1.567(17)  &    &    &    \\  \hline
\multicolumn{5}{|c|}{$aE_l = 0.7040(42)$} \\ \hline
\end{tabular}
\caption{Glueball masses at $\beta=6.0625$  on a $20^4$ lattice
  for all representations, $R$, of the cube, and for both values of parity, P,
  and charge conjugation, $C$. Also the energy $E_l$ of a winding flux tube.}
\label{table_M_b6.0625}
\end{table}

\begin{table}[htb]
\centering
\begin{tabular}{|c|c|c|c|c|} \hline
\multicolumn{5}{|c|}{$aM$ , $\beta=6.235$} \\ \hline
  $R$   & P=+,C=+ & P=-,C=+ &  P=+,C=-   &  P=-,C=-   \\ \hline
$A_1$ & 0.4969(29) & 0.781(13)  &  1.431(20) &  1.512(21)   \\
      & 0.8626(52) & 1.132(11)  &  1.482(25) &  1.588(27)   \\
      & 1.096(10)  & 1.437(20)  &  1.77(4)   &     \\
      & 1.134(11)  & 1.485(22)  &            &     \\
      & 1.054(27)  & 1.597(33)  &            &     \\
      & 1.161(36)  &            &            &     \\ \hline
$A_2$ & 1.082(32)  & 1.404(21)  &  1.095(9)   &  1.340(18)   \\
      & 1.25(6)    & 1.45(13)   &  1.307(18)  &  1.57(3)  \\
      & 1.429(16)  &            &  1.395(19)  &  1.67(4)   \\ \hline
$E$   & 0.7342(45) & 0.926(13)  &  1.306(10)  &  1.186(10)   \\
      & 1.013(6)   & 1.249(8)   &  1.416(13)  &  1.38(6)   \\
      & 1.067(25)  & 1.422(15)  &  1.455(16)  &  1.41(8)   \\
      & 1.122(8)   &            &  1.39(7)    &  1.49(8)   \\
      & 1.206(11)  &            &             &  1.587(21)   \\ \hline
$T1$  & 1.148(7)   & 1.226(30)  &  0.897(10)  &  1.163(24)   \\
      & 1.137(23)  & 1.302(11)  &  1.071(18)  &  1.305(31)   \\
      & 1.407(12)  & 1.337(10)  &  1.169(8)   &  1.490(12)   \\
      & 1.392(51)  & 1.463(65)  &  1.307(33)  &  1.541(16)   \\
      & 1.436(13)  &            &             &     \\ \hline
$T2$  & 0.7374(31) & 0.962(5)   &  1.087(22)  &  1.200(28)   \\
      & 1,0247(52) & 1.219(6)   &  1.258(27)  &  1.283(25)   \\
      & 1.096(18)  & 1.233(36)  &  1.333(12)  &  1.50(6)   \\
      & 1.152(8)   & 1.333(11)  &  1.357(13)  &  1.557(17)   \\
      & 1.145(25)  &            &  1.450(13)  &  1.54(7)   \\ \hline
\multicolumn{5}{|c|}{$aE_l = 0.5435(23)$} \\ \hline
\end{tabular}
\caption{Glueball masses at $\beta=6.235$  on a $26^4$ lattice
  for all octahedral representations, $R$, and for both values of parity, P,
  and charge conjugation, $C$. Also the energy $E_l$ of a winding flux tube.}
\label{table_M_b6.235}
\end{table}

\begin{table}[htb]
\centering
\begin{tabular}{|c|c|c|c|c|} \hline
\multicolumn{5}{|c|}{$aM$ , $\beta=6.3380$} \\ \hline
  $R$   & P=+,C=+ & P=-,C=+ &  P=+,C=-   &  P=-,C=-   \\ \hline
$A_1$ & 0.4276(37) & 0.688(8)  & 1.234(46) &  1.327(17)  \\
      & 0.7452(45) & 0.956(22) & 1.318(62) &  1.277(50)   \\
      & 0.9386(88) & 1.16(5)   & 1.38(9)   &  1.39(6)  \\
      & 0.939(18)  & 1.19(5)   &           &    \\
      & 0.974(17)  &   &   &    \\
      & 0.971(18)  &   &   &   \\  \hline
$A_2$ & 0.992(8)   & 1.239(15) & 0.963(10)  & 1.140(32)   \\
      & 1.084(41)  & 1.411(20) & 1.161(11)  & 1.335(70)   \\
      & 1.232(44)  & 1.43(9)   & 1.152(33)  &    \\
      & 1.191(47)  &           & 1.28(6)    &    \\  \hline
$E$   & 0.6356(27) & 0.8099(63) & 1.111(29)  & 1.008(21) \\
      & 0.8634(84) & 1.061(17)  & 1.238(28)  & 1.193(32) \\
      & 0.934(14)  & 1.179(32)  & 1.261(38)  & 1.280(38) \\
      & 0.961(13)  & 1.230(39)  & 1.243(35)  & 1.33(4)   \\
      & 0.967(16)  &            & 1.30(6)    &    \\  \hline
$T1$  & 0.995(13)  & 1.108(18) & 0.7797(60) & 1.062(15)   \\
      & 0.995(14)  & 1.119(20) & 0.923(13)  & 1.165(17)   \\
      & 1.263(7)   & 1.110(20) & 1.020(5)   & 1.254(26)   \\
      & 1.264(30)  & 1.25(4)   & 1.158(9)   & 1.288(32)   \\
      &            & 1.34(6)   & 1.201(22)  &    \\  \hline
$T2$  & 0.6369(29) & 0.8207(86) & 0.952(11) & 1.056(14)   \\
      & 0.8856(39) & 1.031(15)  & 1.131(22) & 1.110(16)   \\
      & 0.963(12)  & 1.169(19)  & 1.157(7)  & 1.296(37)   \\
      & 0.968(10)  & 1.142(22)  & 1.202(22) & 1.255(30)   \\
      & 0.974(12   & 1.164(21)  & 1.269(28) & 1.29(4)    \\  \hline
\multicolumn{5}{|c|}{$aE_l = 0.4652(21)$} \\ \hline
\end{tabular}
\caption{Glueball masses at $\beta=6.3380$  on a $30^4$ lattice
  for all octahedral representations, $R$, and for both values of parity, P,
  and charge conjugation, $C$. Also the energy $E_l$ of a winding flux tube.}
\label{table_M_b6.3380}
\end{table}

\begin{table}[htb]
\centering
\begin{tabular}{|c|c|c|c|c|} \hline
\multicolumn{5}{|c|}{$aM$ , $\beta=6.50$} \\ \hline
  $R$   & P=+,C=+ & P=-,C=+ &  P=+,C=-   &  P=-,C=-   \\ \hline
$A_1$ & 0.3474(22) & 0.5568(50) & 0.992(24)  & 1.016(47)   \\
      & 0.6036(36) & 0.761(12)  & 1.150(11)  & 1.034(36)   \\
      & 0.7629(52) & 0.948(21)  & 1.092(21)  & 1.037(74)   \\
      & 0.768(10)  & 0.948(17)  & 1.169(34)  & 1.136(25)   \\
      & 0.815(6)   &   &   &    \\
      & 0.842(6)   &   &   &    \\  \hline
$A_2$ & 0.795(10)  & 1.012(26)  & 0.763(11) &  0.915(13)  \\
      & 0.894(17)  & 1.081(33)  & 0.913(8)  &  1.055(10)  \\
      & 1.058(27)  & 1.088(26)  & 0.932(18) &  1.067(34)  \\
      & 1.004(16)  & 1.234(49)  & 1.054(22) &  1.215(13)  \\  \hline
$E$   & 0.5054(32) & 0.657(8)   & 0.916(11) & 0.839(9)   \\
      & 0.6971(70) & 0.837(9)   & 0.908(37) & 0.975(13)   \\
      & 0.779(9)   & 0.984(14)  & 1.058(15) & 1.083(16)   \\
      & 0.785(6)   & 1.004(19)  & 1.040(14) & 1.121(24)   \\
      & 0.810(9)   &   &   &    \\  \hline
$T1$  & 0.813(8)  & 0.894(9)  & 0.636(5)  & 0.859(10)   \\
      & 0.806(12) & 0.914(12) & 0.758(7)  & 0.987(12)   \\
      & 0.943(9)  & 0.910(27) & 0.811(6)  & 0.993(13)   \\
      & 0.951(8)  & 0.996(27) & 0.911(10) & 1.058(18)   \\
      &           & 0.997(35) & 0.959(9)  & 1.034(18)  \\  \hline
$T2$  & 0.5057(19) & 0.6638(63) & 0.734(11)  & 0.854(7)     \\
      & 0.6959(44) & 0.840(8)   & 0.897(10)  & 0.945(9)     \\
      & 0.8083(73) & 0.939(10)  & 0.921(8)   & 0.998(10)   \\  
      & 0.806(9)   & 0.952(8)   & 0.955(10)  & 1.051(9)     \\  
      & 0.859(8)   & 0.976(7)   & 1.000(15)  & 1.125(8)     \\  \hline
\multicolumn{5}{|c|}{$aE_l = 0.3811(19)$} \\ \hline
\end{tabular}
\caption{Glueball masses at $\beta=6.50$  on a $38^4$ lattice
  for all octahedral representations, $R$, and for both values of parity, P,
  and charge conjugation, $C$. Also the energy $E_l$ of a winding flux tube.}
\label{table_M_b6.50}
\end{table}

\clearpage

\begin{table}[htb]
\centering
\begin{tabular}{|c|c|c|c|c|} \hline
\multicolumn{5}{|c|}{$M_G/\surd\sigma$  continuum limit} \\ \hline
  $R$   & P=+,C=+ & P=-,C=+ &  P=+,C=-   &  P=-,C=-   \\ \hline
A1  & 3.405(21) & 5.276(45) & 9.32(28)  &  9.78(46)  \\
    & 5.855(41) & 7.29(13)  &           &  10.39(37) \\
    & 7.515(50) & 9.18(26)  &           &           \\ 
    & 7.38(11)  & 9.37(22)  &           &            \\  \hline
A2  & 7.705(85) & 9.80(22)  & 7.384(90) &  8.96(15)  \\
    & 8.61(20)  & 11.17(30) & 8.94(10)  &  10.21(20) \\ 
    &           &           & 8.90(21)  &             \\  \hline
E   & 4.904(20) & 6.211(56) & 8.77(12)  &  7.91(10)  \\
    & 6.728(47) & 8.23(9)   & 9.03(23)  &  9.39(18)  \\ 
    & 7.49(9)   & 9.47(16)  & 10.39(21) &  10.40(22) \\ 
    & 7.531(60) &           &           &            \\  \hline
T1  & 7.694(81) & 8.48(12)  & 6.065(40) &  8.31(10)  \\
    & 7.72(11)  & 8.57(13)  & 7.21(8)   &  9.30(14)*  \\ 
    &           & 8.66(15)  & 7.82(6)   &  9.72(15)  \\ 
    &           & 9.56(28)  & 8.92(10)  &            \\  \hline
T2  & 4.884(19) & 6.393(45) & 7.22(9)   &  8.198(80) \\
    & 6.814(31) & 8.15(7)   & 8.72(11)  &  8.99(11)*  \\ 
    & 7.716(70)* & 9.23(12)* & 9.06(8)   &  9.69(13)  \\ 
    & 7.677(71) &           &  &   \\  \hline
\end{tabular}
\caption{Continuum limit of glueball masses in units of the string tension,
  for all representations, $R$, of the rotation symmetry of a cube, for
  both values of parity, P,
  and charge conjugation, $C$. Ground states and some ecited states. Stars
  indicate poor fits ($\chi^2$ per degree of freedom greater than 2.5).}
\label{table_MK_R}
\end{table}

\begin{table}[htb]
\centering
\begin{tabular}{|c|c|c|c|c|} \hline
\multicolumn{5}{|c|}{$M_G \, GeV$} \\ \hline
  $R$   & P=+,C=+ & P=-,C=+ &  P=+,C=-   &  P=-,C=-   \\ \hline
A1  & 1.651(23) & 2.599(39) & 4.52(15)  &  4.74(23)  \\
    & 2.840(40) & 3.54(8)   &           &  5.04(19) \\
    & 3.65(6)   & 4.45(14)  &           &           \\ 
    & 3.58(15)  & 4.54(12)  &           &            \\  \hline
A2  & 3.74(7)   & 4.75(13)  & 3.58(7)   &  4.35(9)  \\
    & 4.18(11)  & 5.42(16)  & 4.34(8)   &  4.95(12) \\ 
    &           &           & 4.32(12)  &             \\  \hline
E   & 2.378(31) & 3.012(46) & 4.25(8)   &  3.84(7)  \\
    & 3.26(5)   & 3.99(7)   & 4.38(13)  &  4.55(11)  \\ 
    & 3.63(7)   & 4.59(10)  & 5.04(12)  &  5.04(13) \\ 
    & 3.65(6)   &           &           &            \\  \hline
T1  & 3.73(6)   & 4.11(8)   & 2.942(41) &  4.03(17)  \\
    & 3.74(7)   & 4.16(8)   & 3.50(6)   &  4.51(9)*  \\ 
    &           & 4.20(9)   & 3.79(6)   &  4.71(10)  \\ 
    &           & 4.64(15)  & 4.33(8)   &            \\  \hline
T2  & 2.369(31) & 3.101(44) & 3.50(6)   &  3.98(7) \\
    & 3.30(5)   & 3.95(6)   & 4.23(8)   &  4.36(8)*  \\ 
    & 3.74(6)*  & 4.48(8)*  & 4.39(7)   &  4.70(9)  \\ 
    & 3.72(6)   &           &           &            \\  \hline
\end{tabular}
\caption{Continuum glueball masses in GeV units using the value
  $r_0=472(5)fm$ from \cite{RS-2014} to set the scale. Errors are statistical only.}
\label{table_MGeV_R}
\end{table}

\begin{table}[htb]
\centering
\begin{tabular}{|llc|} \hline
\multicolumn{3}{|c|}{continuum $J^{PC}$ from cubic $R$} \\ \hline
$J^{PC}$    &        &  cubic $R^{PC}$  \\  \hline
$0^{++}$gs  & $\sim$ & $A1^{++}$gs     \\ 
$0^{++}$ex1 & $\sim$ & $A1^{++}$ex1    \\ 
$2^{++}$gs  & $\sim$ & $E^{++}$gs + $T2^{++}$gs    \\  
$2^{++}$ex1 & $\sim$ & $E^{++}$ex1 + $T2^{++}$ex1    \\ 
$3^{++}$gs  & $\sim$* & $A2^{++}$gs + $T1^{++}$gs(ex1) + $T2^{++}$ex3(ex2)    \\ 
$4^{++}$gs  & $\sim$* & $A1^{++}$ex2 + $E^{++}$ex2 +$T1^{++}$ex1(gs) + $T2^{++}$ex2(ex3)   \\ \hline
 $0^{-+}$gs  &  $\sim$  &  $A1^{-+}$gs     \\ 
 $0^{-+}$ex1 &  $\sim$  &  $A1^{-+}$ex1    \\
 $2^{-+}$gs  &  $\sim$  &  $E^{-+}$gs + $T2^{-+}$gs   \\
 $2^{-+}$ex1 &  $\sim$  &  $E^{-+}$ex1 + $T2^{-+}$ex1   \\ 
 $1^{-+}$gs  &  $\sim$  & $T1^{-+}$gs    \\ 
 $1^{-+}$ex1 &  $\sim$* & $T1^{-+}$ex1    \\  
 $1^{-+}$ex2 &  $\sim$* & $T1^{-+}$ex2 \\ \hline
 $2^{+-}$gs  &  $\sim$* & $E^{+-}$gs + $T2^{+-}$ex1 \\
 $1^{+-}$gs  &  $\sim$  & $T1^{+-}$gs   \\ 
 $1^{+-}$ex1 &  $\sim$  & $T1^{+-}$ex2   \\ 
 $3^{+-}$gs  &  $\sim$  & $A2^{+-}$gs + $T1^{+-}$ex1 + $T2^{+-}$gs   \\ 
 $4^{+-}$gs  &  $\sim$** & $A1^{+-}$gs + $E^{+-}$ex1 +$T1^{+-}$ex3 + $T2^{+-}$ex2   \\ \hline
 $2^{--}$gs  &  $\sim$  & $E^{--}$gs + $T2^{--}$gs \\
 $1^{--}$gs  &  $\sim$  & $T1^{--}$gs   \\ \hline
\end{tabular}
\caption{Identification of continuum $J^{PC}$ states from the results for the cubic representations
  in  Table~\ref{table_MK_R}. Ground state denoted by $gs$, $i$'th excited state by $exi$. Where
  there is some ambiguity, a single star denotes 'likely'
  while two stars indicate 'significant uncertainty'.}
\label{table_M_J_R}
\end{table}

\begin{table}[htb]
\centering
\begin{tabular}{|l|c|c|c|c|} \hline
\multicolumn{5}{|c|}{$M_G/\surd\sigma$ continuum limit} \\ \hline
  $J$   & P=+,C=+ & P=-,C=+ &  P=+,C=-   &  P=-,C=-   \\ \hline
 0 gs   & 3.405(21)   & 5.276(45) &           &    \\
 0 ex1  & 5.855(41)   & 7.29(13)  &           &    \\
 2 gs   & 4.894(22)   & 6.32(9)   & 8.74(12)*  &  8.08(15)   \\
 2 ex1  & 6.788(40)   & 8.18(8)   &           &    \\
 1 gs   &             & 8.48(12)  & 6.065(40) &  8.31(10)  \\
 1 ex1  &             & 8.57(13)* & 7.82(6)   &    \\
 1 ex2  &             & 8.66(15)* &           &    \\
 3 gs   & 7.71(9)*    &           & 7.27(12)   &    \\
 4 gs   & 7.60(12)*   &           & 9.02(10)** &    \\ \hline
\multicolumn{5}{|c|}{$r_0\surd\sigma = 1.160(6)$} \\ \hline
\end{tabular}
\caption{Continuum limit of glueball masses, in units of the string tension,
  for those $J^{PC}$ representations we can identify. Ground state denoted by $gs$,
  $i$'th excited state by $exi$. Stars denote uncertainties.
  To change scale from $\surd\sigma$ to $r_0$ use value indicated.}
\label{table_MK_J}
\end{table}

\begin{table}[htb]
\centering
\begin{tabular}{|l|l|l|l|l|} \hline
\multicolumn{5}{|c|}{$M_G \, GeV$} \\ \hline
  $J$   & P=+,C=+ & P=-,C=+ &  P=+,C=-   &  P=-,C=-   \\ \hline
 0 gs   & 1.653(26)   & 2.561(40) &           &    \\
 0 ex1  & 2.842(40)   & 3.54(8)   &           &    \\
 2 gs   & 2.376(32)   & 3.07(6)   & 4.24(8)*  &  3.92(9)   \\
 2 ex1  & 3.30(5)     & 3.97(7)   &           &    \\
 1 gs   &             & 4.12(8)   & 2.944(42) &  4.03(7)  \\ 
 1 ex1  &             & 4.16(8)*  & 3.80(6)   &    \\
 1 ex2  &             & 4.20(9)*  &           &    \\
 3 gs   & 3.74(7)*    &           & 3.53(8)   &    \\
 4 gs   & 3.69(8)*    &           & 4.38(8)** &    \\ \hline
\end{tabular}
\caption{Glueball masses in physical {\bf GeV} units using $r_0=0.472(5)fm$.
  For those glueballs whose $J^{PC}$ we can identify. Ground state denoted by $gs$,
  $i$'th excited state by $exi$. Stars denote a level of uncertainty as in
  Table~\ref{table_J_R}. Errors are statistical only.}
\label{table_MGeV_J}
\end{table}

\begin{table}[htb]
\centering
\begin{tabular}{|rcr|cccccc|} \hline
\multicolumn{3}{|c|}{$Q_I$(2 cools)}  &  \multicolumn{6}{|c|}{num conf with $Q_I$(50 cools)}   \\ \hline
  $Q_I$ & $Q\in$ & num conf & -2  & -1 & 0 & 1 & 2 & 3 \\ \hline
 -2 & [-2.20,-1.25] & 176   &  170  & 6   &  0   &  0  &   0   &  0  \\
 -1 & [-1.25,-0.40] & 474   &   0   & 472 &  2   &  0  &   0   &  0  \\
  0 & [-0.40,0.40]  & 741   &   0   &  1  & 739  &  1  &   0   &  0  \\
  1 & [0.40,1.25]   & 658   &   0   &  0  &  5   & 653 &   0   &  0 \\
  2 & [1.25,2.20]   & 330   &   0   &  0  &  0   &  4  & 325   &  1 \\
  3 & [2.20,3.10]   & 225   &   0   &  0  &  0   &  0  &   1   & 224 \\ \hline
\end{tabular}
\caption{Number fields with topological charge $Q_I$ after 50 cooling sweeps of fields with the specified topological
  charge after 2 cooling sweeps. On a $26^338$ lattice at $\beta=6.50$. Charges $Q_I$ at 2 cools
  assigned when lattice charge $Q$ lies in the intervals shown.}
\label{table_QI2_QI50}
\end{table}

\begin{table}[htb]
\centering
\begin{tabular}{|c|ll|ll||c|ll|} \hline
  &  \multicolumn{2}{|c|}{$\beta=6.50$}  &  \multicolumn{2}{|c||}{$\beta=6.0625$} &  \multicolumn{3}{|c|}{$\beta=6.50$} \\ \hline
cools & $\langle{Q^2}\rangle$ & $\langle{Q_I^2}\rangle$ & $\langle{Q^2}\rangle$ & $\langle{Q_I^2}\rangle$ & cools &  $\langle{Q^2}\rangle$ & $\langle{Q_I^2}\rangle$ \\ \hline
4   & 1.804  & 2.139   & 2.002  &  2.679 & 1  & 1.518  &     \\
8   & 1.931  & 2.137   & 2.192  &  2.684 & 2  & 1.845  & 2.543      \\
12  & 1.975  & 2.136   & 2.257  &  2.669 & 3  & 2.054  & 2.555      \\
16  & 1.998  & 2.135   & 2.290  &  2.657 & 4  & 2.158  &  2.555      \\
20  & 2.012  & 2.135   & 2.312  &  2.649 & 5  & 2.218  &  2.555      \\
    &        &         &        &        & 10  & 2.336  & 2.548     \\
    &        &         &        &        & 20  & 2.400  & 2.548     \\
    &        &         &        &        & 30  & 2.424  & 2.547     \\
    &        &         &        &        & 40  & 2.437  & 2.547     \\
    &        &         &        &        & 50  & 2.444  & 2.544     \\  \hline
error & 0.20 & 0.21    & 0.022  & 0.026 & error & 0.46  &  0.48   \\ \hline
\end{tabular}
\caption{Topological susceptibility versus number of cooling sweeps, using
  the lattice $Q$ and the integer-valued projection, $Q_I$. 
  For two values of $\beta$ with a separate calculation at $\beta=6.50$
  for a wider range of cooling sweeps. The statistical errors are highly correlated
  between cooling sweeps.}
\label{table_Q_QI}
\end{table}

\begin{table}[htb]
\centering
\begin{tabular}{|ll|l|} \hline
\multicolumn{3}{|c|}{$\langle{Q}_{hot}\rangle = Z(\beta) Q_I$} \\ \hline
$\beta$ & $Z(\beta)$ & $Z_{1-loop}(\beta)$ \\ \hline
5.6924 & 0.0636(18) & 0.0266 \\
5.800  & 0.0869(30) & 0.0447 \\
5.990  & 0.130(18)  & 0.0750 \\
6.0625 & 0.147(7)   & 0.0861 \\
6.235  & 0.178(13)  & 0.1113 \\
6.3380 & 0.207(15)  & 0.1258 \\
6.500  & 0.236(23)  & 0.1476 \\ \hline
\end{tabular}
\caption{Multiplicative renormalisation factor, $Z(\beta)$, relating 
  the average lattice topological charge, $Q_{hot}$,  calculated on the rough
  Monte Carlo fields and the integer valued topological charge $Q_I$ calculated
  after 20 `cooling' sweeps of those fields. Also the 1-loop perturbative
  prediction. At the values of $\beta$ shown.}
\label{table_Z}
\end{table}

\begin{table}[htb]
\centering
\begin{tabular}{|lc|cc|cc|} \hline
$\beta$ & lattice & $\langle Q_I^2 \rangle$ & $\langle Q^2 \rangle$
& $\chi^{1/4}_I/\surd\sigma$ & $\chi^{1/4}/\surd\sigma$\\ \hline
5.6924 & $8^316$  & 5.452(29) &  4.034(21) & 0.4016(59)  & 0.3725(54) \\
5.80   & $10^316$ & 5.111(16) &  4.021(13) & 0.4222(10)  & 0.3976(9) \\
5.8941 & $12^316$ & 4.352(43) &  3.570(35) & 0.4289(13)  & 0.4081(12) \\
5.99   & $14^320$ & 4.428(56) &  3.790(48) & 0.4312(21)  & 0.4147(20) \\
6.0625 & $14^320$ & 2.649(26) &  2.312(22) & 0.4281(16)  & 0.4138(15) \\
6.235  & $18^326$ & 2.55(13)  &   2.32(12) & 0.4268(54)  & 0.4169(54) \\
6.3380 & $22^330$ & 2.94(14)  &   2.72(13) & 0.4261(51)) & 0.4178(50) \\
6.50   & $26^338$ & 2.14(23)  &   2.01(22) & 0.4075(106) & 0.4011(97) \\ \hline
$\infty$   &  $\beta\ge 5.8941$   &   &    & 0.4259(28) & 0.4203(26) \\ 
           &  $\beta\ge 5.99$     &   &    & 0.4187(53) & 0.4125(47)  \\ \hline
\end{tabular}
\caption{Fluctuations of the topological charge after 20 cooling sweeps.
  $Q$ is the lattice value while $Q_I$ is the corresponding
  integer value. Also the corresponding susceptibilities in units of
  the string tension, and their continuum limits where the fits cover the
  range of $\beta$ indicated.}
\label{table_QQkhi}
\end{table}

\begin{table}[htb]
\centering
\begin{tabular}{|cc|cc|cc|} \hline
\multicolumn{6}{|c|}{$a\chi^{1/4}_I$} \\ \hline
\multicolumn{2}{|c|}{$\beta=5.8941$} & \multicolumn{2}{|c|}{$\beta=5.99$} & \multicolumn{2}{|c|}{$\beta=6.235$}\\ \hline
 $10^316$   &  0.11258(25)   &  $14^320$  &  0.09478(30)   &  $18^326$ &  0.0640(9)   \\ 
 $12^316$   &  0.11201(28)   &  $18^4$    &  0.09435(17)   &  $26^4$   &  0.0632(8)   \\ 
 $14^316$   &  0.11203(23)   &            &                &           &              \\ \hline
\end{tabular}
\caption{The topological susceptibility, $\chi_I$, on different space-time volumes at the
values of $\beta$ shown.}
\label{table_khiV}
\end{table}

\begin{table}[htb]
\centering
\begin{tabular}{|lc|cc|} \hline
$\beta$ & lattice & $\tau_{|\Delta Q|=1}$ & $\tilde{\tau}_{|\Delta Q|=1}$ \\ \hline
6.0625 & $14^320$ & 36(2)    &  37(2)  \\
6.235  & $18^326$ & 123(2)   &  130(2) \\
6.3380 & $22^330$ & 255(4)   &  231(4) \\
6.50   & $26^338$ & 1323(50) &  1382(52) \\ \hline
\end{tabular}
\caption{Average number of sweeps between changes in the topological charge
  by one unit, $\tau_{|\Delta Q|=1}$, on lattice volumes shown;
  $\tilde{\tau}_{|\Delta Q|=1}$ is corresponding value scaled to a standard lattice volume
  $V=(3/\surd\sigma)^4$ at each value of $\beta$.}
\label{table_tau}
\end{table}

\clearpage

\begin{figure}[htb]
\begin	{center}
\leavevmode
\input	{plot_EeffGllA1++_l18.tex}
\end	{center}
\caption{Effective energies of the lightest $A1^{++}$ states from the basis of the usual
  single loop operators, $\bullet$, and from the same basis but extended with our
  ditorelon operators, $\circ$. Horizontal line is the energy of the lightest ditorelon state
  neglecting interactions between the torelons. On a $18^326$ lattice
  at $\beta=6.235$. }
\label{fig_EeffGllA1++_l18}
\end{figure}

\begin{figure}[htb]
\begin	{center}
\leavevmode
\input	{plot_EeffGllE++_l18.tex}
\end	{center}
\caption{Effective energies of the lightest $E^{++}$ states from the basis of the usual
  single loop operators, $\bullet$, and from the same basis but extended with our
  ditorelon operators, $\circ$.  Horizontal line is the energy of the lightest ditorelon state
  neglecting interactions between the torelons. On a $18^326$ lattice
  at $\beta=6.235$.}
\label{fig_EeffGllE++_l18}
\end{figure}

\begin{figure}[htb]
\begin	{center}
\leavevmode
\input	{plot_EeffGllA1++_l26.tex}
\end	{center}
\caption{Effective energies of the lightest $A1^{++}$ states from the basis of the usual
  single loop operators (filled points) and from the same basis but extended with our
  ditorelon operators (unfilled points). Horizontal line is the energy of the lightest
  ditorelon state neglecting interactions between the torelons. On a $26^326$ lattice
  at $\beta=6.235$.}
\label{fig_EeffGllA1++_l26}
\end{figure}

\begin{figure}[htb]
\begin	{center}
\leavevmode
\input	{plot_EeffGllE++_l26.tex}
\end	{center}
\caption{Effective energies of the lightest $E^{++}$ states from the basis of the usual
  single loop operators (filled points) and from the same basis but extended with our
  ditorelon operators (unfilled points). Horizontal line is the energy of the lightest
  ditorelon state neglecting interactions between the torelons. On a $26^326$ lattice
  at $\beta=6.235$.}
\label{fig_EeffGllE++_l26}
\end{figure}

\clearpage

\begin{figure}[htb]
\begin	{center}
\leavevmode
\input	{plot_MJ2ppKb_cont.tex}
\end	{center}
\caption{Continuum extrapolations of the lightest two $J^{PC}=2^{++}$ glueballs,
  as obtained from the lightest two $E^{++}$ states ($\bullet$, each doubly degenerate)
  and the lightest two $T2^{++}$ states ($\circ$, each triply degenerate).
  Hence the five components of each of the two $J=2$ states.}
\label{fig_MJ2ppK_cont}
\end{figure}

\begin{figure}[htb]
\begin	{center}
\leavevmode
\input	{plot_MJ4ppKb_cont.tex}
\end	{center}
\caption{Continuum extrapolation of the lightest $J^{PC}=4^{++}$ glueball.
  Obtained from the second excited $A1^{++}$ ($\bullet$), the doubly degenerate
  second excited $E^{++}$ ($\circ$), the triply degenerate first excited $T1^{++}$
  ($\lozenge$),  and the triply degenerate second excited $T2^{++}$ ($\star$):
  hence the nine components of the $J=4$ ground state.}
\label{fig_MJ4ppK_cont}
\end{figure}

\begin{figure}[htb]
\begin	{center}
\leavevmode
\input	{plot_MJ2mpKb_cont.tex}
\end	{center}
\caption{Continuum extrapolations of the lightest two $J^{PC}=2^{-+}$ glueballs,
  as obtained from the lightest two $E^{-+}$ states ($\bullet$, each doubly degenerate)
  and the lightest two $T2^{-+}$ states ($\circ$, each triply degenerate).
  Hence the five components of each of the two $J=2$ states.}
\label{fig_MJ2mpK_cont}
\end{figure}

\clearpage

\begin{figure}[htb]
\begin	{center}
\leavevmode
\input	{plot_Q_p_b6.50.tex}
\end	{center}
\caption{Histogram of lattice $Q$ after 4 cooling sweeps, $\circ$,
  and after 20 cooling sweeps, $\bullet$. From 40000 measurements
  at $\beta=6.50$ on a $26^338$ lattice.
  Points corresponding to $N(Q)=0$ are suppressed.}
\label{fig_Q_p_b6.50}
\end{figure}

\begin{figure}[htb]
\begin	{center}
\leavevmode
\input	{plot_Qlong_p_b6.50.tex}
\end	{center}
\caption{Histogram of lattice $Q$  on fields after 2 cooling sweeps, $\circ$,
  and after 20 cooling sweeps, $\bullet$. From 2500 measurements at $\beta=6.50$
  on a $26^338$ lattice. Points corresponding to $N(Q)=0$ are suppressed.}
\label{fig_Qlong_p_b6.50}
\end{figure}

\end{document}